\documentclass[12pt]{article}
\usepackage[utf8]{inputenc}
\usepackage[english]{babel}
\usepackage{amsmath, amssymb,amsfonts,fullpage,amsthm}
\usepackage{array}
\usepackage[dvips]{graphicx}
\usepackage{epsfig,float,color}
\usepackage{subfigure, subfig} 

\usepackage{verbatim, url}
\usepackage{bm,latexsym,mathrsfs,setspace, booktabs}
\usepackage{multirow}
\usepackage{pdflscape}
\usepackage[dvipsnames]{xcolor}
\usepackage[normalem]{ulem} 

\usepackage[authoryear]{natbib}

\usepackage{xr-hyper} 
\usepackage{xr} 

\makeatletter
\newcommand*{\addFileDependency}[1]{
	\typeout{(#1)}%
	\@addtofilelist{#1}
	\IfFileExists{#1}{}{\typeout{No file #1.}}
}
\makeatother

\newcommand*{\myexternaldocument}[1]{%
	\externaldocument{#1}%
	\addFileDependency{#1.tex}%
	\addFileDependency{#1.aux}%
}
\myexternaldocument{supplemental_material}

\newcommand{\calP}{{\mathcal P}}

\newcommand{\ind}{\stackrel{\mathrm{ind}}{\sim}}
\DeclareMathOperator{\Be}{Beta}

\DeclareMathOperator{\logit}{logit}

\newcommand{\sal}[1]{\textcolor{PineGreen}{#1}}
\newcommand{\fq}{\color{black}\rm  }   
\newcommand{\qf}{\color{black}\rm  }
\newcommand{\defi}{\stackrel{\mathrm{def}}{=}}
\newtheorem{proposition}{Proposition}

\begin{document}

	\title{Informed Random Partition Models with Temporal Dependence}
	\author{Sally Paganin\\
		Department of Statistics, The Ohio State University, Columbus, Ohio \\
		and  Garritt L. Page \\
		Department of Statistics, Brigham Young University, Provo, Utah\\
		and 
		Fernando Andr\'es Quintana, Departamento de Estad\'{i}stica \\ Pontificia Universidad Cat\'{o}lica de Chile, Santiago
	}
	
	\date{June, 2025}
	
	\maketitle
	\begin{abstract}
Model-based clustering is a powerful tool often used to discover hidden structure in data by grouping observational units that exhibit similar response values. Recently, clustering methods have been developed that allow the inclusion of an ``initial'' partition of the data informed by expert opinions, starting from a probability distribution on the space of partitions. Then, using some similarity criteria,  partitions different from the initial one are down-weighted, i.e. they are assigned reduced probabilities. We take a different perspective and model the probability that each unit follows the initial partition via auxiliary variables. Our \emph{informed partition model} provides flexibility to include varying levels of uncertainty to any subset of the partition (i.e., locally weighted prior information). Additionally, it can accommodate settings with multiple dependent partitions, such as temporal or multi-view data. Theoretical properties of the proposed construction are explored, which can be useful for prior elicitation. We illustrate the gains in prior specification flexibility via simulation studies and an application to a dataset concerning spatio-temporal evolution of ${\rm PM}_{10}$ measurements in Germany. 		
	\end{abstract}

	
	\noindent%

	\doublespacing

\section{Introduction}\label{sect:intro}

\sloppy 

\fq Cluster analysis 
has become a popular tool for discovering latent data structures. \qf Several approaches that implement cluster analysis have emerged.  One is purely algorithmic, based on assigning individuals/units to groups based on a distance metric (e.g., $k$-means or hierarchical clustering).  A second one incorporates probabilistic group assignment by way of a finite/infinite mixture model.  Using finite or infinite mixtures to carry out clustering has received considerable attention in the statistical literature. A prototypical application of this strategy from a Bayesian perspective involves the adoption of a model that can be expressed as the convolution of a continuous kernel $k(y\mid\theta)$ \fq indexed by a parameter $\theta$, \qf and a discrete random probability measure $G$ that can be represented as $G(\cdot)=\sum_{h=1}^H w_h\delta_{\theta_h}$ with $1< H\le \infty$. Here, $\theta_1,\ldots,\theta_H$ are i.i.d. from some distribution $G_0$ supported on a suitable space $\Theta$, and are independent of the weights $\{w_h\}$, which are required to be a.s. non-negative and to satisfy $\Pr(\sum_{h=1}^H w_h= 1)=1$. The mixture model that so arises adopts the form $p(y\mid \{w_h\}, \{\theta_h\})=\int_{\Theta}k(y\mid\theta)\,dG(\theta)=\sum_{h=1}^H w_h k(y\mid\theta_h)$. Models based on such constructed mixtures have become quite popular in the Bayesian clustering literature, as mixture components are used to define different subsets in a partition. Under this approach, distributions on partitions are {\it induced} from the (in)finite mixture model. Some recent applications are given in \cite{ni-mueller-etal:20}, \cite{lijoi-prunster-rigon:23} and references therein.  For recent reviews in Bayesian cluster analysis, see \cite{wadeReview} and  \cite{grazian2023review}. 

A third Bayesian approach consists of modeling the partitioning of units into groups directly by way of a random partition model \citep[see, e.g., Chapter 8 in][]{MullerQuintana:2015}. This procedure allows the user to define probability models directly on partitions, compared to, e.g., cases where a partition is induced by the adoption of a discrete random probability measure as described above. This modeling approach has been shown to provide great benefit in, for example, the temporal evolution of random partition distributions when a sequence of partitions is available (a case we consider here);  see a discussion of this in \cite{page_dahl_quintana:22}. 
Apart from being seemingly more coherent (i.e., the object of inferential preference is that which receives modeling attention), Bayesian random partition models allow practitioners to include prior information in the clustering more directly.  

An early example of of a random partition modeling is the product partition model (PPM), described in \cite{HAR90}. This model can be specified in a way that its random partition distribution coincides with the well-known clustering properties arising from the popular Dirichlet process (DP) of \cite{FE73}. The PPMx, discussed in \cite{MQR:11} is a modification of the PPM to include covariate dependence in the prior partition distribution, similar to nonparametric regression. The modification is driven by cohesion functions that encourage homogeneity of subsets in terms of covariate values.  A related construction appears in \cite{park&dunson:10}. \cite{argiento2022modelbased} extended the PPMx to clustering distributions that arise from normalized completely random measures \citep{regazzini-completely}. Other options of random partition models include the Ewens-Pitman attraction approach by \cite{dahl-day-tsai:17}, where pairwise similarities are used to specify cluster allocations. \cite{franzolini2023conditional} introduce the class of telescopic clustering models,  based on the notion of conditional partial exchangeability, that allows for dependence among partitions of a given set of subjects but based on different features, which is particularly suitable for multi-view or longitudinal data.  \cite{giampino2024localleveldynamicrandom} propose local-level dynamic random partition models for detecting change points in the context of multivariate time series. Their approach is built on a state equation that describes the time evolution of partitions in a simple Markovian dynamic where the current partition is either identical to or independently sampled from the previous one. 

Recently, \cite{smith_allenby:2020} and \cite{paganin_etal:2021} developed for the first time methods that permit users to inform the prior ``location'' on the partition space by including an initial guess on what the clustering might be. \fq Motivation for these works comes from applications in economy \citep{smith_allenby:2020} and epidemiology \citep{paganin_etal:2021}, where domain knowledge about groupings is available and considered useful for interpretation. Indeed, it is often the case that partial information about  sub-populations is available from previous studies or domain knowledge, but until recently, the lack of appropriate tools prohibited its inclusion in modeling. For example, in the U.S. some states can be confidently grouped based on their historical political leanings {\it a priori}, while more uncertainty accompanies the grouping of swing states. In medical settings, observational data use disease classification codes based on organ systems that can be used to group patients at different levels of detail. However, there is uncertainty in this grouping, as codes may not be consistent across hospital databases \citep{shi2021} or they may not reflect relevant biological information \citep{straub2024hierarchical}. Or, when modeling data with spatial structure, it is natural to group observations on the basis of their relative distances, but again, uncertainty accompanies spatially coherent groupings.  In addition, data in all these examples are often collected over time or across multiple views, motivating the development of methods that are contained herein.\qf 

In this work, we add to the budding literature of \emph{informed partition modeling}, that is, modeling strategies that can accommodate available prior information on a given partition. In particular, we build on ideas in \cite{page_dahl_quintana:22}, who developed a flexible joint probability model for a sequence of temporally correlated partitions. We extend their method to include the notion of an ``initial'' partition that informs the clustering when modeling either one or multiple dependent partitions. Our approach is from a completely different perspective compared to \cite{smith_allenby:2020} and \cite{paganin_etal:2021}, which results in a number of innovations.  \cite{paganin_etal:2021} explores the idea of adjusting partition probabilities based on the distance from the initial partition as defined by some loss function (they employ the variation of information loss by \citealt{meila2007}). This approach is available for any random partition distribution but results in needing to evaluate a normalizing constant that becomes intractable even for relatively small sample sizes.   In addition, it is not possible to customize prior uncertainty for subsets of the initial partition.  \cite{smith_allenby:2020} instead measure proximity to the initial partition by enumerating units that are co-clustered in the estimated partition and in the initial partition.  Their development only considers a particular random partition distribution and depends on the order in which units are allocated.  \fq This approach, too, is not able to treat subsets of the initial partition with more uncertainty than others. Recently, \cite{dahl2023} generalized \cite{smith_allenby:2020}'s approach so that any random partition model can be employed and introduces a permutation parameter that removes dependence on order. Our approach is flexible in the sense that it can be built on any random partition model, does not depend on the order in which units are clustered, and scales relatively well. \qf An additional key contribution we make is that our approach permits eliciting prior uncertainty at the unit level.  
The appeal of this property stems from the desire to flexibly express varying degrees of prior certainty on various subsets of the initial partition.  Informing the clustering of each unit individually allows us to put more prior weight on the co-clustering of a given subset of units compared to others for which less information is available.
\fq Lastly,  our approach facilitates the incorporation of initial partition information for a sequence of partitions that may exhibit temporal dependence or even no trend at all. \qf

The rest of this article is organized as follows. Section~\ref{sect:model} describes the general model for a single \fq partition and its extension to multiple partitions, particularly the case of a sequence of partitions that evolve over time. \qf
Some properties and special cases of interest are also stated and
discussed.  Guidelines for prior elicitation are also presented for some of the particular cases we describe.  Section~\ref{sect:simul} describes and summarizes the
results of extensive simulation studies carried out to explore
various model aspects including how the prior distribution is
specified, the effect of prior information on the temporal
progression, and a comparison with alternative approaches to
reflect prior information on partitions. Section~\ref{sect:application} describes the results of applying the proposed approach to a sequence of measures of particulate matter over a number of time periods in Germany, where the relative strengths of various potential model specifications are assessed, especially regarding the role of prior spatial information. Finally, Section~\ref{sect:disc} concludes with
a summary and some additional considerations regarding model construction and performance.  Additional simulations, data analysis results, and plots are presented in a Supplementary Materials file.

\section{Description of the Informed Partition Prior}\label{sect:model}

In this section, we first provide details of our method in the context of a single partition.  Then we extend ideas to the case of a sequence of partitions and highlight various ways in which the initial partition can inform clustering. 

\subsection{Informed Partition model for a single partition}\label{sect:priorsingle}

We introduce the notation and basic ideas from methods described in \cite{page_dahl_quintana:22}, focusing on the special case where we observe one partition conditionally on a \fq initial \qf guess. We review this material
to the extent required to present our own contribution. \fq For any integer $m\ge 1$, let $[m]=\{1,\ldots,m\}$ and let $\rho=\{S_1,\ldots,S_{k}\}$ denote a partition of $[m]$ where $k$ is the number of clusters and $S_1,\ldots,S_{k}$ the corresponding non-empty and mutually exclusive subsets of $[m]$ comprising $\rho$. We will assume throughout the convention that the subsets in any given partition are listed
in lexicographic order. 
Alternatively, we can specify $\rho$ through cluster membership indicators $c_1,\ldots,c_m$ such that $c_i=j \Leftrightarrow i\in S_j$.  We will use $\rho_0 = \{S_{01},\ldots,S_{0k_0}\}$ to denote an initial partition supplied by the user with $c_{01},\ldots,c_{0m}$ being the corresponding cluster labels. 
\qf 

We construct a random probability model by specifying a prior distribution on $\rho$ that accounts for the information in $\rho_0$. Following \cite{page_dahl_quintana:22}, we introduce a vector of auxiliary variables $\bm{\gamma} = (\gamma_1, \ldots, \gamma_m)$ that determines which units are allowed to be reallocated in $\rho$ with respect to $\rho_0$. Specifically, we introduce a collection of binary variables $\gamma_{1},\ldots,\gamma_m$ defined as follows: 

\fq
\begin{align}
\gamma_{i} & =
\left\{
\begin{array}{c l}
 1 & \mbox{if unit $i$ is {\it not to be} reallocated with respect to $\rho_0$}   \\
 0 &  \mbox{otherwise},
\end{array}
\right.
\end{align}
and assume that
\begin{align}
\gamma_i \mid \alpha_i & \stackrel{ind}{\sim} \mbox{Bern}(\alpha_i), \quad \alpha_i \sim \Be(a_i, b_i) \quad i=1, \ldots, m.   \nonumber
\end{align}
Under this construction, $\alpha_i$ controls the probability that each unit follows the initial partition. In other words, the values of $\alpha_i$ will reflect our prior beliefs in the initial guess, allowing for different degrees of uncertainty among observations. These prior beliefs can be included by selecting $a_i$ and $b_i$ in $\alpha_i \sim \Be(a_i, b_i)$, which provides quite a bit of flexibility to the user. Indeed, varying degrees of confidence in the strength of those subsets contained in $\rho_0$ can be expressed under this general framework. \qf

To illustrate this point,  consider the simple case of clustering six units based on the initial partition $\rho_0 = \{\{1,3\}, \{2,4,5\}, \{6\}\}$.  Further, suppose that an expert is quite certain that units 1 and 3 should be grouped together and not grouped with unit 6, but less sure about units 2, 4, and 5. A straightforward way to account for this information is by setting $a_1 = a_3 = a_6 = 99$ and $b_1 = b_3 = b_6 = 1$ along with $a_2 = a_4 = a_5 = 1$ and $b_2 = b_4 = b_5 = 9$. Extensions of this prior specification to more general scenarios with $m$ units can be done following a similar strategy. Assume that, without loss of generality, it is believed that elements in subsets $S_{01},\ldots, S_{0\ell}$ with $1\le\ell<k_0$ are highly likely to be grouped as specified by $\rho_0$, but that there is considerably less certainty on the remaining subsets. A prior
specification that reflects such information would consider $a_i=99$ and $b_i=1$ for all $i\in S_{01}\cup\cdots\cup S_{0\ell}$ and $a_i=1$ and $b_i=9$ for all $i\in S_{0,\ell+1}\cup\cdots \cup S_{0k_0}$. This effectively implies that elements in the ``highly likely clusters'' will a priori remain unmoved with a probability around $0.99$. In contrast, the others will be freed for reallocation with a prior probability around $0.9$.

\fq 
The prior distribution for the partition $\rho$ takes the following form  
\begin{equation}\label{eq:rhoprior1}
        p(\rho \mid  \rho_0, \bm{\alpha}) = \sum_{\bm{\gamma} \in \Gamma}   p(\rho \mid  \rho_0, \bm{\gamma}) p(\bm{\gamma}\mid \bm{\alpha}), 
\end{equation}
where $\Gamma=\{0,1\}^m$ denotes the collection of all possible vectors of zeros and ones of size $m$, $\bm{\alpha} = (\alpha_1, \ldots, \alpha_m)$, and 
\begin{equation}\label{eq:gammaprior}
p(\bm{\gamma} \mid \bm{\alpha}) = \prod_{i = 1}^m \alpha_i^{\gamma_i}( 1- \alpha_i)^{1 - \gamma_i}. 
\end{equation}
We remark here that \eqref{eq:rhoprior1} constitutes a slight abuse of notation in the sense that it includes a ``conditioning'' on $\rho_0$. Nevertheless, this must be understood as our way to emphasize the role played by $\rho_0$ in the prior construction. In what follows, we maintain this notation.

The partition probability
$\Pr(\rho = \lambda \mid  \rho_0, \bm{\gamma})$ 
relies on the definition of \emph{compatibility} introduced in \cite{page_dahl_quintana:22},
%
as discussed next. \qf We say that partitions $\rho$ and $\rho_0$ are compatible with respect to $\bm{\gamma}$, if $\rho$ can be obtained from $\rho_0$ by reallocating items as indicated by $\bm{\gamma}$, i.e., those items $i$ such that $\gamma_i = 0$ for $i = 1, \ldots, m$. To check if $\rho$ and $\rho_0$ are compatible with respect to $\bm{\gamma}$ we reason
as follows. Let $\mathfrak{R} = \{i : \gamma_{i} = 1\}$ be the collection of units that are fixed according to the initial partition $\rho_0$ and let $\mathfrak{R}^C= \{i : \gamma_{i} = 0\}$ be
the collection of units that are not. Denote by $\rho^{\mathfrak{R}}$ the ``reduced'' partition that remains after removing all items in $\mathfrak{R}$ from the subsets of $\rho$.  Similarly, let $\rho^{\mathfrak{R}}_{0}$ be the reduced initial partition. \fq Then $\rho$ and $\rho_0$ are {\em compatible with respect to} $\bm{\gamma}$ if and only if $\rho^{\mathfrak{R}}_{0} = \rho^{\mathfrak{R}}$. 
Now let $\calP$ denote the set of all partitions of $m$ units and let 
$\calP_{C}(\rho_0, \bm{\gamma}) = \{\rho \in \calP : \rho^{\mathfrak{R}}_{0} = \rho^{\mathfrak{R}}\}$  
be the collection of partitions that are compatible with $\rho_{0}$ based on $\bm{\gamma}$.  The conditional distribution
$p(\rho \mid \rho_{0}, \bm{\gamma})$ is a random partition distribution with support $\calP_C(\rho_0, \bm{\gamma})$, so that
\begin{equation}\label{eq:rhoprior2}
    \Pr(\rho = \lambda \mid  \rho_0, \bm{\gamma}) = \frac{\Pr(\rho = \lambda) \mathbb{I}\{\lambda \in \calP_{C}(\rho_0, \bm{\gamma})\}}{\sum_{\lambda^\prime \in\calP} \Pr(\rho = \lambda^\prime) \mathbb{I}\{\lambda^\prime \in \calP_{C}(\rho_0, \bm{\gamma})\}}. 
\end{equation}
Here $\Pr(\rho = \lambda)$ can, in principle, be any random partition model a practitioner may wish to adopt. However, for the results to be developed next, it will be convenient to assume from now on that
$\Pr(\rho = \lambda)$ arises from an exchangeable partition probability function (EPPF), i.e. $\Pr(\rho=\{S_1,\ldots,S_k\})=\mathfrak{p}(|S_1|,\ldots,|S_k|)$ where $\mathfrak{p}$ is a function of sequences of positive integers that is symmetric in its arguments and satisfies the sample size consistency property \citep[see, e.g.][]{pitman95EPPF}. To simplify the notation, let $\bar{p}(\rho)$ denote the probability mass function of a partition
$\rho=\{S_1,\ldots,S_k\}$ as evaluated from the EPPF, namely $\bar{p}(\rho)=\mathfrak{p}(|S_1|,\ldots,|S_k|)$.

\fq Notice that, even if the definition of $\Pr(\rho = \lambda \mid  \rho_0, \bm{\gamma})$ depends on the concept of compatibility, the support for $\rho$ after marginalization with respect to $\bm{\gamma}$ corresponds to the entire partition space $\mathcal{P}$, provided all $\bm{\gamma} \in \Gamma$ have non-zero probability. Since $\gamma_i \mid \alpha_i \stackrel{ind}{\sim} \mbox{Bern}(\alpha_i)$, $i = 1, \ldots, m$, this is true when no more than one element in $\bm{\alpha}$ is set to one. If at least two of these probabilities are set to 1, then the support of $\rho$ is restricted to a subset of $\mathcal{P}$. Moreover, \qf 
if $\alpha_i = 1$, $\forall i$, then $p(\rho \mid \rho_0) = \delta_{\rho_0}$, the prior collapses on the prior partition $\rho_0$, while if $\alpha_i = 0$, $\forall i$, then $p(\rho \mid \rho_0) = \bar{p}(\rho)$, so that the prior reduces to the chosen EPPF,  i.e., $\rho_0$ plays no role in the prior specification. The prior probability mass function of $\rho$ given the initial partition
$\rho_0$ marginalized over $\bm{\gamma}$ is given in the following proposition. \qf
\begin{proposition}\label{prop:marginalrho1}
Given $\rho$, $\rho_0$, $\bm{\gamma}$ and the concept of compatibility just described, \fq the prior distribution of $\rho$ given the initial partition $\rho_0$ marginalized over $\bm{\gamma}$ is 
\begin{align}\label{eq:rho}
p(\rho \mid \rho_0, \bm{\alpha}) & = \sum_{\bm{\gamma} \in \Gamma} p(\rho \mid  \rho_0, \bm{\gamma}) p(\bm{\gamma} \mid \bm{\alpha}) \nonumber \\ 
& = \sum_{\bm{\gamma} \in \Gamma}\frac{\bar{p}(\rho)}{\bar{p}(\rho^{\mathfrak{R}})}I[\rho^{\mathfrak{R}}_{0} = \rho^{\mathfrak{R}}]\,p(\bm{\gamma} \mid \bm{\alpha})
\end{align}
where $\Gamma$ was defined after \eqref{eq:rhoprior1}, and
$p(\bm{\gamma} \mid \bm{\alpha})$ is given in \eqref{eq:gammaprior}.
\qf
\end{proposition}
\begin{proof}
The proof follows from Proposition 3 of \cite{page_dahl_quintana:22}. 
\end{proof}

\fq We observe that the ratio $\bar{p}(\rho) / \bar{p}(\rho^{\mathfrak{R}})$ in \eqref{eq:rho} corresponds to the predictive distribution derived from the chosen EPPF, which is particularly useful when formulating computational strategies (see Section \ref{sect:computation}). \qf
In what follows, we will use the notation $\rho \sim iCRP(\rho_0, \bm{\alpha}, M)$ to denote the special case in which $\rho$ is distributed according to an informed partition model \fq for which $p(\rho\mid\rho_0, \bm{\alpha})$ in \eqref{eq:rho} arises when $\bar{p}(\rho)$ is given by the Chinese Restaurant Process (CRP) prior whose EPPF $\mathfrak{p}_M$ is provided in \eqref{eq:crpEPPF}  
\citep[see][for a discussion on this name]{pitman96SSM}:
\begin{align}\label{eq:crpEPPF}
   \bar{p}(\rho)&= \Pr(\rho = \{S_1, \ldots, S_k\} \mid M) = \mathfrak{p}_M(|S_1|,\ldots,|S_k|) \nonumber \\
   &= \frac{M^k}{\prod_{i=1}^{m}(M+i-1)}\prod_{i=1}^k(|S_i| - 1)!,
\end{align}
where $M>0$.
A popular extension of \eqref{eq:crpEPPF} is the Pitman-Yor (PY), or two-parameter Poisson-Dirichlet process \citep{Pitman-Yor:97} and that has EPPF given in closed form by
\begin{equation}\label{eq:PYEPPF}
\mathfrak{p}_{\vartheta,M}\left(|S_1|, \ldots, |S_k|\right)=\frac{\left(\prod_{\ell=1}^{k-1}(M+\ell \vartheta)\right)\left(\prod_{i=1}^k[1-\vartheta]_{|S_i|-1}\right)}{[1+M]_{m-1}}
\end{equation}
where $[x]_n=\prod_{j=1}^n(x+j-1)$, and we restrict our attention to the case $M>-\vartheta$ and $0\le\vartheta<1$. When $\vartheta=0$, \eqref{eq:PYEPPF} becomes \eqref{eq:crpEPPF}. When the PY EPPF $\mathfrak{p}_{\vartheta,M}$ is employed to define our prior $p(\rho\mid\rho_0, \bm{\alpha})$ in \eqref{eq:rho}, we will use the notation $\rho \sim iPY(\rho_0, \bm{\alpha}, \vartheta, M)$.
\qf

\subsection{Properties of the Informed Partition Model}
In this section, we provide some technical results that, among other things, connect $a_i$ and $b_i$ to the {\it a priori} co-clustering probabilities.  This will facilitate building intuition associated with the uncertainty connected to elicitation of the initial partition. The Rand Index \citep{rand:1971,hubert&arabie:1985} is often used as a metric to measure the ``distance'' between two partitions (e.g., $\rho_0$ and $\rho$).  The form of the Rand index is given by
\begin{equation}\label{rand.index}
	R(\rho_0,\rho)=\frac{\upsilon_1+\upsilon_2}{\binom{m}{2}},
\end{equation}
where $\upsilon_1$ is the number of pairs $(i,j)$ with $i,j\in[m]$, $i\ne j$ that co-cluster in both $\rho_0$ and $\rho$, and similarly, $\upsilon_2$ is the number of such pairs that do not co-cluster in both $\rho_0$ and $\rho$. 
It is easy to see that, in general, $0\le R(\rho_0,\rho)\le 1$. Before considering \eqref{rand.index} more rigorously, note first that $\rho_0$ is fixed so all randomness in \eqref{rand.index} arises from $\bm{\gamma}$ and the reconstruction of $\rho$ starting from $\rho_0^{\mathfrak{R}}$, using \fq model \eqref{eq:rho}, in turn induced by the chosen EPPF $\bar{p}(\rho)$. \qf Another key observation is
that the conditional probability $\Pr(c_{i}=c_{j}\mid \bm{\gamma})$ for  $i\ne j$ depends only on $\rho_0^{\mathfrak{R}}$ and the values of $\gamma_{i}$ and $\gamma_{j}$, that is, whether $i$ and $j$ are up for reallocation or not.  Some new notation is necessary before proceeding. \fq For $i,j\in [m]$ with $i\ne j$, let $\mathfrak{T}_{i,j}:\,{\cal P}\rightarrow {\cal P}$ be the mapping over the set of all partitions of $[m]$ that moves unit $j$ from its current cluster and reallocates it to the cluster where unit $i$ belongs, that is, $\mathfrak{T}_{i,j}(\rho)$ has $c_i=c_j$, leaving all other units in their original locations. For instance, with $m=4$ and $\rho=\{\{1,2 \},\,\{3\},\, \{4\}\}$ we get  $\mathfrak{T}_{1,2}(\rho)=\rho$ and $\mathfrak{T}_{1,3}(\rho)= \{\{1,2,3 \},\, \{4\}\}$. Similarly, define $\mathfrak{T}_{i,j}^{+\ell}:\,{\cal P}\rightarrow {\cal P}$ so that $\mathfrak{T}_{i,j}^{+\ell}(\rho)$ is the partition resulting from removing both units $i$ and $j$ from their current groups and putting them together in the $\ell$ th group, that is, $c_i=c_j=\ell$, where again we assume the groups are numbered in lexicographic order, and as usual, we allow $i$ and $j$ to form a separate group from those left after their removal.
The result of applying $\mathfrak{T}_{i,j}$ or $\mathfrak{T}_{i,j}^{+\ell}$  to a reduced partition, such as $\rho_0^{\mathfrak{R}}$ is defined analogously, with the caveat that for $\mathfrak{T}_{i,j}(\rho^{\mathfrak{R}})$ to make sense, it is implicit that $\gamma_i$ must be equal to one, that is, unit $i$ is not removed from the partition, and in the case of $\mathfrak{T}_{i,j}^{+\ell}$, both $i$ and $j$ must be removed from $\rho$. \qf 
Furthermore, let $\bm{\gamma}^{-(i,j)}$ denote $\bm{\gamma}$ with $\gamma_{i}$ and $\gamma_{j}$ removed. \fq Then the conditional probability that unit $i$  co-clusters with unit $j$ in $\rho$, starting from $\rho_0$, and conditional on a reallocation pattern $\bm{\gamma}$ is given by
%
\begin{eqnarray}
	\Pr(c_{i}=c_{j}\mid \bm{\gamma}^{-(i,j)},\gamma_{i}=\gamma_{j}=1)&=&\left\{\begin{array}{ll}
		0 & \mbox{if $c_{0i}\ne c_{0j}$} \label{eq:prcicj1}\\
		1 & \mbox{if $c_{0i}= c_{0j}$},
	\end{array}\right. \\
	\Pr(c_{i}=c_{j}\mid \bm{\gamma}^{-(i,j)},\gamma_{i}=1,\gamma_{j}=0)&=& 
	\frac{\bar{p}\left(\mathfrak{T}_{i,j}(\rho_0^{\mathfrak{R}})\right)}{\bar{p}\left(\rho_0^{\mathfrak{R}}\right) } \label{eq:prcicj2}\\
	\Pr(c_{i}=c_{j}\mid \bm{\gamma}^{-(i,j)},\gamma_{i}=0,\gamma_{j}=1)&=&
	\frac{\bar{p}\left(\mathfrak{T}_{j,i}(\rho_0^{\mathfrak{R}})\right)}{\bar{p}\left(\rho_0^{\mathfrak{R}}\right) } \label{eq:prcicj3}\\
	\Pr(c_{i}=c_{j}\mid \bm{\gamma}^{-(i,j)},\gamma_{i}=0,\gamma_{j}=0)&=& \sum_{\ell=1}^{|\rho_0^{\mathfrak{R}}|+1}
    \frac{\bar{p}\left(\mathfrak{T}_{i,j}^{+\ell}(\rho_0^{\mathfrak{R}})\right)}{\bar{p}\left(\rho_0^{\mathfrak{R}}\right) } 
    \label{eq:prcicj4}
	\end{eqnarray} 
where, $|\rho_0^{\mathfrak{R}}|$ denotes the number of clusters in the reduced partition $\rho_0^{\mathfrak{R}}$ and as before, $\bar{p}$ is the EPPF-induced probability mass function for partitions. 
\vspace{0.5cm}

\noindent {\bf Remark.}
When $\bar{p}$ arises from the PY EPPF \eqref{eq:PYEPPF}, \eqref{eq:prcicj2}
-- \eqref{eq:prcicj4} can be obtained explicitly. For instance, denote the reduced partition $\rho_0$ by $\rho_0^{\mathfrak{R}}=\{S_{01}^{\mathfrak{R}},\ldots,S_{0 k_0}^{\mathfrak{R}}\}$, and the number of elements in $\rho_0^{\mathfrak{R}}$ by $m^{\mathfrak{R}}$. We then find that \eqref{eq:prcicj2} becomes $(|S_{0c_i}^{\mathfrak{R}}|-\vartheta)/(|m^{\mathfrak{R}}|+M)$, \eqref{eq:prcicj3} becomes $(|S_{0c_j}^{\mathfrak{R}}|-\vartheta)/(|m^{\mathfrak{R}}|+M)$, and \eqref{eq:prcicj4} is given by
$$\sum_{\ell=1}^{|\rho_0^{\mathfrak{R}}|}\frac{(|S_{0\ell}^{\mathfrak{R}}|-\vartheta)(|S_{0\ell}^{\mathfrak{R}}|+1-\vartheta)}{(m^{\mathfrak{R}}+M)(m^{\mathfrak{R}}+M+1)} + \frac{(M+|\rho_0^{\mathfrak{R}}|\vartheta)(1-\vartheta)}{(m^{\mathfrak{R}}+M)(m^{\mathfrak{R}}+M+1)}.$$
As before, $\vartheta=0$ corresponds to the CRP case.

With \eqref{eq:prcicj1} -- \eqref{eq:prcicj4} we are able to establish the expected Rand index, which is provided in the following proposition.
\qf
\begin{proposition}\label{prop:expRandIndex}
Let ${\cal M}_{i,j}(\bm{\gamma}^{-(i,j)},\gamma_{i},\gamma_{j})$ denote the formulas \eqref{eq:prcicj1} -- \eqref{eq:prcicj4}  according to the values of $\gamma_{i}$ and $\gamma_{j}$. Then, the expected Rand index
between $\rho_0$ and $\rho$ is given by 
\begin{align}
\fq E(R(\rho_0,\rho)\mid\bm{\alpha}) = \qf &\fq \frac{1}{\binom{m}{2}}\sum_{\bm{\gamma}\in\Gamma}\sum_{1\le i<j\le m}  \left\{{\cal M}_{i,j}(\bm{\gamma}^{-(i,j)},\gamma_{i},\gamma_{j}) I\{c_{0i}=c_{0j}\}\right. \qf
\nonumber \\
&\fq \left.+(1-{\cal M}_{i,j}(\bm{\gamma}^{-(i,j)},\gamma_{i},\gamma_{j})) I\{c_{0i}\ne c_{0j}\}\right\}  p(\bm{\gamma}\mid \bm{\alpha}). \qf \label{eq:expRI}
\end{align}
\end{proposition}
\begin{proof}
The proof follows by conditioning on $\bm{\gamma}$ and considering
\eqref{eq:prcicj1} -- \eqref{eq:prcicj4}.
\end{proof}

\noindent {\bf Remark.}
It is easy to show that when $\alpha_{i}=1$ for all $i\in[m]$, \eqref{eq:expRI} becomes 1 when $\rho_0=\rho$. \fq On the other hand, when $\alpha_{i}=0$ for all $i\in[m]$ then $\Pr(\bm{\gamma}=(0,\ldots,0)\mid\bm{\alpha}=\bm{0})=1$ and $\rho^{\mathfrak{R}}=\emptyset$ which results in  ${\cal M}_{i,j}(\bm{\gamma}^{-(i,j)},\gamma_{i},\gamma_{j})=\Pr(c_i = c_j)$ and  $1-{\cal M}_{i,j}(\bm{\gamma}^{-(i,j)},\gamma_{i},\gamma_{j})=\Pr(c_i \ne c_j)$ for all $\gamma_i, \gamma_j$ pairs. \qf
Further, notice that when $\alpha_{i}=0$ for all $i\in[m]$ then $\sum_{1\le i<j\le m} I\{c_{0i}=c_{0j}\}=\sum_{\ell=1}^{k_0} \binom{|S_{0\ell}|}{2} I\{|S_{0\ell}|\ge 2\}$.  Thus, when $\alpha_{i}=0$ for all $i\in[m]$ the expected Rand index becomes
\begin{align}\label{eq:ERIindep}
\fq E(R(\rho_0,\rho)\mid\bm{\alpha}=\bm{0}) \qf =&\Pr(c_i = c_j)\frac{\sum_{\ell=1}^{k_0}	\binom{|S_{0\ell}|}{2} I\{|S_{0\ell}|\ge 2\}}{\binom{m}{2}} \nonumber \\
& + \Pr(c_i \ne c_j) \left(1- \frac{\sum_{\ell=1}^{k_0}	\binom{|S_{0\ell}|}{2} I\{|S_{0\ell}|\ge 2\}}{\binom{m}{2}}\right).
\end{align}
\fq Equation \eqref{eq:ERIindep} represents the expected Rand index between $\rho_0$ and $\rho$ in the case where $\rho_0$ does not play a role in the distribution of $\rho$, and the expectation is completely driven by randomness in $\rho$. \qf Note that the quantity
$\binom{m}{2}^{-1}\sum_{\ell=1}^{k_0}
\binom{|S_{0\ell}|}{2} I\{|S_{0\ell}|\ge 2\}$ balances the EPPF's proclivity to
merge or separate, as represented by $\Pr(c_i = c_j)$ and $\Pr(c_i \ne c_j)$,
respectively. If $\rho_0$ contains one large cluster (i.e. of size close to $m$), the ratio will be large (close to $1$) and so \eqref{eq:ERIindep} will be driven mostly by the value of $\Pr(c_i = c_j)$, e.g. $1/(M+1)$ for the CRP or $(1-\vartheta)/(M+1)$ for the PY process. The opposite case occurs when $\rho_0$ contains many small clusters, resulting in a small ratio and \eqref{eq:ERIindep} will be driven mainly by the value of $\Pr(c_i \ne c_j)$. \fq The previous discussion suggests that when $\rho_0$ consists of a few large clusters and $a_i\ll b_i$ so that $\alpha_i$ is close to zero with probability almost 1, the similarity between $\rho_0$ and $\rho$, as measured by $E(R(\rho_0,\rho))$ will increase if the EPPF favors grouping elements, thus increasing the chance of mimicking $\rho_0$. In the iCRP and iPY cases this is achieved with a small value of $M$. 
\qf

To further illustrate the results described above, consider the $m=2$ case, for which $\rho_0$ can be either $\{\{1,2\}\}$ or $\{\{1\},\{2\}\}$.  Now $R(\rho_0,\rho)$ becomes $1$ if $\rho_0=\rho$ and $0$ otherwise. The expected Rand index in this case is provided in the following proposition.
\begin{proposition}\label{prop:m=2}
When $m=2$ we have
$$\fq E(R(\rho_0,\rho)\mid\bm{\alpha}) \qf=\left\{\begin{array}{ll}
\zeta_1      & \mbox{if $\rho_0=\{\{1,2\}\}$} \\
\zeta_2   & \mbox{if $\rho_0=\{\{1\},\{2\}\}$}
\end{array}\right.
$$
where 
\fq
\begin{eqnarray*}
\zeta_1\defi\Pr(c_{1}=c_{2}\mid\rho_0=\{\{1,2\}\}, \bm{\alpha})&=&\alpha_{1}\alpha_{2}+\bar{p}(\{\{1,2\}\})(1-\alpha_{1}\alpha_{2})   \\
\zeta_2\defi\Pr(c_{1}\ne c_{2}\mid \rho_0=\{\{1\},\{2\}\}, \bm{\alpha})&=&\alpha_{1}\alpha_{2}+\bar{p}(\{\{1\},\{2\}\})(1-\alpha_{1}\alpha_{2}),
\end{eqnarray*}
 where, as pointed out earlier, the notation 
$\Pr(c_{1}=c_{2}\mid\rho_0=\{\{1,2\}\}, \bm{\alpha})$
and $\Pr(c_{1}\ne c_{2}\mid \rho_0=\{\{1\},\{2\}\}, \bm{\alpha})$ is here employed to emphasize the dependence on
the initial partition $\rho_0$. \qf
\end{proposition}
\begin{proof}
The proof follows straightforwardly by conditioning on $(\gamma_{1},\gamma_{2})$.
\end{proof}

To facilitate prior elicitation (i.e., selecting values for $a_i$ and $b_i$ in \fq $\alpha_i \sim \Be(a_i, b_i)$) \qf it would be useful to express the co-clustering probabilities marginalized over $\bm{\alpha}$.  Thus, we can further compute expectations with respect to $\bm{\alpha}$ to express these quantities in terms of $\{a_{i}\}$ and $\{b_{i}\}$. For instance, the expected co-clustering probabilities at the end of Proposition~\ref{prop:m=2} become
\begin{eqnarray*}
E\left(\zeta_1\right)&=& \fq \bar{p}(\{\{1,2\}\})+\left(1-\bar{p}(\{\{1,2\}\})\right) \qf \frac{a_{1}a_{2}}{(a_{1}+b_{1})(a_{2}+b_{2})}  \\
E\left(\zeta_2\right)&=& \fq \bar{p}(\{\{1\},\{2\}\})+\left(1-\bar{p}(\{\{1\},\{2\}\})\right) \qf \frac{a_{1}a_{2}}{(a_{1}+b_{1})(a_{2}+b_{2})}.
\end{eqnarray*}
These simple formulas can be used to guide prior elicitation for $\{a_{i}\}$ and $\{b_{i}\}$. For a given choice of EPPF, the values of \fq $\bar{p}(\{\{1,2\}\})$
and $\bar{p}(\{\{1\},\{2\}\})$ \qf are determined, and so, one may attempt solving for
the values of $a_1/(a_1+b_1)$ and $a_2/(a_2+b_2)$ from these equations, by providing suitable specifications to the expected values of the Rand Index $E\left(\zeta_1\right)$
and $E\left(\zeta_2\right)$.

\subsection{Informed partition model for a sequence of random partitions}\label{sec:model_for_seq}

Consider now a sequence of partitions $\bm{\rho}=(\rho_1,\ldots,\rho_T)$ of $[m]$ where $\rho_t = \{S_{t1}, \ldots, S_{tk_t}\}$. Alternatively, cluster labels can be employed so that $c_{ti} =j$ implies that unit $i$ is allocated to $S_{tj}$.  Based on ideas in \cite{page_dahl_quintana:22} we construct a prior
model $p(\bm{\rho}\mid\rho_0) =p(\rho_1,\ldots,\rho_T \mid \rho_0)$  that potentially incorporates serial dependence among random partitions and is informed by $\rho_0$. Alternatively, we may adopt a multi-view data framework with partitions being related to the different available variables.  Our construction extends the idea underlying \eqref{eq:rhoprior1} by introducing a matching $(T\times m)$-dimensional sequence of binary indicators $\bm{\gamma}$ where for $t=1,\ldots,T$ and $i=1,\ldots,m$ we introduce 
\begin{align}
\gamma_{ti} 
 =
\left\{
\begin{array}{c l}
 1 & \mbox{if unit $i$ is {\it not to be} reallocated for variable or at time $t$}   \\
 0 &  \mbox{otherwise}.
\end{array}
\right.
\end{align}
\fq Further, we have $\gamma_{ti} \mid \alpha_{ti} \sim \mathrm{Bern}(\alpha_{ti})$ with $\alpha_{ti} \sim \mbox{Beta}(a_{ti}, b_{ti})$.
In the context of multiple partitions, we use $\mathbf{A}$, a matrix of $T$ rows and $m$ columns, to refer to the collection $\{\alpha_{ti}\}$.
There is substantial flexibility in specifying $p(\bm{\rho}\mid\rho_0,\mathbf{A})$ and the prior for $\mathbf{A}$, which is why specific constructions are typically case-dependent. \qf We discuss first potentially useful alternatives for $p(\bm{\rho}\mid\rho_0,\mathbf{A})$, and describe prior choices for $\mathbf{A}$ in Section \ref{sec:model_alpha}. \fq In what follows, we use the notation $\bm{\alpha}_{t\bullet}=(\alpha_{t1},\ldots,\alpha_{tm})$ for each $1\le t\le T$ so that the transpose of $\bm{A}$ is $\mathbf{A}^{\prime} = (\bm{\alpha}_{1\bullet}, \ldots, \bm{\alpha}_{T\bullet})$ \qf
\begin{description}
\item[Conditional independence. ] In this case, we model the partitions in $\bm{\rho}$ as \fq
$$p(\bm{\rho}\mid\rho_0, \mathbf{A})=p(\rho_1\mid\rho_0,\bm{\alpha}_{1\bullet})\times\cdots \times p(\rho_T\mid\rho_0,\bm{\alpha}_{T\bullet}),$$
that is, partitions are conditionally independent given the initial $\rho_0$, and
where each of $p(\rho_t\mid\rho_0, \bm{\alpha}_{t\bullet})$ is specified as in $\eqref{eq:rhoprior1}$. A
convenient default choice is to assume $\rho_t\stackrel{ind}{\sim} iCRP(\rho_0,\bm{\alpha}_{t\bullet},M)$, \qf as
defined in Section~\ref{sect:priorsingle}. This
case is useful when it is judged that there is no time trend in the $\rho_t$'s so that
the influence of $\rho_0$ is constant across time. This would be a case where any potential time trend is learned from the data, and not imposed by the prior. Another scenario where this approach may prove beneficial is in a multiview setting (\citealt{dombowsky2024, franzolini2023conditional}). In this case, $t$ would not index time but rather a variable measured from each unit and $\gamma_{ti} = 1$ if unit $i$ is not to be reallocated with respect to $\rho_0$ for variable $t$. Depending on the prior specifications for the $(T\times m)$-dimensional parameter $\mathbf{A}$ we could have a model that includes a conditional i.i.d. specification (e.g. $a_{ti}=a_i$ and $b_{ti}=b_i$ for all $t$) or varying degrees of shrinkage towards $\rho_0$, e.g. as implied by a sequence of prior mean values $a_{ti}/(a_{ti}+b_{ti})$ that decrease with $t$ for every $i$. 
\item[Markovian dependence.] In this case, we follow \cite{page_dahl_quintana:22} and consider a Markovian structure of the form \fq 
\begin{equation}\label{eq:priormarkov}
	p(\bm{\rho} \mid \rho_0,\mathbf{A})=p(\rho_1 \mid \rho_0,\bm{\alpha}_{1\bullet})p(\rho_2 \mid \rho_1,\bm{\alpha}_{2\bullet})\cdots p(\rho_T\mid\rho_{T-1},\bm{\alpha}_{T\bullet}),
\end{equation}
where time dependence in the sequence of partitions is accounted for by using the partition at time $t-1$ as the center to inform that at time $t$. A default choice here would be to specify $\rho_t\mid\rho_{t-1},\bm{\alpha}_{t\bullet}
\sim iCRP(\rho_{t-1},\bm{\alpha}_{t\bullet},M)$. That is, centering is based on the partition from the previous time point and $\rho_1 \sim iCRP(\rho_0,\bm{\alpha}_{1\bullet},M)$. \qf Note that $\gamma_{ti} = 1$ if unit $i$ is not to be reallocated when moving from time  $t-1$ to $t$. This creates a Markovian structure for the sequence of partitions that teamed with the flexibility inherent to the definition of the $\mathbf{A}$ parameters,
facilitates propagating
various degrees of dependence on the partition information supplied by $\rho_0$. 
\end{description}

\fq
\begin{proposition}\label{prop:m=2+sequence}
When $m=2$ and we are under the Markovian dependence prior
\eqref{eq:priormarkov} we have
$$E(R(\rho_0,\rho_t)\mid\mathbf{A})=\left\{\begin{array}{ll}
\zeta_{t1}      & \mbox{if $\rho_0=\{\{1,2\}\}$} \\
\zeta_{t2}   & \mbox{if $\rho_0=\{\{1\},\{2\}\}$}
\end{array}\right.
$$
where, for $t\ge 1$
\begin{eqnarray*}
\zeta_{t1}&=&\prod_{\ell=1}^t\{\alpha_{\ell 1}\alpha_{\ell 2}\}+\bar{p}(\{\{1,2\}\})\left(1-\prod_{\ell=1}^t\{\alpha_{\ell 1}\alpha_{\ell 2}\}\right)   \\
\zeta_{t2}&=&\prod_{\ell=1}^t\{\alpha_{\ell 1}\alpha_{\ell 2}\}+\bar{p}(\{\{1\},\{2\}\})\left(1-\prod_{\ell=1}^t\{\alpha_{\ell 1}\alpha_{\ell 2}\}\right).
\end{eqnarray*}
\end{proposition}
\begin{proof}
The proof follows again by conditioning on the entire collection of $\{(\gamma_{\ell 1},\gamma_{\ell 2}):\,\ell=1,\ldots,t\}$ indicators. For instance, when all $\gamma$'s are equal to 1, the entire sequence $\rho_1,\ldots,\rho_t$ is fixed at $\rho_0$. Otherwise, to obtain $\rho_t=\rho_0$, it is required that $\rho_0$ is formed when reallocating element(s) at the latest time $1\le\ell\le t$ for which $\gamma_{\ell i}=0$ or $\gamma_{\ell 2}=0$.
\end{proof}

We observe that the conclusion of Proposition~\ref{prop:m=2+sequence} reduces to Proposition~\ref{prop:m=2} when $t=1$. Furthermore, when all the $\alpha_{ti}$'s are i.i.d. with mean $\mu$ then
\begin{eqnarray*}
E(\zeta_{t1})&=&\mu^{2t}+(1-\mu^{2t})\bar{p}(\{\{1,2\}\}) \\
E(\zeta_{t2})&=&\mu^{2t}+(1-\mu^{2t})\bar{p}(\{\{1\},\{2\}\})
\end{eqnarray*}
In nontrivial cases we have $0<\mu<1$ so that for large values of $t$, $E(\zeta_{t1})$ and $E(\zeta_{t2})$ approach $\bar{p}(\{\{1,2\}\})$ and $\bar{p}(\{\{1\},\{2\}\})$, respectively. 
\qf

\subsection{Models for the Reallocation Probabilities} \label{sec:model_alpha}
In this section, we discuss a number of possible prior models for $\mathbf{A}$. We first describe possible priors that are of the conjugate variety and then introduce two possibilities that incorporate more structure.  We emphasize that many more options exist depending on the specific prior features the practitioner wishes to impose. 

A natural starting point to specifying a prior for $\mathbf{A}$ would be the conjugate Beta distribution as introduced briefly in Section \ref{sect:priorsingle}.  The most flexible case would assume that $\alpha_{ti} \sim \Be(a_{ti}, b_{ti})$, $t = 1, \ldots, T$, $i = 1, \ldots, m$.  This prior specification (referred to as ``time $\times$ unit local'') maximizes the user's control over each unit's probability of being reallocated across time or measured variables. This flexibility comes at a potential cost, as the posterior distribution for each $\alpha_{ti}$ is informed by a single $\gamma_{ti}$, which amplifies the influence of the prior specification. Due to this, it may be appealing to borrow strength across units and/or time.  Borrowing strength across time is possible by assuming $\alpha_{ti} = \alpha_i$ for all $t = 1, \ldots, T$ and $\alpha_{i} \sim \Be(a_i, b_i)$, so that all time points are used to inform the probability of reallocation for each unit.  We refer to this prior as the ``unit local'' prior.  Similarly, borrowing-strength across units is achieved assuming $\alpha_{ti} = \alpha_t$ for all $i = 1, \ldots, m$ and $\alpha_t \sim \Be(a_t, b_t)$. Now, all units within a time period are used to inform the probability of reallocation for each unit at a specific time point.  We refer to this prior specification as the ``time local'' prior.  Finally, borrowing strength across time and units is achieved by assuming $\alpha_{ti} = \alpha$ for all $t = 1, \ldots, T$ and $i = 1, \ldots, m$ and $\alpha \sim \Be(a,b)$. This prior is referred to as the ``global'' prior.  A summary of prior distributions that employ the $\Be$ distribution is provided in Table \ref{tab:alpha_models}.
\begin{table}[tb]
 \centering
 \caption{Conjugate type priors for $\mathbf{A}$. We index units using $i$ for $i = 1, \ldots, m$ and time/variable using $t$ from $t = 1, \ldots, T$.}
\begin{tabular}{l l l l} \toprule
 name & description: $\mathbf{A}=(\alpha_{ti})$ &  model & borrowing-of-strength \\\midrule 
 global & $\alpha_{ti}=\alpha$, constant in $i,t$ & $\alpha \sim \Be(a,b)$ & across time and units \\
 time local &  $\alpha_{ti}=\alpha_t$, constant in $i$ & $\alpha_t \sim \Be(a_t,b_t)$ & across units \\
 unit local & $\alpha_{ti}=\alpha_i$, constant in $t$ & $\alpha_i \sim \Be(a_i,b_i)$ & across time \\
 time$\times$unit  local & unrestricted & $\alpha_{ti} \sim \Be(a_{ti},b_{ti})$ & none \\ \bottomrule
 \end{tabular}
 \label{tab:alpha_models}
 \end{table}
Of the prior models in Table \ref{tab:alpha_models}, only the unit local and time$\times$unit local priors permit users to emphasize (or de-emphasize) particular subsets of the initial partition as described in Section \ref{sect:priorsingle}. 

It is possible to incorporate more structure in the prior for $\mathbf{A}$ if, for example, smoother temporal trends of $\alpha_{ti}$ are desired {\it a priori}.  We briefly describe two ways in which this can be carried out.

\begin{description}
\item[Latent Autoregressive Process.] A possible temporally structured prior construction considers a time series-like model for $\{\alpha_{ti}\}$. There are many options, but a relatively simple approach would consider dynamic evolution in the logit scale. Letting $\xi_{ti}=\logit(\alpha_{ti})$ we propose
\begin{equation}\label{eq:priorAR}
\xi_{ti}=\beta_{0i} + \beta_{1i} \xi_{t-1,i}+\epsilon_{ti},
\end{equation}
where $\epsilon_{ti}\ind N(0,\kappa_i^2)$, a suitable prior is placed on $\beta_{0i}$, $\beta_{1i}$, and $\kappa_i^2$, and $\xi_{0i}$ is user-specified. As a default, we recommend using $\xi_{0i}=\logit\left(a_{0i}/(a_{0i} + b_{0i})\right)$ where $a_{0i}$ and $b_{0i}$ are elicited such that $E(\alpha_{0i}) = a_{0i}/(a_{0i} + b_{0i})$. This setting can be used to express different types of prior information directly on the reallocation probabilities. For instance, by choosing a large initial value for $\xi_{0i}$ (e.g., $\xi_{0i}=2.5$ or $3.0$) and an informative prior on $\kappa_i^2$ that concentrates on small values, then with high prior probability unit $i$ remains in its initial cluster. Similarly, a high prior probability of reallocation can be expressed with a negative initial value, e.g. $\xi_{0i}=-2.5$ or $-3.0$.  We consider this extension in the prior simulations in Section~\ref{sect:simul} (see Figures~\ref{fig:mark_indep} and ~\ref{fig:pairwise.prob.lag}) and the data application in Section~\ref{sect:application}. 

\item[Time trends.] A second prior construction involves deterministic time trends for $\{a_{ti}\}$ and $\{b_{ti}\}$. Starting from given values $a_0$ and $b_0$ (e.g. $a_0=b_0=1$) we consider
$$a_{ti}=h_i^a(t),\qquad b_{ti}=h_i^b(t),$$
where $h_1^a,\ldots,h_m^a$ and $h_1^b,\ldots,h_m^b$ are deterministic and user-specified functions on the positive numbers such that $h_i^a(0)=a_0$ and $h_i^b(0)=b_0$. In the simplest case, $h_i^a=h^a$ and $h_i^b=h^b$ for all $i=1,\ldots,m$, but unit-specific choices are easily specified. Dropping subindices to simplify notation, a linear trend that starts at $a_0$ for $t=0$ and ends at $A_0$ for $t=T$
is given by $h^a(t)=a_0+(A_0-a_0)\times \frac{t}{T}$. The growing trend is obtained when $A_0>a_0$, and a decreasing one follows when $A_0<a_0$. An asymptotic alternative could be a sigmoid curve that reaches a certain limit as $T\longrightarrow\infty$, e.g. $h^a(t)=2(A_0-a_0)/(1+\exp(-t/A_1))+2a_0-A_0$, where $A_0>a_0$ for a growing trend and $A_0<a_0$ for a decreasing one, $A_0$ is the asymptote, and where $A_1>0$ is a given scale parameter to speed up/down the trend (e.g. $A_1=1$ for a moderate pace). Similar definitions can be adopted for $h^b$. We observe that time trends can be combined with either conditional independence or with Markovian dependence, depending on the desired prior structure. 

\end{description}

\subsection{Posterior Computation} \label{sect:computation}
With a prior distribution for $(\rho_1, \ldots, \rho_T)$ specified, it remains to determine a model for the response variable which we denote by $Y_i$ with $i = 1, \ldots, m$ for $T=1$ or $Y_{ti}$ for $T > 1$.   The modeling decisions associated with $Y_{ti}$ should be driven by its characteristics.  For example, the data scenarios in Sections \ref{sect:simul} and \ref{sect:application} assume that a conditionally independent Gaussian distribution for $\bm{Y} = (Y_{11}, \ldots, Y_{Tm})$ is appropriate.  Once specified, posterior sampling of $\bm{\rho} \mid \bm{Y}$  can be achieved by employing the algorithm described in \cite{page_dahl_quintana:22}.  The algorithm requires checking the compatibility between
$\rho_{0}$ and $\rho_1$ for $T = 1$. 
$\rho_{0}$ and $\rho_t$ 
must be checked for each $t > 1$ under the conditionally independent model and for $\rho_{t-1}$ and $\rho_t$ under the Markovian structure model.  Once compatibility is ensured, $\bm{\rho}$ is updated by sampling the cluster membership allocations $c_{ti}$ using Algorithm 8 of \cite{neal:2000}. In addition, sampling from the full conditional for the $\mathbf{A}$ parameters reduces to a simple Beta-Binomial update, while the binary $\bm{\gamma}$ indicators are easily handled. 
The exception to this is in the case of the latent autoregressive model \eqref{eq:priorAR}, for which a simple Metropolis update is employed for the $\xi_{ti}=\logit(\alpha_{ti})$ parameters. 


For more details see \cite{page_dahl_quintana:22} and its supplementary material.  The computer codes that employ the MCMC algorithm to carry out model fitting in Sections \ref{sect:simul} and \ref{sect:application} are available in the {\tt drpm} {\tt R}-package that is located at \url{https://github.com/gpage2990/drpm}. 




\section{Simulation Studies}\label{sect:simul}

In this section, we conduct a number of prior and posterior simulation studies to illustrate the behavior of our informed random partition model. The prior simulations aim to provide intuition on the effect of different prior model choices when modeling either a single partition or a sequence of dependent partitions. In the first case, we use a toy example to empirically demonstrate the effect of our proposed informed partition model and other priors recently proposed in the literature \citep{paganin_etal:2021, smith_allenby:2020}. In the second case, we examine different model choices as illustrated in Section~\ref{sec:model_for_seq}, along with the latent autoregressive model presented in Section~\ref{sec:model_alpha}. 
\fq In Section~\ref{sec:post.sim.onetime.point} \qf we conduct a set of posterior simulations to compare the effect of different prior parameter choices when modeling a single partition, \fq including the case in which we use a misspecified model. Finally, we also compare posterior results using other informative priors. \qf It is important to note that these comparisons must be done carefully, as aligning models based on certain prior properties can be difficult. Additional results are provided in Sections S1 and S2 of the Supplementary Material.

\subsection{Prior Simulation: comparison with other informative priors}

To gain some intuition about the effect of our informed partition model, we consider a simple toy example where we can easily compute and represent the probabilities assigned to each partition. We also compare our proposal with the probability distribution induced by the Centered Partition Process (CPP) proposed in \cite{paganin_etal:2021} and the Location-Scale Partition distribution (LSP) in \cite{smith_allenby:2020}. We briefly summarize these two priors in the following.

The CPP defines a class of priors on the  partition space $\calP$  that directly penalizes a chosen baseline EPPF $p_0(\cdot)$ to shrink the prior probability mass towards a known partition $\rho_0$
\begin{equation*}
  p(\rho\mid\rho_0, \psi) \propto p_0(\rho) \exp\{ -\psi d(\rho, \rho_0)\},
\end{equation*}
where $\psi > 0$ is a penalization parameter and $d(\rho, \rho_0)$ a distance metric between partitions. There are many possible choices for the baseline EPPF and of the distance metric; here we consider the CRP distribution in \eqref{eq:crpEPPF} with concentration parameter $M$ and the Variation of Information (VI), discussed in \cite{meila2007}, and denote such distribution as $CPP(\rho_0,\psi, M)$. The penalization parameter $\psi$ controls the amount of shrinkage towards $\rho_0$ with higher values resulting in higher probability mass on partitions close to  $\rho_0$.

The LSP \citep{smith_allenby:2020} is a distribution on $\calP$ that is characterized by a location partition ($\rho_0$) and a scale parameter $\nu > 0$, denoted as $LSP(\rho_0, \nu)$. This last controls the probability mass concentration around $\rho_0$, with small values favoring partitions closer to $\rho_0$. The LSP distribution builds on the covariate-dependent PPMs presented in \cite{park&dunson:10} and \cite{MQR:11}, treating the location partition $\rho_0$ as a covariate to inform $\rho$. The probability mass function induced on $\calP$ can be factored into a sequence of conditional probabilities
\begin{equation*}
  p(\rho\mid\rho_0, \nu) \propto \prod_{i = 1}^m p(c_i \mid \boldsymbol{c}_{1:{(i -1)}}, \rho_0, \nu)
\end{equation*}
where $\boldsymbol{c}_{1:{(i- 1)}}$ denotes the vector of cluster membership indicator for unit 1 to $i - 1$. \fq We have that for the first indicator  $\Pr(c_1 = 1) = 1$ and then
\begin{equation*}
  \Pr(c_i = k\mid \boldsymbol{c}_{1:{(i -1)}}, \rho_0, \nu) \propto
  \begin{cases}
      \displaystyle\frac{\nu + \sum_{l = 1}^{i-1} \mathbb{I}\{c_l = k\} \mathbb{I}\{c_{0l} = c_{0i}\} } {\nu K_0^{(i-1)} + \nu + n_k^{(i - 1)}}, & k = 1, \ldots, K^{i -1} \\
      \displaystyle\frac{\nu + \mathbb{I}\{c_{0i} = K_0^{i-1}\}} {\nu K_0^{(i-1)} + \nu + 1}, & k = K^{i -1} + 1
  \end{cases}
\end{equation*}
\qf where $K^{i-1} = \text{max}(c_1,\ldots,c_{i -1})$,  $K_0^{i-1} = \text{max}(c_{01},\ldots,c_{0 (i-1)})$, and $n_k^{(i - 1)} =\sum_{l = 1}^{i - 1} \mathbb{I}\{c_l = k\}$ is the number of items in group $k$ among the first $i -1 $ items. We refer to \cite{smith_allenby:2020} for details on the construction and properties of the LSP distribution.

\begin{figure}[H]
\includegraphics[width = 0.32\textwidth]{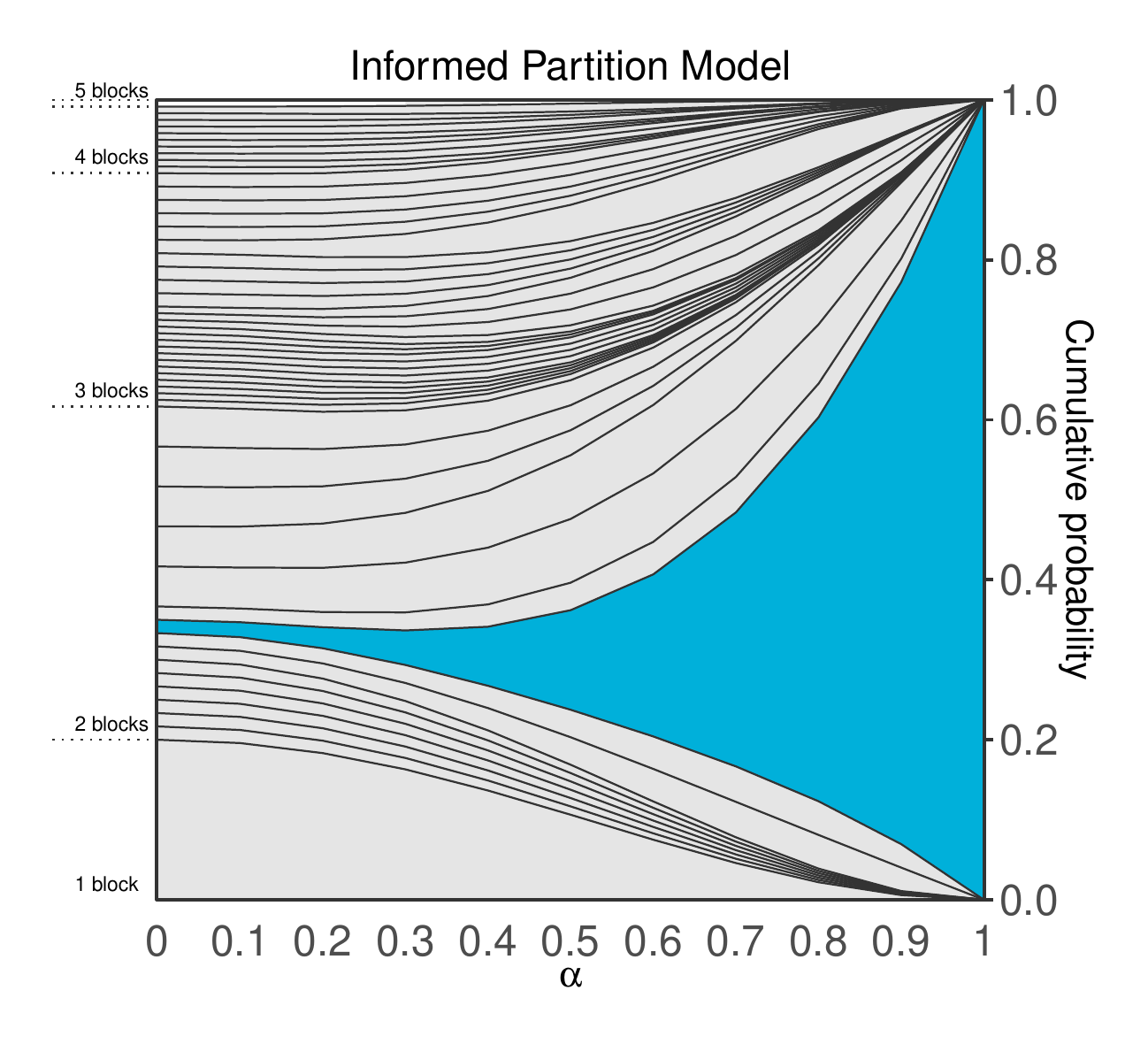}
\includegraphics[width = 0.32\textwidth]{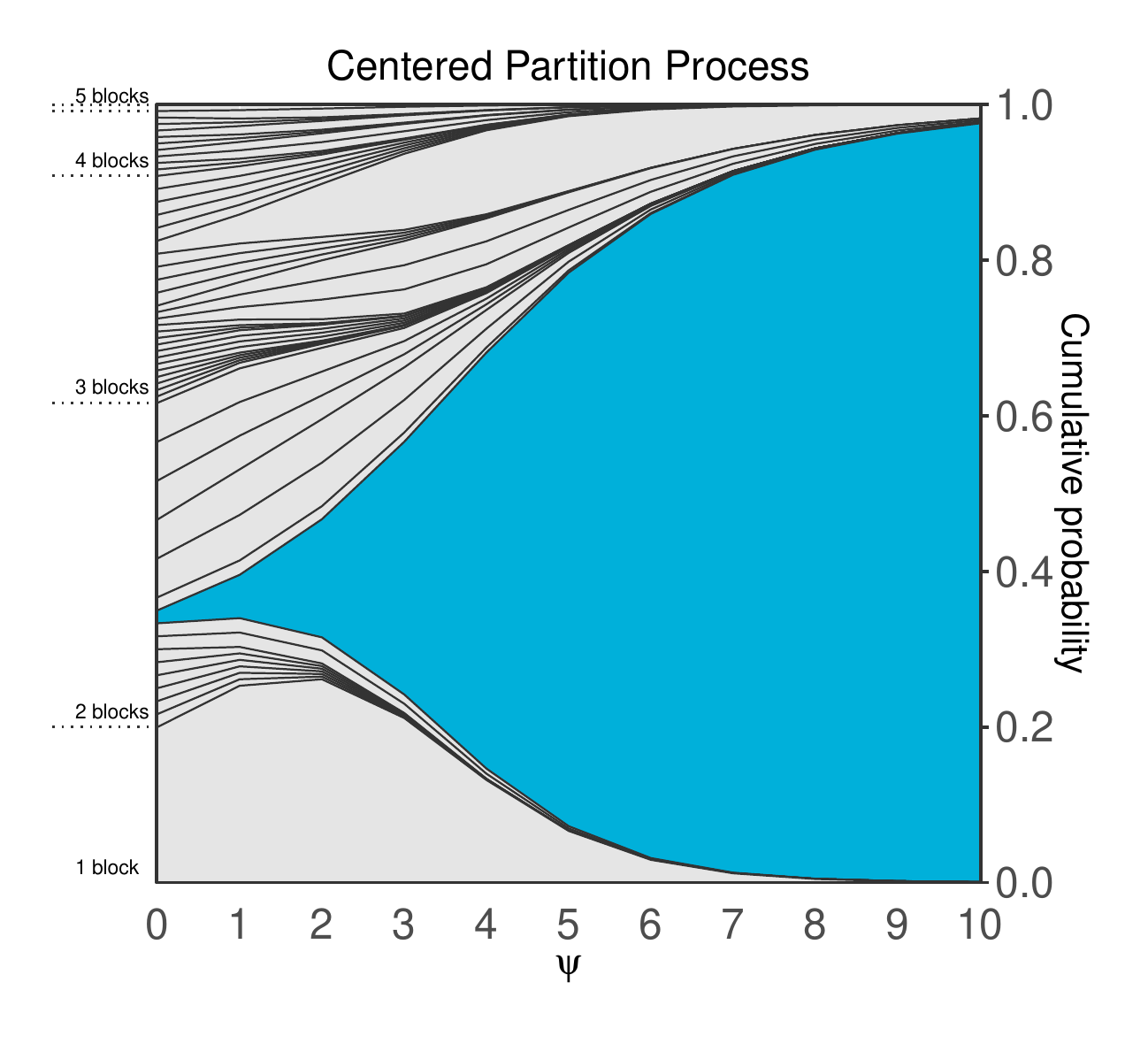}
\includegraphics[width = 0.32\textwidth]{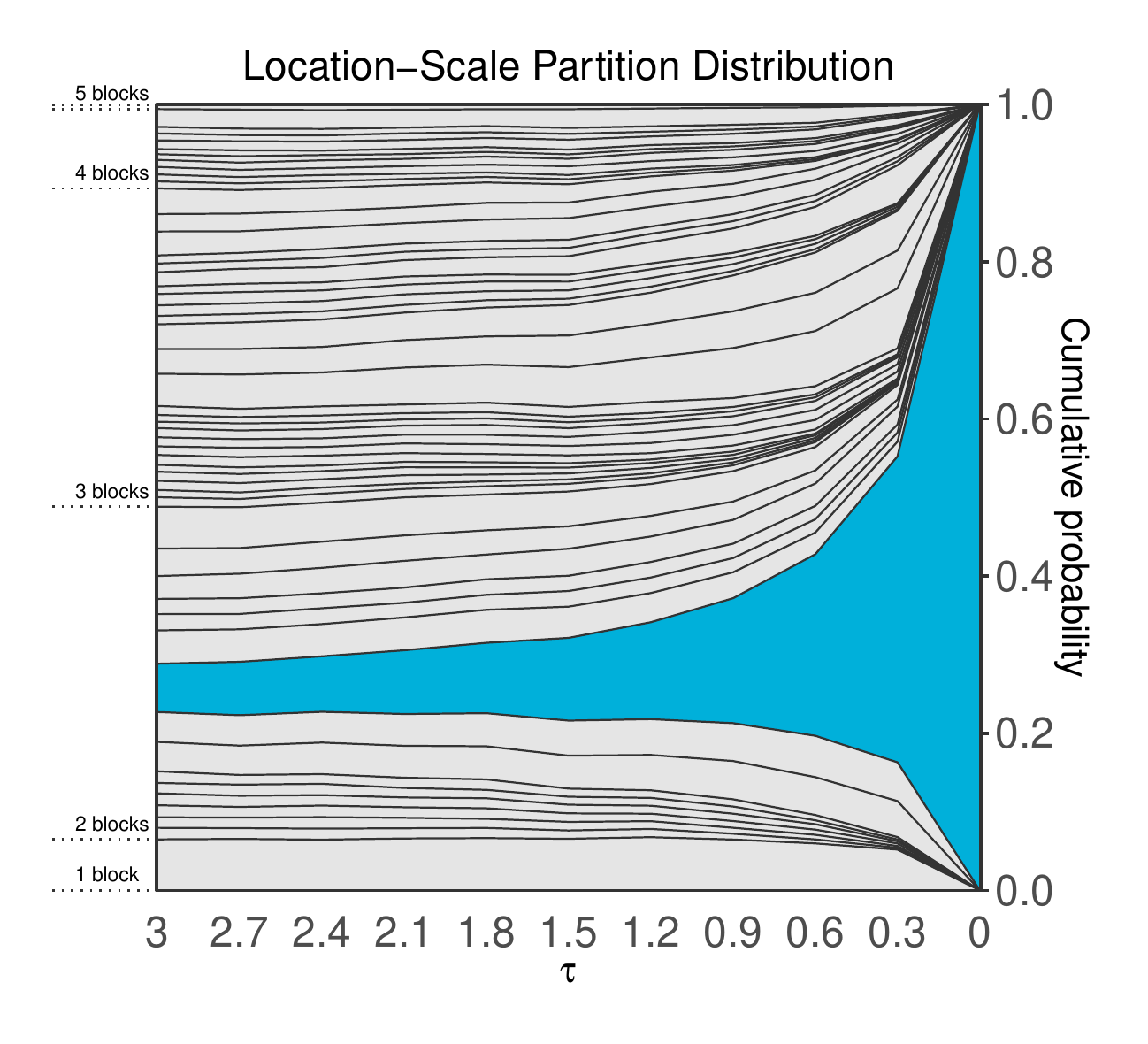}
\caption{iCRP vs CPP vs LSP.  Prior probabilities for each of the $52$ partitions when the prior guess is ${\rho}_{0} = \{\{1,2\}, \{3, 4, 5\}\}$ (highlighted in blue). The cumulative probabilities across different values of the penalization parameters are joined to form the curves, while the probability of a given partition corresponds to the area between the curves. Notice that the x-axis in the third panel is reversed for ease of interpretation.}
\label{fig:inducedPrior_comparison}
\end{figure}

Consider the following toy example: for $m = 5$, there are $52$ possible set partitions, and we can easily compute the probability of each partition under different prior distributions. Figure~\ref{fig:inducedPrior_comparison} shows the prior probability for each partition induced by the iCRP, the CPP, and the LSP prior, considering a CRP with $M = 1$. We choose as an initial partition $\rho_{0} = \{\{1,2\},\{3, 4, 5\}\}$, and use different values for the respective distribution shrinkage parameters.  For the iCRP prior we fix a global $\alpha$ considering values between $(0,1)$. For the CPP, we take the parameter $\psi \in (1, 10)$, while for the LSP prior $\nu \in (0, 3)$. 

It is important to remark that the values of the tuning parameters for the three distributions are not directly comparable, as they have different interpretations and scales. Regardless, Figure~\ref{fig:inducedPrior_comparison} gives some insights into the different behavior of the three prior distributions and the effect of their shrinkage parameters. 
The iCRP model and LSP prior converge to $\rho_0$, when values of their shrinkage parameter are $\alpha = 1$ and $\nu = 0$ respectively; under the CPP convergence to $\rho_0$ is formally obtained for $\psi \rightarrow\infty$, although in this example for $\psi = 15$ the probability assigned to $\rho_0$ is roughly $0.99$. The iCRP and CPP also include the CRP as a limiting distribution when their shrinkage parameters are equal to zero. Overall, the three priors show different characteristics in how they increase the probability weight of the initial partition, with the iCRP and LSP showing a ``concave'' shape and the CPP a ``convex'' one.   

The informed partition model can use different $\alpha_{i}$ values for each unit, allowing a more flexible representation of the partition distribution than the one induced by the CPP and LSP priors. We refer to Figure~\ref{fig:enter-label} in the Supplementary Material for an example where we consider different values of these parameters at the unit level. 


\subsection{Prior simulation: Multiple time points}
\label{sec:prior.sim.mult.point}

We next consider $m = 20$ observations and $T = 9$ time points and use Monte Carlo simulation to generate sequences of dependent partitions following models described in the previous section: conditional independence and Markovian models for $\rho$, and the latent autoregressive model for $\mathbf{A}$. In all cases, we set $\rho_0$ to a partition with four clusters each with five units, and use $5,000$ simulations. For the conditional independent and Markovian models, we use the unit local prior and use $\bm{\alpha}_{t\bullet} = (0.25\bm{1}_5, 0.5\bm{1}_5, 0.75\bm{1}_5,0.95\bm{1}_5)$ for all $t$, where $\bm{1}_r$ is a vector filled with $r$ ones and set $M = 1$. This means that we are giving higher prior weight to the fourth and third clusters. We induce a similar prior for the latent autoregressive model case, setting  $\bm{\xi_{0}} = (-2.5\bm{1}_5, -1\bm{1}_5, 1\bm{1}_5, 2.5\bm{1}_5)$. We fix $\beta_{0i} = 0$ and $\kappa_i = 0.5$, for $i = 1, \ldots, m$ and consider the cases where the process is stationary (AR, S) with $\beta_{1i} = 0.8$ and non-stationary (AR, NS)  with $\beta_{1i} = 1.5$, $i = 1, \ldots, m$.

\begin{figure}[H]
  \centering
\includegraphics[trim={0 0cm 0 0cm}, clip,width = 1\textwidth]{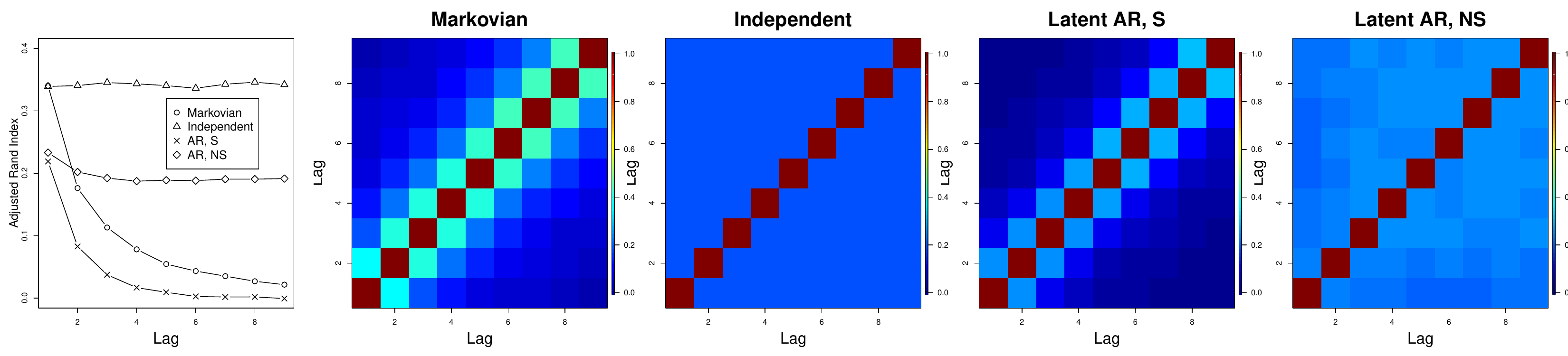}
  \caption{The first panel from the left shows the adjusted Rand index (ARI) between $\rho_t$ and $\rho_0$ for each $t$ under different models. The other panels show the pairwise ARI between $\rho_t$ and $\rho_{t'}$ for $t \ne t'$ for each model. 
  }
  \label{fig:mark_indep}
\end{figure}

For each Monte Carlo simulation, we computed the adjusted Rand index (ARI) \citep{hubert&arabie:1985} between $\rho_t$ and $\rho_0$ denoted by $ARI(\rho_t, \rho_0)$ and the lagged ARI between each time point $ARI(\rho_t, \rho_{t'})$. Figure \ref{fig:mark_indep} shows results averaged across $5,000$ simulations from the prior. As expected, the time lagged-ARI with $\rho_0$ decreases under Markovian dependence while remaining constant under the conditional independence (first panel of Figure \ref{fig:mark_indep}). The latent autoregressive process shows different behaviors depending on whether the process is stationary or not. 
For the pairwise time-lagged ARI under different models, the Markovian and stationary latent regression models display intuitive temporal decay as the time between partitions increases, while the conditionally independent model remains constant. Finally, the nonstationary autoregressive latent process shows a smoother dependence between partitions.

\begin{figure}[H]
  \centering
  \includegraphics[width = 1\textwidth]{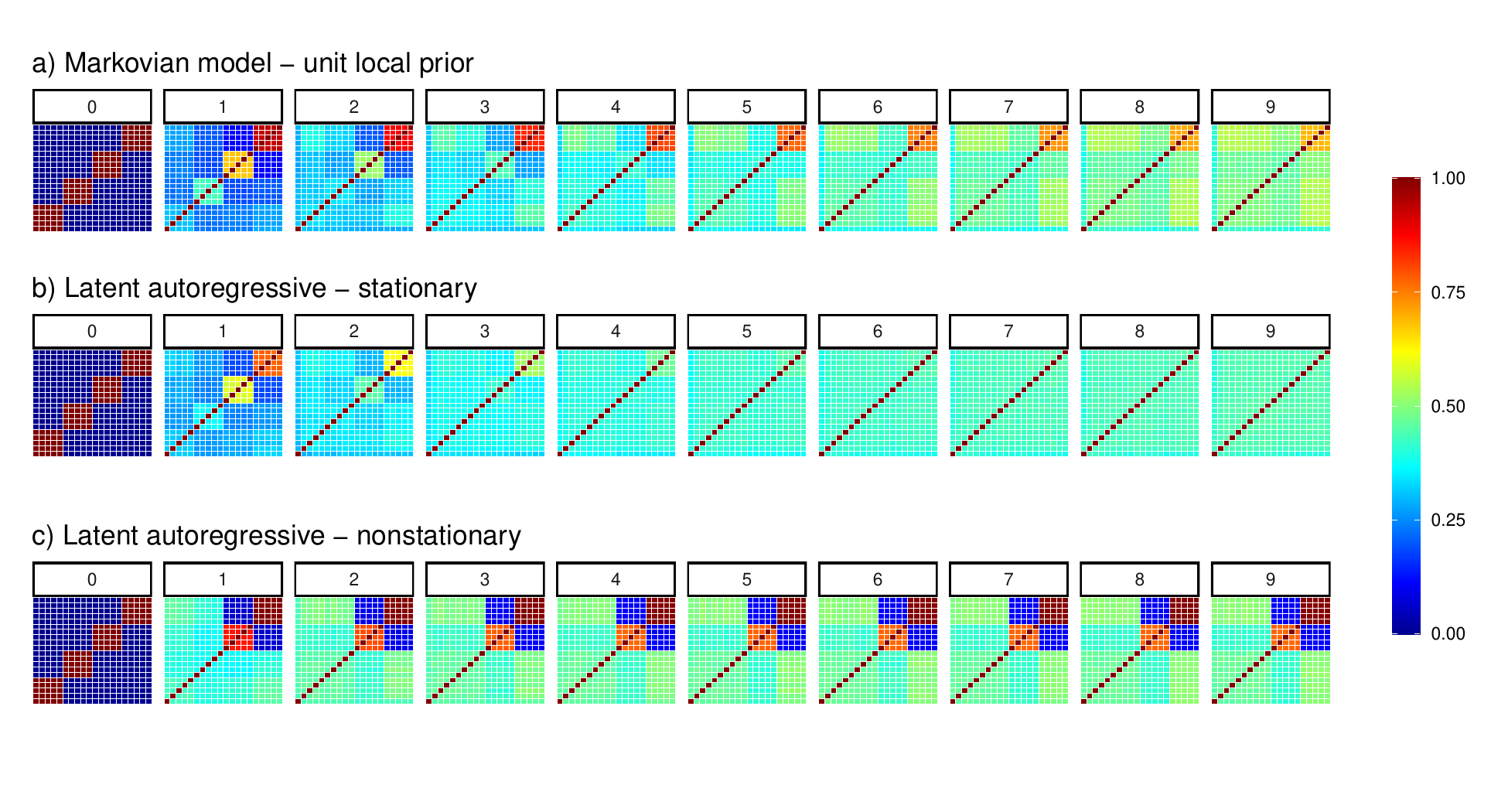}
  \caption{Monte Carlo estimate of pairwise probability matrices over time using the Markovian and latent autoregressive (stationary, nonstationary) models. The initial partition $\rho_0$ is displayed in panel 0 of each sequence. }
  \label{fig:pairwise.prob.lag}
\end{figure}

In Figure \ref{fig:pairwise.prob.lag} we display the $20\times 20$ pairwise co-clustering probability or similarity matrices at each time, for the Markovian and latent autoregressive models. The first panel in each sequence shows the initial partition. For all models, as time increases the weight of $\rho_0$ on the pairwise probabilities decreases particularly for those clusters whose $\alpha$ value is small. The cluster that corresponds to $\alpha = 0.95$ has pairwise probabilities that persist (i.e., remain greater than 0.5) with the extent of persistence depending on the model specification.

\subsection{Posterior simulation: Proof of Concept With One Time Point} \label{sec:post.sim.onetime.point}

In this simulation study, we explore how our prior construction is able to recover the true partition under different scenarios. To this end, we generate data sets of $100$ observations, with four clusters each comprised of 25 observations, whose values are randomly generated from a \fq Gaussian or a $t_3$ distribution. The latter is included to study the performance of the iCRP under a misspecified model.  \qf  To investigate how cluster separation impacts partition estimates, we set cluster means to $(-h, 0, h, 2h)$, for $h = 1, 2, 3$.   Using a standard deviation of 0.5 in all data-generating scenarios,  $100$ replications are generated per scenario.

We consider the following model where $\rho$ is expressed using cluster labels $c_1, \ldots, c_m$, 
\begin{equation} \label{eq:simple.model}
\begin{aligned}
Y_{i} \mid  \bm{\mu}^{\star}, \bm{\sigma}^{2\star}, \bm{c} & \stackrel{ind}{\sim} N(\mu^{\star}_{c_{i}}, \sigma_{c_{i}}^{2\star}), \ i = 1, \ldots, m    \\
(\mu_{j}^{\star}, \sigma^{\star}_{j})\mid \theta, \tau^2 & \stackrel{ind}{\sim} N(\theta, \tau^2) \times UN(0,A_{\sigma}), \ j = 1, \dots, k ,\\
(\theta, \tau) & \stackrel{iid}{\sim} N(m_0, s_0^2) \times UN(0, A_{\tau}),\\
\rho \mid  \bm{\alpha} & \sim iCRP(\rho_0, \bm{\alpha}, M).\\
\end{aligned}
\end{equation}
Recall that in this case with one partition, $\bm{\alpha}$ reduces to the vector $(\alpha_1, \ldots, \alpha_m)$.
Values of the hyperparameters are chosen as $A_{\sigma} = \mbox{sd}(Y)/2$, $m_0 = 0$, $s^2_0 = 100^2$, $A_{\tau} = 100$, and $M = 1$. We first investigate the effect of our prior construction for fixed values of $\bm{\alpha}$, while in a second set of simulations, we explore the effect that the prior specification for $\bm{\alpha}$ has on posterior inference. For the initial partition $\rho_0$, we consider the following: 
\begin{itemize}
  \item $\rho_0=\rho_{\text{true}}$: a partition that corresponds to that which generated the data,
  \item $\rho_0=\rho_{\text{merge}}$: a partition that contains only two clusters each with 50 units that result in merging clusters with means  $(-h, 0)$ and also those with means $(h, 2h)$, 
  \item $\rho_0=\rho_{\text{split}}$: a partition that contains eight clusters by evenly randomly splitting the four original clusters. 
    \fq \item $\rho_0=\rho_{\text{kmeans}}$: a partition that is empirically produced using the k-means algorithm with four clusters.  \qf
\end{itemize}

\fq Generally, we interpret the initial partition as the summary of domain knowledge, for example, coming from previous experiments. In the last case, $\rho_{\text{kmeans}}$ is defined using the response in model \eqref{eq:simple.model}. We include the case for comparison, but we would discourage this as a general approach, as it uses the response twice. In this simulation, note also that $\rho_{\text{kmeans}}$ has an inherent advantage over $\rho_{\text{merge}}$ and $\rho_{\text{split}}$ as it is computed using the correct number of clusters and changes with each simulated data set. An example of each of the $\rho_0$ partitions (save $\rho_{\text{kmeans}}$) \qf is provided in the first row of Figure~\ref{fig:init_part} of the Supplementary Material. 

\begin{figure}
  \centering
  \includegraphics[width=1\textwidth]{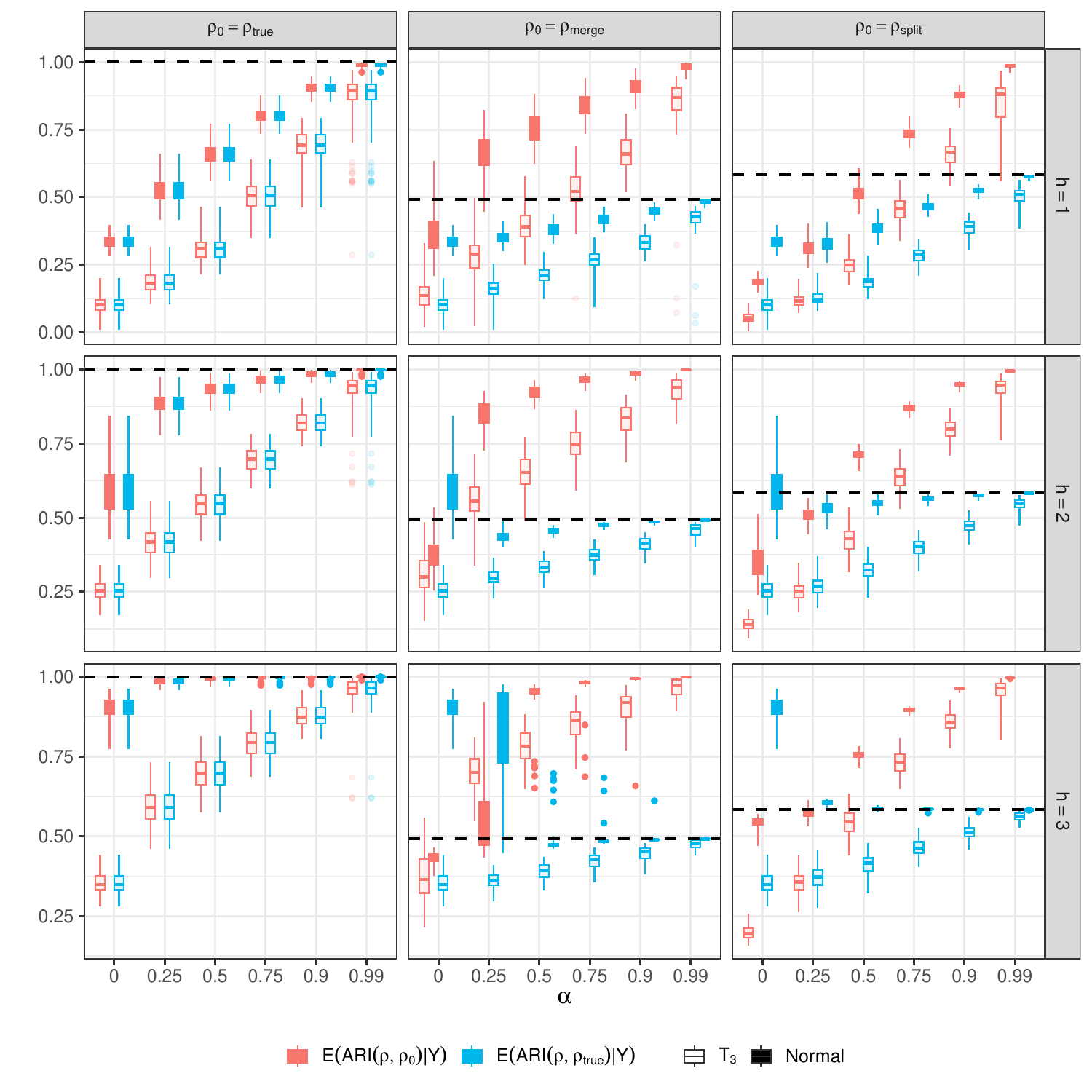}
    \caption{Results for fixed global $\alpha$. Distribution of $E(ARI(\rho, \rho_{0})\mid \bm{Y})$ (red boxplots) and $E(ARI(\rho, \rho_{true})\mid\bm{Y})$ (blue boxplots) across $100$ replicated data sets, for each value of $\alpha \in \{0, 0.25, 0.5, 0.75, 0.9, 0.99\}$ are displayed. \fq Solid and transparent fill distinguish the data-generating scenario (Normal and $t_3$ distribution). \qf  Each panel shows results for different combinations of the cluster mean separation values used in the data-generating process and the type of initial partition $\rho_0$. The black dashed line is $ARI(\rho_{\text{true}}, \rho_0)$.
    }
   \label{fig:post.sim.merge}
\end{figure}

To each simulated data set, we fit model \eqref{eq:simple.model} fixing $\alpha_{i}=\alpha$ for all $i = 1, \ldots, m$ and considering values for $\alpha \in \{0, 0.25, 0.5, 0.75, 0.9, 0.99\}$.  Model fitting was carried out by collecting 1,000 MCMC samples after discarding the first 10,000 as burn-in and thinning by 10 (resulting in a total of 20,000 MCMC samples).  To measure the accuracy of the partition estimates, we employ the adjusted Rand index $ARI(\rho, \rho_0)$.  Results are provided in Figure \ref{fig:post.sim.merge}.

In Figure \ref{fig:post.sim.merge} the dashed line corresponds to $ARI(\rho_0, \rho_{true})$, the \fq solid \qf red boxplots correspond to $E(ARI(\rho, \rho_{0}) \mid \bm{Y})$ (i.e., $\frac{1}{B} \sum_{b = 1}^B ARI(\rho^{(b)}, \rho_0)$ where $B$ is the number of posterior samples) \fq for data generated using a Normal distribution. The transparent red boxplots display results for data generated using a $t_3$ distribution. \qf  Similarly, the \fq solid \qf blue boxplots correspond to the $E(ARI(\rho, \rho_{true})\mid \bm{Y})$ \fq for normal data and transparent blue boxplots to $t_3$ data. We do not include results for $\rho_{\text{kmeans}}$, since the initial partition is not the same across replicated data sets. As a result, the definition of $ARI(\rho, \rho_{0})$ is not the same as in the other cases. \qf  Notice that as $\alpha$ increases, $E(ARI(\rho, \rho_{0})\mid \bm{Y})$ approaches 1 and $E(ARI(\rho, \rho_{true})\mid \bm{Y})$ approaches $ARI(\rho_0, \rho_{true})$ for both simulation scenarios. This illustrates the expected behavior that the prior carries more weight in the estimation of $\rho$ as $\alpha$ increases.  This, as expected, can negatively affect model fit if $\rho_0$ is ``poorly'' specified.  To see this, we provide Figure \ref{suppmat_fig:waic_simstudy2} in the Supplementary Material.  This figure displays WAIC values (smaller is better) associated with model fit. It turns out that $\rho_0 = \rho_{\text{merge}}$ results in the largest WAIC value.  This is due to the fact that clusters in $\rho_{\text{merge}}$ group units with dissimilar measurements, unlike the case for $\rho_{\text{split}}$ or $\rho_{\text{true}}$.  \fq As expected, the misspecified model ($t_3$) performed worse than the correctly specified model. \qf

Using the same synthetic data, we also fit a version of model \eqref{eq:simple.model} that treats $\bm{\alpha}$ as an unknown parameter and employs the global and unit local prior as described in Table~\ref{tab:alpha_models}. This will permit studying how $\alpha$ (or $\bm{\alpha} = (\alpha_1, \ldots, \alpha_m)$ for the unit local prior) are learned from the data.   To this end, for the global prior we consider   $\alpha \sim \Be(1,1)$, $\alpha \sim \Be(1,9)$, or $\alpha \sim \Be(10,10)$, and for the unit local prior $\alpha_i \stackrel{iid}{\sim} \Be(1,1)$, $\alpha_i \stackrel{iid}{\sim} \Be(1,9)$, or $\alpha_i \stackrel{iid}{\sim} \Be(10,10)$.   Since the prior on $\sigma^{\star}_j$ also impacts clustering, (clusters are {\it a posteriori} more homogeneous for smaller values of $A_{\sigma}$), we considered $A_{\sigma} \in \{0.5, 1, 1.5\}$.  
%
%
Values of the other hyperparameters are the same as in the previous simulation, where $m_0 = 0$, $s^2_0 = 100^2$, $A_{\tau} = 100$, and $M = 1$. In addition to exploring different initial partitions $\rho_0 \in \{\rho_{\text{true}}, \rho_{\text{merge}}, \rho_{\text{split}}, \rho_{\text{kmeans}}\}$, we also fit the model in \eqref{eq:simple.model} without an initial partition, i.e., assume that $\rho \sim CRP(M=1)$ in \eqref{eq:crpEPPF}. This allows us to understand under which settings using prior information aids in performance.

We use the same MCMC settings as for the previous simulations. For each replicated data set we recorded the posterior mean of the $\alpha$ parameter when using the global prior, and the average of the posterior means of $\alpha_i$ for $i = 1, \ldots, m$ when using the unit local prior.  In addition, we recorded $E(ARI(\rho, \rho_0) \mid \bm{Y})$ and the log pseudo marginal likelihood (LPML) as defined in \cite{geisser&eddy:79}.    

As expected, the posterior distribution of $\alpha$ is less sensitive to prior specification under the global prior compared to the unit local prior.  Under the unit local prior, the posterior mean of each $\alpha_i$ is pulled towards the prior mean because a single $\gamma_i$ is informing $\alpha_i$, for $i = 1, \ldots, m$ .
However, the trends and comparisons between the different $\rho_0$ partitions associated for each prior are similar.  For this reason here we provide results for the global prior in Figures \ref{fig:mean_alpha_global} - \ref{fig:lpml_global} and those for unit local prior in the Supplementary Material (Figures \ref{fig:mean_alpha_local} - \ref{fig:lpml_local}).  

\begin{figure} 
  \centering
  \includegraphics[width=1\textwidth, page=1]{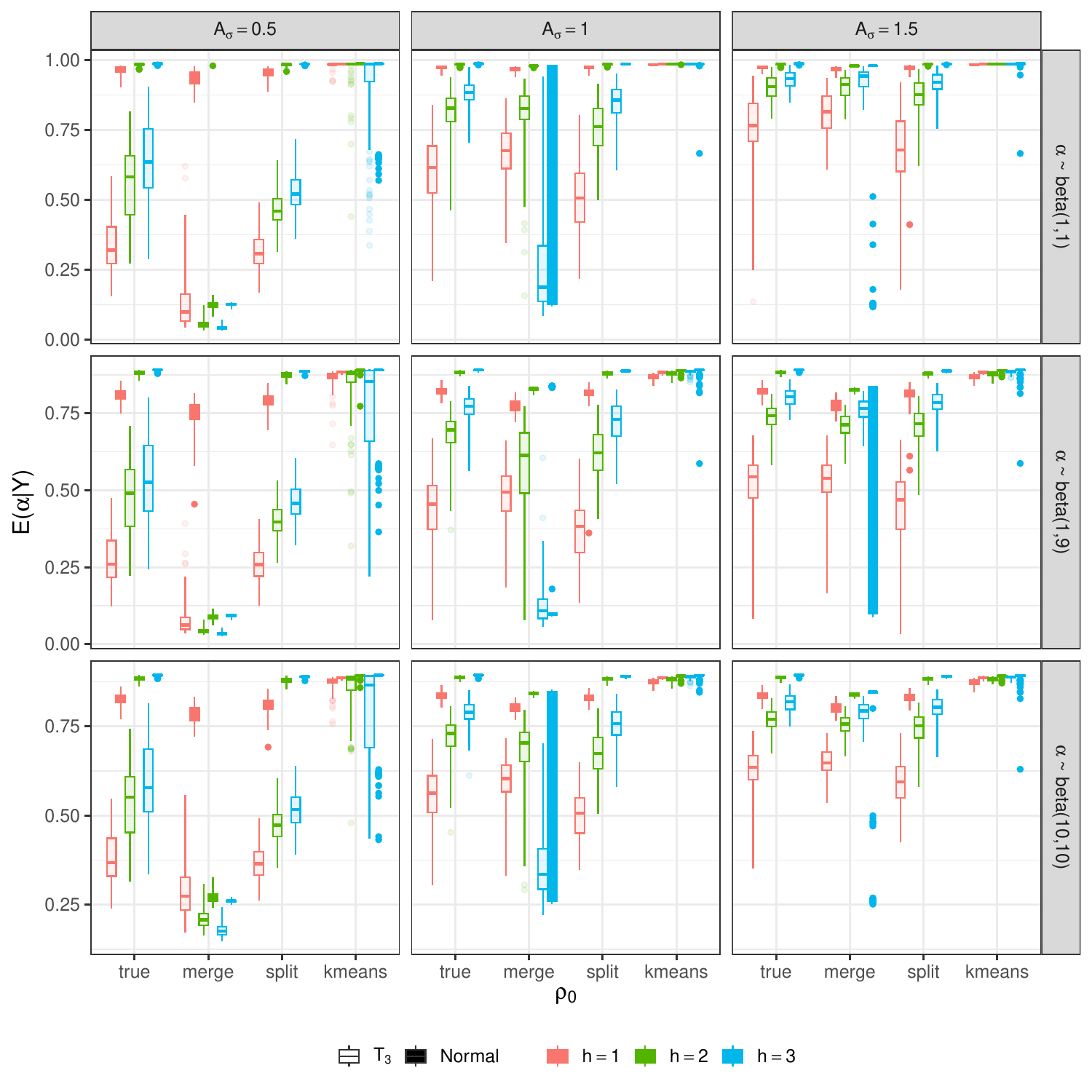}
    \caption{ Results for random global $\alpha$. Distribution of the posterior mean of $\alpha$ for different choices of $\rho_0$ (x-axis) across $100$ replicated data sets using different values for the cluster means separation (boxplot colors). \fq Solid and transparent fill distinguish the data-generating scenario (Normal and $t_3$ distribution). \qf Each panel shows results for different combinations of prior choices for $\alpha$ and $A_{\sigma}$.
    }
   \label{fig:mean_alpha_global}
\end{figure}

Figure \ref{fig:mean_alpha_global} displays the distribution of  $E(\alpha \mid \bm{Y})$ across $100$ replicates  for the global prior \fq for both types of data\qf. Notice that the estimated value of $\alpha$ depends quite heavily on an interaction between the initial partition $\rho_0$, $A_{\sigma}$, and the informativeness of cluster membership in the data (i.e., the value of $h$ when generating data).  Recall that as $A_{\sigma}$ decreases, clusters become more homogeneous {\it a posteriori}.  
Thus, if the initial partition $\rho_0$ groups units with response values that are quite different, then the estimated value of $\alpha$ must be small to allow many units to be reallocated.
We observe this phenomenon in Figure \ref{fig:mean_alpha_global} for $\rho_0 = \rho_{\text{merge}}$ and $A_{\sigma} = 0.5$.  However, since $\rho_{\text{split}}$ does not contain clusters with heterogeneous response values, the posterior estimate of $\alpha$ is close to $1$ for all values of $A_{\sigma}$,  indicating that a relatively small number of units are being moved away from $\rho_0$.  This results in a reduced ARI values (see Figure \ref{fig:ari_global}) as the data are not informative enough to discredit the formation of the extraneous clusters that exist in $\rho_0$. Evidence of this can be seen in Figure \ref{fig:lpml_global} where the LPML values for $\rho_0 = \rho_{\text{split}}$ are quite competitive even though the partition estimate is biased \fq(regardless of the type of data generated)\qf.  

With regards to $ARI(\rho, \rho_{true})$, Figure \ref{fig:ari_global} shows that if cluster membership is not well informed by the data (i.e., $h = 1$), then employing any of the $\rho_0$ we consider  provides benefit.  However, as cluster membership becomes more informed by data (i.e., $h = 2,3$), employing an initial partition that allocates units with similar response values into different groups (e.g., $\rho_{\text{merge}}$ or $\rho_{\text{split}}$) can adversely affect the partition estimate. This is exacerbated when $A_{\sigma}$ is large, which allows for a large likelihood variance favoring heterogeneous clusters. Thus, it seems that the data can ``overcome'' more readily a misspecified initial partition $\rho_0$ that has too few clusters (e.g., $\rho_{\text{merge}})$, than one that has too many (e.g., $\rho_{\text{split}}$) if units with similar responses are allocated to different clusters {\it a priori}.
This behavior is somewhat mitigated when using the unit local prior with $\alpha_i \sim \Be(1,9)$ which expresses uncertainty with the $\rho_0$ employed (see Figure \ref{fig:ari_local} of the Supplementary Material).

\begin{figure}
  \centering
  \includegraphics[width=1\textwidth, page=3]{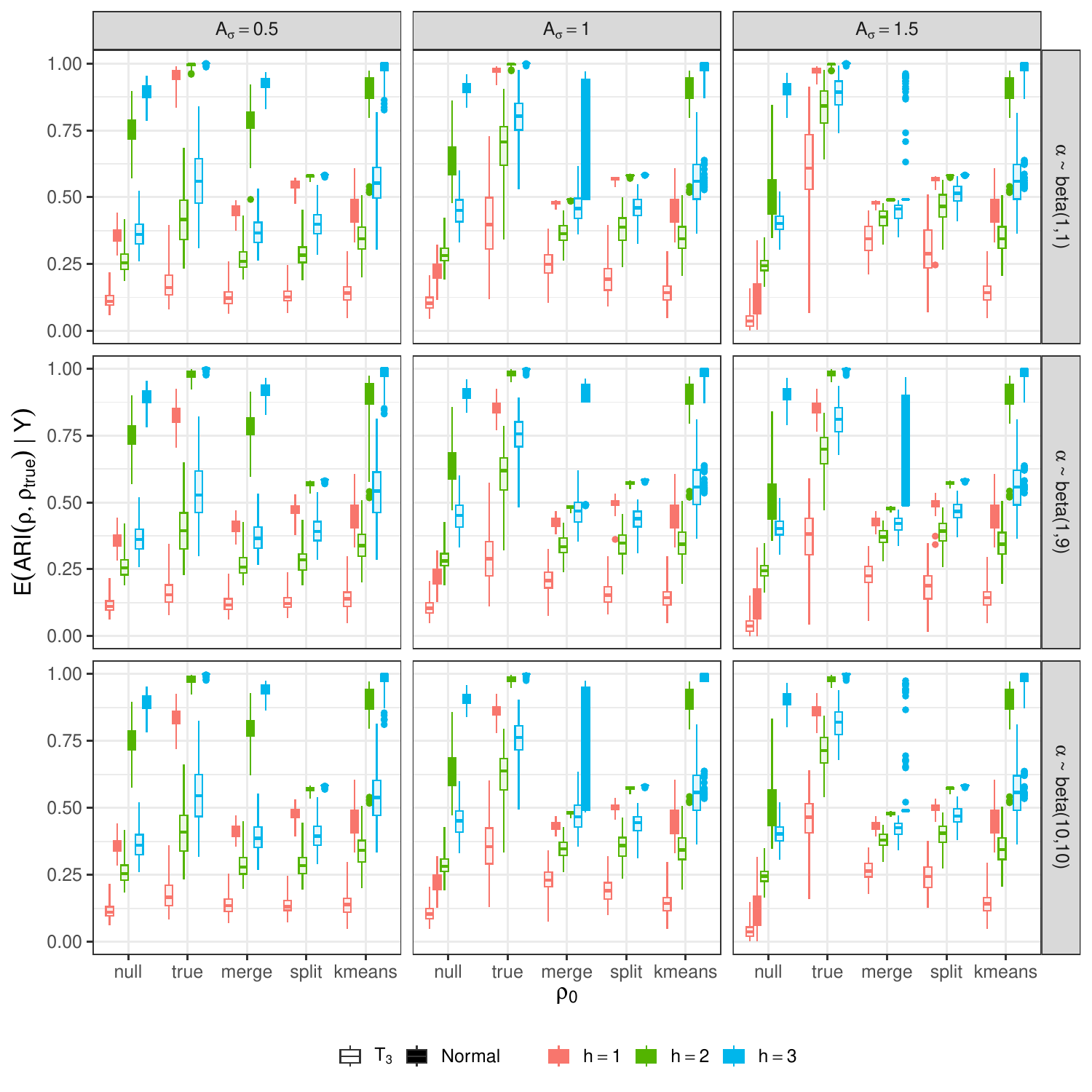}
    \caption{Results for random global $\alpha$. Distribution of $E(ARI(\rho, \rho_{true})\mid \bm{Y})$ for different choices of $\rho_0$ ($x$-axis) across $100$ replicated data sets using different values for the cluster means separation (boxplot colors). \fq Solid and transparent fill distinguish the data-generating scenario (Normal and $t_3$ distribution). \qf Each panel shows results for different combinations of prior choices for $\alpha$ and $A_{\sigma}$. Here $\rho_0 = \mbox{null}$ corresponds to a model that does not include an initial partition. Notice that when $\rho_0 = \mbox{null}$ results do not change for different values of $\alpha$, as that parameter is not included in the model. }
   \label{fig:ari_global}
\end{figure}

With regards to LPML, Figure \ref{fig:lpml_global} shows that including $\rho_0$ only produces adverse effects when the initial partition groups units that produce quite different response values (e.g., $\rho_{\text{merge}}$). Additionally, this only occurs for higher values of the hyperparameter $A_{\sigma}$, that induce a large likelihood variance.  A large likelihood variance permits clusters to be quite heterogeneous and as a result, accommodates a $\rho_0$ that has heterogeneous clusters. Thus, the use of an initial partition has to be accompanied by careful elicitation of the hyperparameters of the priors for the cluster parameters. \fq It is worth nothing that $\rho_{\text{kmeans}}$ results in the best fit according to LPML, particularly when there is large cluster overlap ($h=1$).  But this result is expected as $\rho_{\text{kmeans}}$ groups data into the correct number of clusters {\it a priori}\qf. Overall, the simulations in this section highlight two main results: i) a reasonable initial partition provides benefit in both, partition estimation and model fit; and  ii) the choice of moderately informative hyperparameters for the cluster parameters will have more influence on the posterior estimate of $\rho$ compared to a diffuse prior (which is often the case in Bayesian models). \fq Although our simulation does provide information about how our proposed procedure behaves for a few $\rho_0$ vectors, we would recommend eliciting $\rho_0$ on a case-by-case basis, as dictated by the specific application. \qf

\begin{figure} 
  \centering
  \includegraphics[width=1\textwidth, page=9]{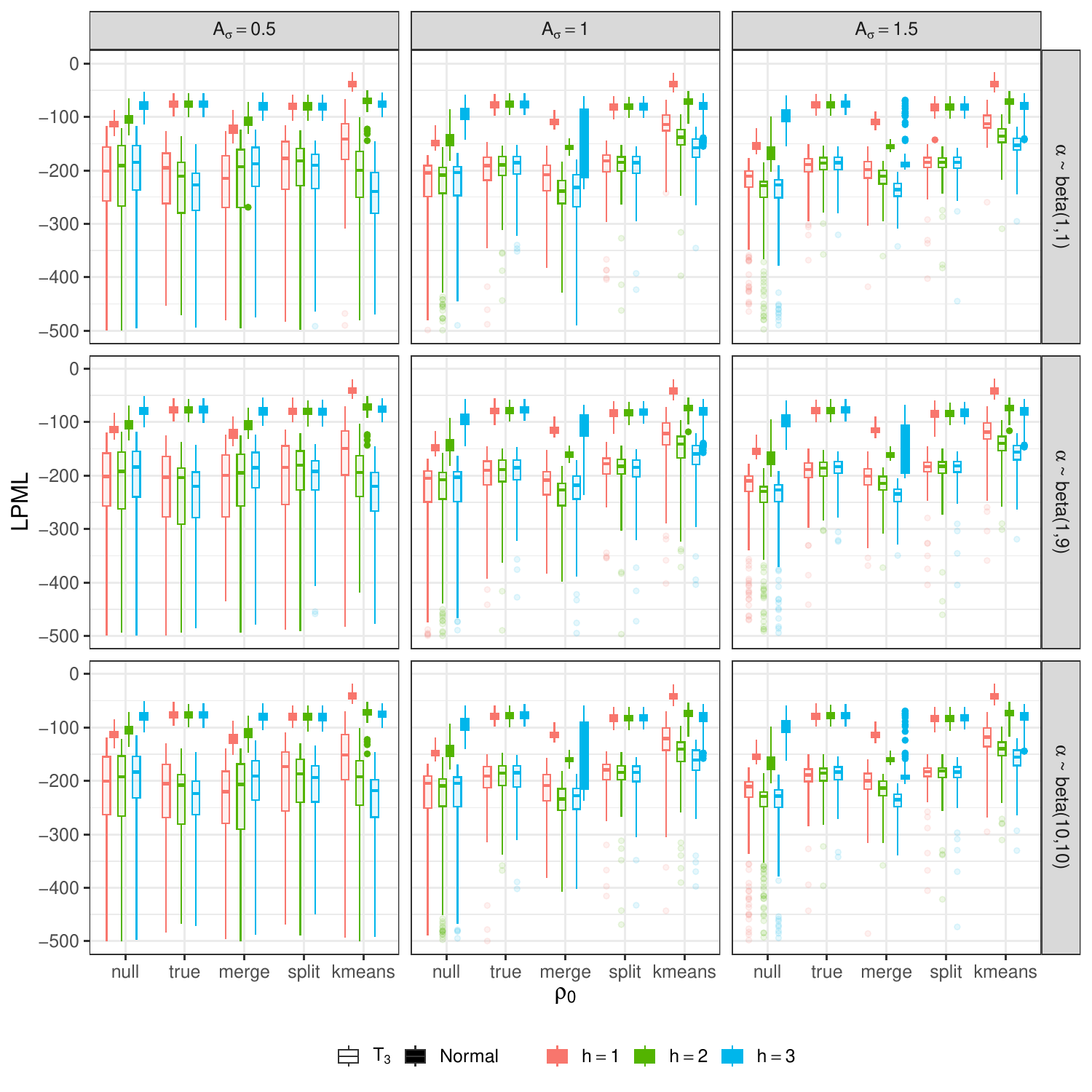}
    \caption{Results for random global $\alpha$. Distribution of LPML for different choices of $\rho_0$ ($x$-axis) across $100$ replicated data sets using different values for the cluster means separation (boxplot colors). \fq Solid and transparent fill distinguish the data-generating scenario (Normal and $t_3$ distribution). \qf Larger values of LPML indicate a better fit.  Each panel shows results for different combinations of prior choices for $\alpha$ and $A_{\sigma}$. Here $\rho_0 = \mbox{null}$ corresponds to a model that does not include an initial partition. Notice that when $\rho_0 = \mbox{null}$ results do not change for different values of $\alpha$, as that parameter is not included in the model. 
    }
   \label{fig:lpml_global}
\end{figure}

%

\subsection{Posterior simulation: comparison with other informative priors} \label{seq:post_sim_comparison}

To compare our construction with those of \cite{paganin_etal:2021} and \cite{smith_allenby:2020}, we consider a small simulation study with data being generated from Gaussian mixtures with different component means and known variance equal to $1$.
As in Section~\ref{sec:post.sim.onetime.point}, we simulate $100$ observations divided into four clusters of equal size, with varying values of the cluster means $(-h, 0, h, 2h)$, for $h = 1, 2, 3$, and use $100$ replications for each scenario.

To focus our investigation only on the effect of partition prior, we consider the following simplified version of the model in \eqref{eq:simple.model}
\begin{equation} \label{eq:simpler.model}
\begin{aligned}
Y_{i} \mid  \bm{\mu}^{\star}, \bm{c} & \stackrel{ind}{\sim} N(\mu^{\star}_{c_{i}},1), \ i = 1, \ldots, m    \\
\mu_{j}^{\star}\mid \theta, \tau^2 & \stackrel{ind}{\sim} N(\theta, \tau^2), \ j = 1, \dots, k, \\
\rho \mid\rho_0 & \sim p(\rho \mid \rho_0 ), 
\end{aligned}
\end{equation}
where $p(\rho \mid \rho_0 )$ indicates an informed prior for the random partition (e.g., iCRP, CPP, or LSP) and hyperparameters are fixed to $\theta = 0$ and $\tau^2 = 10$ across all the choices of partition priors. We considered different choices for the initial partition, with $\rho_0 \in \{ \rho_{\text{true}}, \rho_{\text{merge}}, \rho_{\text{split}} \}$. 
When using our informed partition model, we set $p(\rho \mid \rho_0) = iCRP(\rho_0, \bm{\alpha}, M)$ in \eqref{eq:simpler.model}, with $M = 1$ and fix $\alpha_{i} = \alpha$ for $i = 1, \ldots, m$ to the following values $\alpha \in \{0, 0.25, 0.5, 0.75, 0.9, 0.99\}$. When using the CPP, we set $p(\rho \mid\rho_0 ) = CPP(\rho_0,\psi, M)$ in \eqref{eq:simpler.model} with $M = 1$ and $\psi \in \{0, 10, 20, 50, 80, 100\}$. For each data set, we ran both models with different values of their tuning parameters for $10,000$ iterations and collected $5,000$ samples after burn-in. Finally, for the LSP prior we set $p(\rho \mid\rho_0 ) = LSP(\rho_0,\nu)$ and consider different values for the variance $\nu \in \{10, 5, 1, 0.05, 1/(m\log(m)), 0.1/(m\log(m))\}$. Recall that for the LSP prior smaller values represent more informative priors. Since posterior sampling from a model including the LSP prior uses a Metropolis-Hastings algorithm, we consider $50,000$ iterations and discard half of the samples as a burn-in with thinning equal to $5$.

\begin{figure}
  \centering
  \includegraphics[width=1\textwidth]{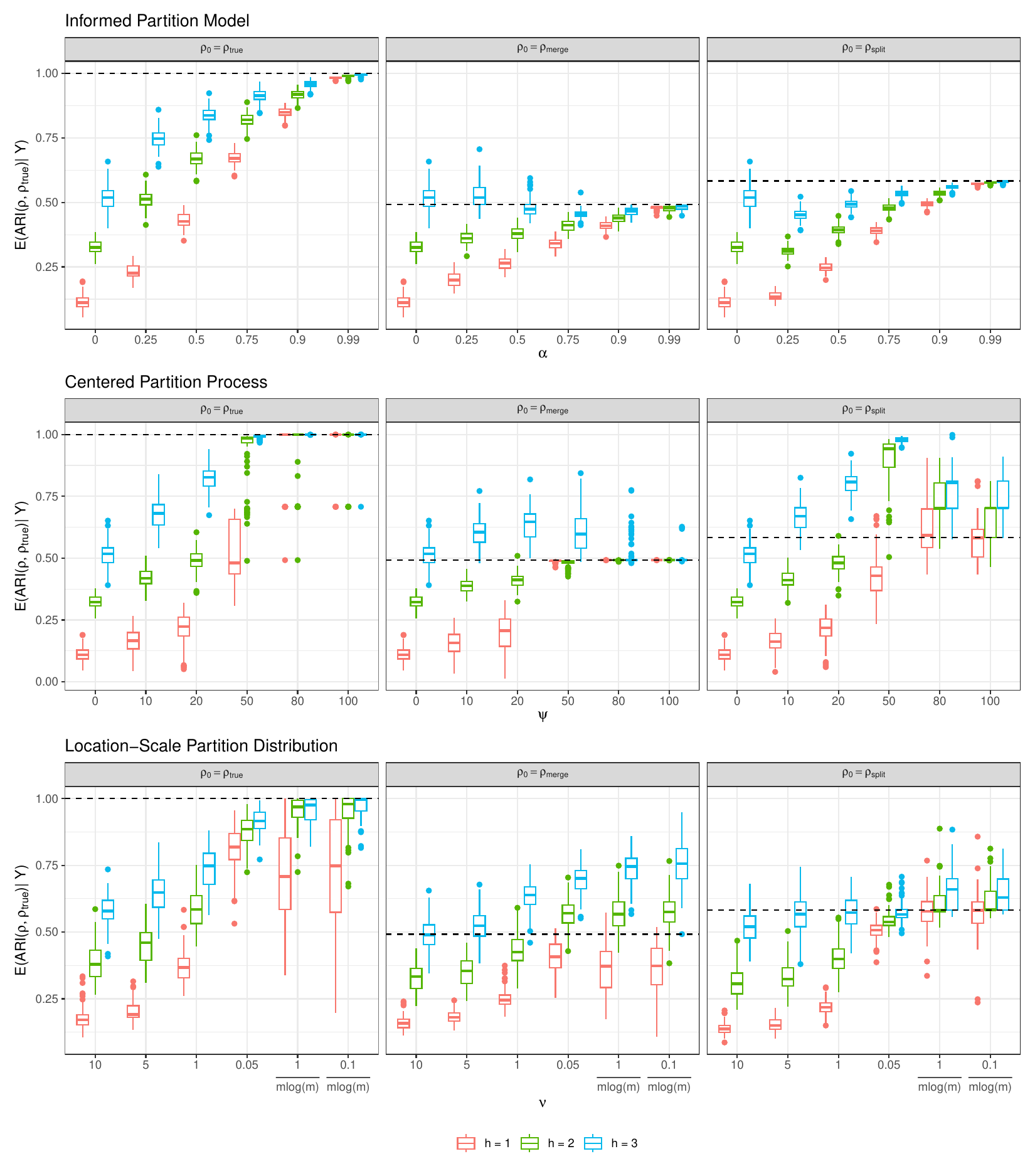}
    \caption{Comparison of posterior results using the iCRP, CPP, and LSP priors for different values of their tuning parameters. Each boxplot represents the distribution of $E(ARI(\rho, \rho_0)\mid \bm{Y})$ across the $100$ generated data sets, with colors distinguishing between data-generating scenarios. The black dashed line is $ARI(\rho_{\text{true}}, \rho_0)$. Notice that the values of the tuning parameters are not directly comparable.}
    \label{fig:posterior_comparison}
\end{figure}

Figure~\ref{fig:posterior_comparison} shows the distribution of $E(ARI(\rho, \rho_0) \mid \bm{Y})$ under the three partition priors for the different simulation scenarios.  Although values of the tuning parameters are not directly comparable, as the priors become more informative, the $E(ARI(\rho, \rho_0) \mid \bm{Y})$ values for the iCRP and CPP models converge to $ARI(\rho_{\text{true}}, \rho_0)$, i.e. the posterior distribution for $\rho$ becomes more and more concentrated on the initial partition $\rho_0$. This also happens for the LSP prior, but not for all the cases. In particular, when  $\rho_0 = \rho_{\text{merge}}$ the posterior converges to different partitions for each degree of mean separation $h$. This also happens when $\rho_0 = \rho_{\text{true}}$ and $h = 1$.

Overall, under the CPP and LSP priors, there is larger variability in $E(ARI(\rho, \rho_0) \mid \bm{Y})$ compared to the iCRP prior and each degree of mean separation $h$, with $h = 1 $ corresponding to the case of largest variability. This can be interpreted as beneficial when using a misspecified initial partition, as in some cases the posterior is concentrated on partitions that are closer to the true one (e.g., $\rho_0 = \rho_{\text{split}}$ and $h \in (2,3)$ for the CRP, or $\rho_0 = \rho_{\text{merge}}$ and $h = 3$ for the LSP). However, for these two priors, there is higher uncertainty also when the $\rho_0 = \rho_{\text{true}}$, especially when $h = 1$ for the LSP prior which is not desirable. Instead under the iCRP, informative priors lead to small uncertainty in the posterior distribution for the partition, regardless of the initial partition. It is worth mentioning that in this simulation we use a global $\alpha$ parameter for the iCRP, but specifying different degrees of informativeness for each unit can mitigate cases of misspecification of the initial partition.

\section{PM$_{10}$ Application}\label{sect:application}

In this section, we consider an environmental science application to further illustrate the utility of our informed partition procedure. As in Section \ref{sect:simul}, we first consider the case with one time point to focus on the initial partition's impact on partition estimation and model fit and then describe an approach that employs all time points.  We now briefly introduce the data.

The rural background PM$_{10}$ dataset is publicly available in the {\tt gstat} package (\citealt{gstat:2016}) of {\tt R} \citep{R}.  This dataset consists of daily measurements of particulate matter with a diameter of less than 10$\mu$m from 60 atmospheric monitoring stations in rural Germany. Measurements are recorded daily for a number of years, but we focus on average monthly PM$_{10}$ measurements from 2005. In addition to PM$_{10}$ measurements, the longitude and latitude of each monitoring station are recorded.

\subsection{PM$_{10}$ Dataset with One Time Point} \label{sec:pm10.one.time.point}
We begin by focusing only on the average PM$_{10}$ measurements for the month of January. To these measurements, we fit \fq a number of \qf models all of which are based on the model detailed in Section~\ref{sec:post.sim.onetime.point} (see equation \eqref{eq:simple.model}).  \fq For context, we consider a model for which no initial partition is supplied so that a CRP process is used to model $\rho$.  This model will be referred to as the ``baseline model''.  All other models contain one of two possible initial partitions. Details of both initial partitions are provided shortly, but they are displayed in the top left plot of Figures \ref{fig:pm10.ip} and \ref{fig:pm10.ip_kmeans} of the supplementary material.  We also consider the global model for $\bm{\alpha}$ and two versions of the unit local model for $\bm{\alpha}$ (see Table \ref{tab:alpha_models}), where $\bm{\alpha}= (\alpha_1, \ldots, \alpha_m$) 
which will be detailed shortly. \qf 

\fq Due to the spatial nature of atmospheric measurements like PM$_{10}$, both intial partitions we consider contain spatially consistent clusters. The first is comprised of nine clusters with cluster membership being established by determining to which of the nine quadrants displayed in the top left plot of Figure \ref{fig:pm10.ip} a particular station's latitude and longitude belong. The top left plot of Figure \ref{fig:pm10.ip} displays the initial partition with the nine clusters being distinguished by color, numbers indicate the monitoring station, and size is proportional to the monitoring station's PM$_{10}$ measurement.  This initial partition will be referred to as $\rho_{\text{quadrant}}$.  The second initial partition is also comprised of nine clusters that were created using K-means based on each monitoring station's longitude and latitude.  This initial partition is displayed in the top left plot of Figure \ref{fig:pm10.ip_kmeans} of the supplemental material.  This initial partition will be referred to as $\rho_{\text{kmeans}}$.  Note that both initial partitions are based on spatial information, which is not included in the data model. \qf

\begin{figure} 
  \centering
   \includegraphics[width=1\textwidth]{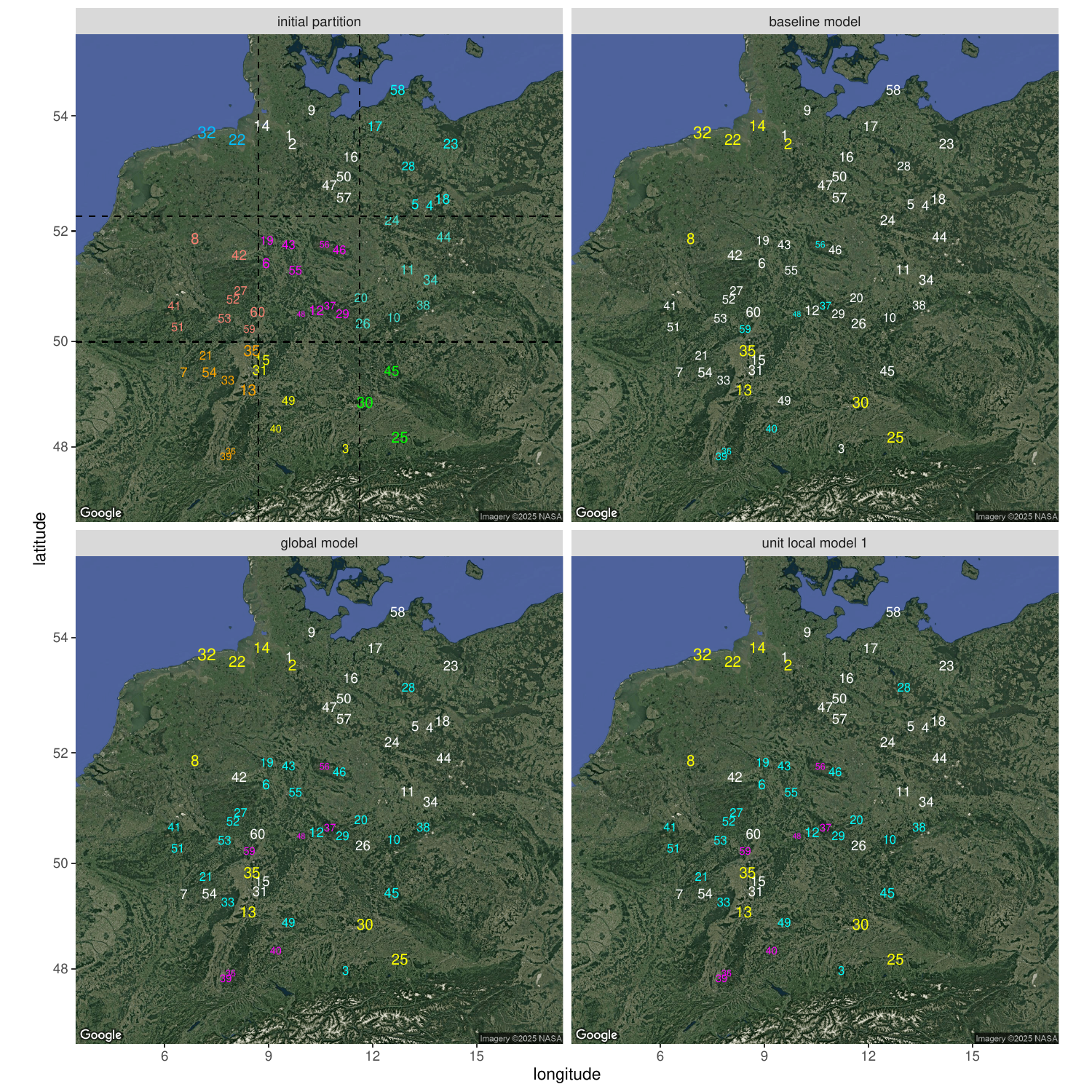}
    \caption{PM$_{10}$ data with number corresponding to station number.  The color in each figure corresponds to cluster membership.  The top left plot displays the initial partition. The top right displays a partition estimate based on a model with no initial partition (baseline model). The bottom left displays a partition estimate based on a model using the initial partition and a global prior on $\bm{\alpha}$ (global model).  The bottom right displays the partition estimate from a model that includes an initial partition and a local prior for $\bm{\alpha}$ \fq(unit local 1 model \qf). }
   \label{fig:pm10.ip}
\end{figure}

Regarding details associated with $\bm{\alpha}$, for the global model where $\alpha_i = \alpha, \forall i$, we assume that $\alpha \sim \Be(a, b)$ with $a=1$ and $b=9$ so that {\it a-priori} the probability that a monitoring station moves away from the initial partition is 0.1.  For the unit local models, we assume $\alpha_i \sim \Be(a_i, b_i)$, allowing different degrees of uncertainty about the prior partition. \fq We specify $a_i$ and $b_i$ in two ways. For the first (unit local 1), note from Figure \ref{fig:pm10.ip} that monitors 36 and 39 are both on the same slope of a small mountain located in the south of Germany, while monitors 22 and 32 are in northern Germany near the border of the North Sea. As a result, there are geographical reasons for desiring these stations to be in the same cluster. Thus, we set $a_i=9$ and $b_i=1$ for stations 36 and 39 in addition to 22 and 32.  For all other stations, we set $a_i = 1$ and $b_i=9$.  The second approach (unit local 2) of setting $a_i$ and $b_i$ is also based on spatial proximity.  The monitoring stations in the central cluster of the partitions displayed in the top left plot of Figures \ref{fig:pm10.ip} and \ref{fig:pm10.ip_kmeans} are less separated than the other clusters.  Thus, for all monitoring stations in the central cluster, we set  $a_i=9$ and $b_i=1$; for the remaining monitoring stations, we set $a_i = 1$ and $b_i=9$. This prior specification favors an entire cluster to remain intact, while the former prior specification favors a pair of stations only to adhere to $\rho_0$.  \qf

Each of the models described was fit by collecting 1,000 MCMC samples after discarding the first 10,000 as burn-in and thinning by 100.  For the hierarchical model specifications we set $A_{\sigma} = 2.5$, $A_{\tau} = 100$, $m_0 = 0$, $s^2_0 = 100$, and $M=1$. To compare the model fits, we computed the LPML and WAIC values of each model. The results are provided in Table \ref{tab:PM10_fits_Jan}.  \fq WAIC indicates that all the models that incorporate $\rho_0$ fit better than those that do not.  LPML is more mixed in this regard, but still favors models with $\rho_0$. It also appears that the added flexibility of the first unit local prior for $\bm{\alpha}$ based on either initial partition results in the best fit.  Due to this, the remainder of the discussion is focused on the unit local 1 model. \qf

 \begin{table}[tb]
 \centering
 \caption{Model fit metrics and posterior co-clustering probabilities between station 22 and 32 and between 36 and 39 along with LPML (larger is better) and WAIC (lower is better) for all models that were fit to data from the month of January in the PM$_{10}$ data.}
\begin{tabular}{l l cc cc} \toprule
$\rho_0$ & model for $\bm{\alpha}$ & LPML & WAIC & $Pr(c_{22} = c_{32}\mid \bm{Y})$  &  $Pr(c_{36} = c_{39}\mid\bm{Y})$ \\\midrule 
none                     & baseline       & -192.65 & 315.37 & 0.71 & 0.69 \\ \midrule
$\rho_{\text{quadrant}}$ & global         & -190.90 & 286.87 & 0.53 & 0.71 \\
$\rho_{\text{quadrant}}$ & unit local 1   & -185.79 & 280.49 & 0.76 & 0.98 \\
$\rho_{\text{quadrant}}$ & unit local 2   & -202.39 & 306.93 & 0.46 & 0.84 \\ \midrule
$\rho_{\text{kmeans}}$   & global         & -199.75 & 290.59 & 0.55 & 0.69 \\
$\rho_{\text{kmeans}}$   & unit local 1   & -189.35 & 280.31 & 0.75 & 0.98 \\
$\rho_{\text{kmeans}}$   & unit local 2   & -205.51 & 294.26 & 0.54 & 0.75 \\\bottomrule
 \end{tabular}
 \label{tab:PM10_fits_Jan}
 \end{table}

\begin{figure}
  \centering
   \includegraphics[width=1\textwidth]{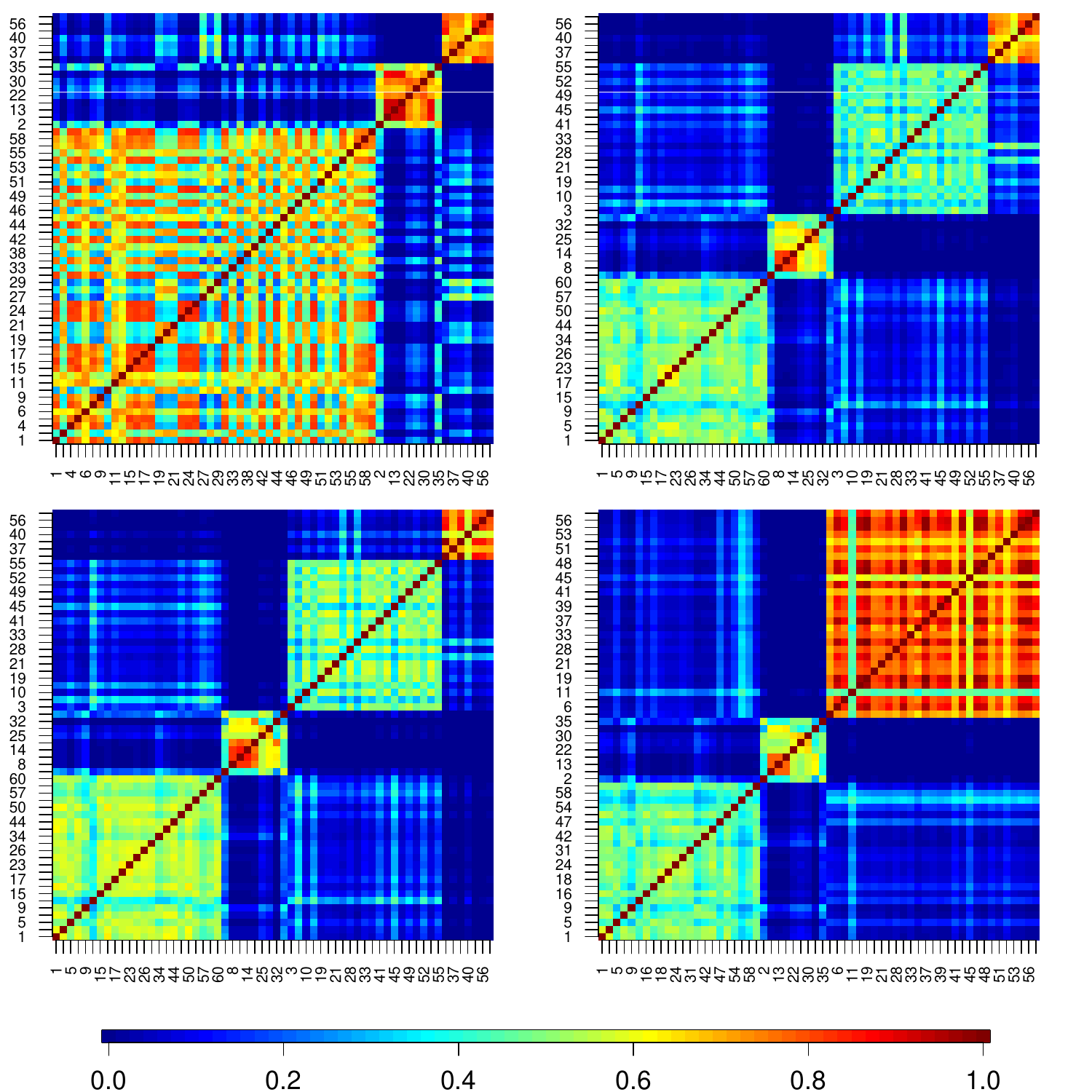}
    \caption{ Posterior co-clustering probabilities for baseline model (top left), global model (top right),  unit local 1 model (bottom left), and unit local 2 model (bottom right).}
   \label{fig:co.clus.prob}
\end{figure}

To obtain a point estimate of the partition for the different models, we minimize the posterior expectation of the variation of information loss (\citealt{meila2007}) using the SALSO method of \cite{dahl_etal:2021} as implemented in the {\tt salso} {\tt R} package (\citealt{dahl_etal:2022}).
\fq The partition estimates for $\rho_{\text{quadrant}}$ are provided in Figure \ref{fig:pm10.ip} and those for $\rho_{\text{kmeans}}$ are provided in Figure \ref{fig:pm10.ip_kmeans}. Since the partition estimates are quite similar, we focus attention on that for $\rho_{\text{quadrant}}$.  \qf Notice that for the baseline model, we estimate one large cluster and two smaller ones, which reflects a clustering behavior typical of the CRP.  For the global and unit local $\bm{\alpha}$ models, the partition estimates are equal and are comprised of four clusters, two of which are medium-sized and two that are a bit smaller.  Differences in the posterior distributions of the partition between the two models do exist, however.   These differences can be seen by way of the co-clustering probabilities, provided in Figure \ref{fig:co.clus.prob}.  The plots are organized based on partition estimates.  That is, the first set of monitoring stations in each of the plots belong to the first cluster, and then those that belong to the second cluster, etc.  Notice first that including $\rho_0$ in the partition model produces partition posterior distributions that seem to be less variable, in the sense that the co-clustering probabilities of monitoring stations within a cluster are all above 0.5, which is not the case for the baseline model.  Co-clustering probabilities from the global and unit local 1 model  are similar, but it seems that unit local 1 model  produces co-clustering probabilities among grouped stations that are closer to one compared to those from the global model. \fq The co-clustering probabilities for unit local 2 model are quite different, which provides insight into its poor model fit. \qf     Table \ref{tab:PM10_fits_Jan} lists the co-clustering probabilities for stations 22 and 32 in addition to stations 36 and 39. \fq  Notice that both co-cluster probabilities are largest for the unit local 1 model.  The unit local 2 model has the smallest co-clustering probability for stations 22 and 32 with the global model producing similar results. \qf This  highlights how users can have varying levels of certainty associated with subsets of the initial partition.


\subsection{PM$_{10}$ Dataset with Multiple Time Points}


Next we consider all $T = 12$ months of average PM$_{10}$ measurements from the year 2005.  To incorporate the idea of temporal dependence in the partition model, we consider the following hierarchical model

\begin{equation}\label{dat.gen.mech}
  \begin{aligned}
Y_{ti} \mid  \bm{\mu}^{\star}_{t}, \bm{\sigma}_t^{2\star}, \bm{c}_{t} & \stackrel{ind}{\sim} N(\mu^{\star}_{c_{ti}t}, \sigma_{c_{ti}t}^{2\star}), \ i = 1, \ldots, m \ \mbox{and} \ t=1, \ldots, T,   \\
(\mu_{jt}^{\star}, \sigma^{\star}_{jt})\mid \theta_t, \tau_t^2 & \stackrel{ind}{\sim} N(\theta_t, \tau_t^2) \times UN(0,A_{\sigma}), \ j = 1, \dots, k_t ,\\
(\theta_t, \tau_t) & \stackrel{iid}{\sim} N(\phi_0, \lambda^2) \times UN(0, A_{\tau}), \ t = 1, \ldots, T,\\
(\phi_0, \lambda) & \sim N(m_0, s^2_0) \times UN(0, A_{\lambda}), \\
\rho_t \mid \rho_{t-1} & \sim \fq iCRP(\rho_{t-1}, \bm{\alpha}_{t\bullet},  M),  \ t = 1, \ldots, T. \qf 
  \end{aligned}
\end{equation}

Here $UN(\cdot)$ denotes a uniform distribution and for hyper-parameters we used $A_{\sigma}=2.5$, $A_{\tau}=100$, $A_{\lambda}=5$, $m_0=0$, $s^2_0=100$, and $M=1$.  We consider different prior specifications for the $\mathbf{A}$ matrix.

To the 12 months of data, we fit the baseline version of the model detailed in \eqref{dat.gen.mech}, where no $\rho_0$ is supplied, \fq and one that considers $\rho_{\text{quadrant}}$ as an initial partition $\rho_0$. Since the behavior between $\rho_{\text{quadrant}}$ and $\rho_{\text{kmeans}}$ was similar in the previous section, here we do not consider $\rho_{\text{kmeans}}$.   \qf
{Our main aim here is to observe how the effect of the initial partition described earlier propagates over time. Thus, for} both of these models, we consider the global and time local models for $\mathbf{A}$ described in Table \ref{tab:alpha_models}. We refer to these four models as \emph{baseline global}, \emph{baseline time local}, \emph{global} and \emph{time local}.  When using a global model for  $\mathbf{A}$, we set  $a=1$ and $b=9$, while for the time local we use $a_t = 1$ and $b_t = 9$ for all $t$.
For the model that includes $\rho_0$, we consider four other prior specifications for $\mathbf{A}$.  Two of these correspond to the priors detailed in Table \ref{tab:alpha_models} (\emph{unit local}, and \emph{unit$\times$time local}) while the other two use the latent autoregressive process (\emph{time local auto-regressive}, \emph{unit$\times$time local auto-regressive }). For the \emph{unit local} model we set $a_i = 1$ and $b_i = 9$ for all $i$ except for $i \in \{22,32,36,39\}$ for which $a_i = 9$ and $b_i=1$ (as was done in the Section \ref{sec:pm10.one.time.point}), while for the \emph{unit$\times$time local} we set $a_{ti} = 1$ and $b_{ti} = 9$ for all $t$ and $i$ except that for $i \in {22,32,36,39}$ we set $a_{ti} = 9$ and $b_{ti} = 1$ for all $t$. When using the latent autoregressive process we assume $\alpha_{ti}$ unrestricted and set $\xi_{0} = \logit(a_{1}/(a_{1}+b_{1}))$ (\emph{unit$\times$time local auto-regressive}). We also consider the case where $\alpha_{ti} = \alpha_t$ (\emph{time local auto-regressive}) and we set $\xi_{0} = \logit(a_{1i}/(a_{1i}+b_{1i}))$. For both, diffuse conjugate priors for $\beta_{0}$, $\beta_{1}$, and $\kappa^2$ are employed. Using these 8 different prior specifications, we fit model \eqref{dat.gen.mech} by collecting 2000 MCMC samples after discarding the first 50,000 as burn-in and thinning by 50 (a total of 150,000 MCMC iterates were sampled).

The LPML and WAIC for each model are provided in Table \ref{tab:PM10_fits_all}.  According to WAIC the \emph{unit$\times$time local} and \emph{baseline global} priors in the worst model fit.  There is a clear indication that including $\rho_0$ provides value in terms of model fit as both the \emph{global} and \emph{time local} models perform better than their baseline counterparts in terms of WAIC and LPML. The added value of including $\rho_0$ diminishes however, if prior on $\mathbf{A}$ is too flexible. This can be seen by the fact that the \emph{unit $\times$ local auto-regressive} model displays marked improvement in model fit compared to the \emph{unit $\times$ local} model.  It appears that WAIC indicates that borrowing strength across time also improves model fit as the \emph{unit local} model fits the best. 

\begin{figure}
  \centering
  \includegraphics[width=1\textwidth]{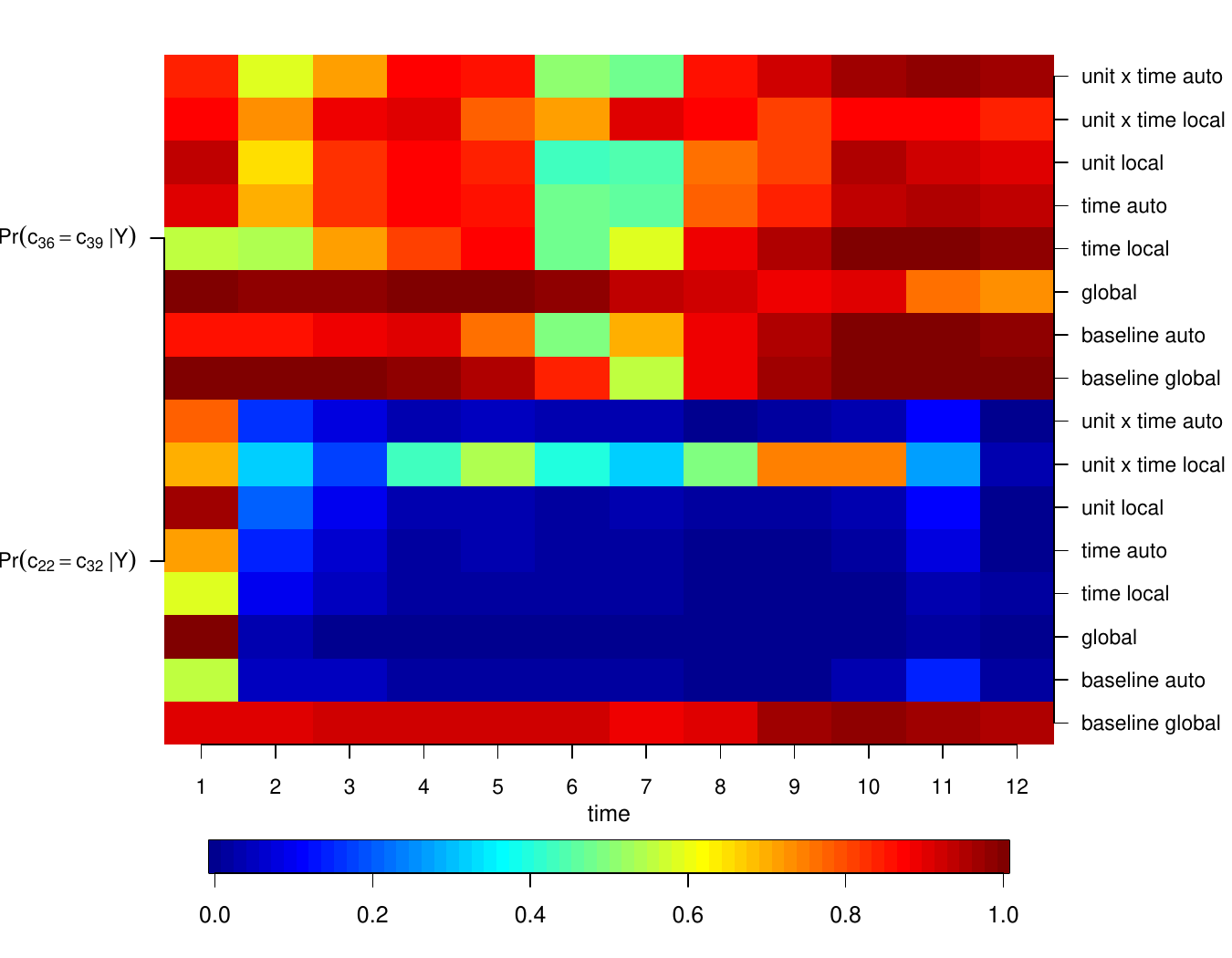}
    \caption{Posterior estimates of the co-clustering probabilities across time between stations 22 and 32 and between stations 36 and 39 for each of the eight models fit to the PM$_{10}$ data. }
   \label{fig:co.clus.prob.alltime}
\end{figure}

Figure~\ref{fig:co.clus.prob.alltime} illustrates the co-clustering probabilities between stations 22 and 32 and stations 36 and 39 across time for each model.  The left margin labels the co-clustering probability  (either $Pr(c_{22} = c_{32}\mid\bm{Y})$ or $Pr(c_{36} = c_{39}\mid\bm{Y})$) and the right margin indicates to which model each row corresponds.  Notice that $Pr(c_{36} = c_{39}\mid\bm{Y})$ is relatively high for most time points across all models. There is a slight dip between June and July for all models besides the \emph{baseline global} \emph{global}, and \emph{unit $\times$ local} priors.  Additionally, it seems that the autoregressive priors  result in a smoother changes over time.  Conversely,  $Pr(c_{22} = c_{32}\mid\bm{Y})$  seems to decrease over time for all models save the \emph{baseline global} and \emph{unit $\times$ local} models.  This seemed counterintuitive at first glance since the prior on $\mathbf{A}$ for these models together with $\rho_0$ encourages stations 22 and 32 to co-cluster.  However, upon further inspection of the data, it appears that station 32's PM$_{10}$ measurements are fairly different from those from the other stations (and in particular station 22) which explains why $Pr(c_{22} = c_{32}\mid\bm{Y})$ is small for all models save the baseline global model.  This is explored further in Section 3 of the Supplementary Material. 


\begin{table}[tb]
 \centering
 \caption{ LPML (larger is better) and WAIC (smaller is better) for four models fit to the data. }
\begin{tabular}{l cc } \toprule
 Prior model for $\mathbf{A}$ & LPML & WAIC \\\midrule 
 baseline (no $\rho_0$) global             & -1953 & 3359 \\ 
 baseline (no $\rho_0$) time local         & -1852 & 2997 \\ 
 global                                    & -1708 & 3002 \\ 
 time local                                & -1659 & 2990 \\ 
 time local auto-regressive                & -1952 & 2981 \\ 
 unit local                                & -2385 & 2974 \\ 
 unit$\times$time local                    & -2674 & 3414 \\ 
 unit$\times$time local auto-regressive    & -2189 & 2996 \\  \bottomrule
 \end{tabular}
 \label{tab:PM10_fits_all}
 \end{table}

\begin{figure}[H] 
  \centering
  \includegraphics[scale=0.83]{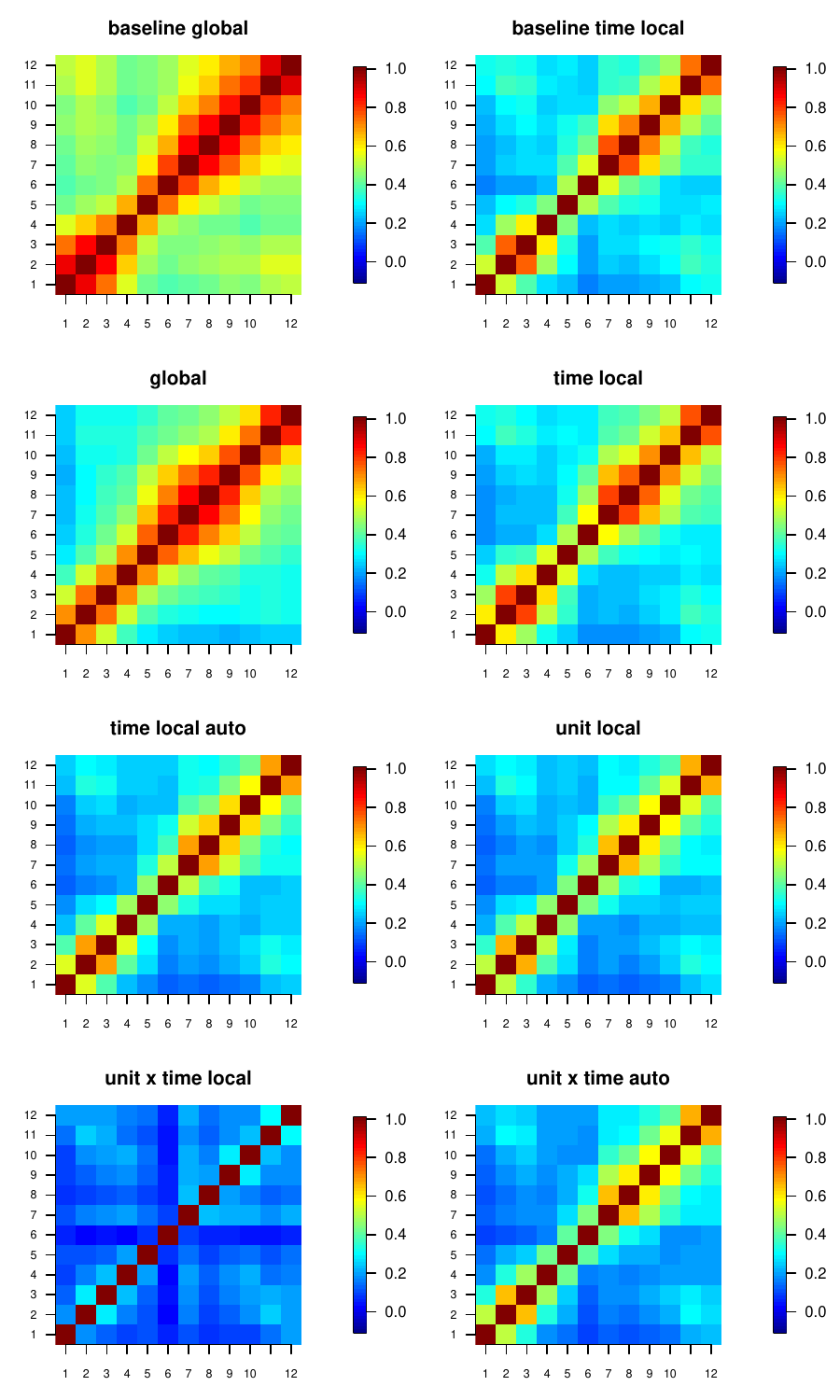}
    \caption{Posterior means for the time-lagged adjusted Rand index associated with each of the eight models fit to the PM$_{10}$ data.}
   \label{fig:ARI}
\end{figure}

Lastly, Figure \ref{fig:ARI} displays the time-lagged ARI value for each model. It appears that the \emph{baseline global} model produces partition estimates whose temporal dependence decays more slowly compared to the models that include $\rho_0$ (see the top left plot of Figure \ref{fig:ARI}). This is because the baseline model produces the highest estimate value for $\alpha$ due to the borrowing-of-strength across time and unit and not having to ``distance'' itself from $\rho_0$ as it was not supplied.  This resulted in less reallocation of stations. 
Conversely, the \emph{unit $\times$ time local} model displays essentially no temporal dependence among the partitions, even if $\rho_0$ is supplied. This model introduces too much flexibility, and as a result, the posterior estimate of each $\alpha_{it}$ is highly dependent on the prior, whose mean is around 0.1.  The other models that include $\rho_0$ demonstrate varying degrees of temporal dependence decay between partition estimates. The decay is faster when $\alpha$ is not fixed across time. The time-lagged ARI results change slightly when comparing the \emph{time local} to the \emph{unit local}model, with the former displaying slightly more persistent temporal dependence.  Interestingly, the \emph{time local autoregressive}, \emph{unit local}, and \emph{unit $\times$ time autoregressive} models display similar dependence.

\section{ Summary and Discussion }\label{sect:disc}

In this work, we propose a probabilistic approach to model one or more dependent random partitions
of $[m]$, with the ability to incorporate prior information in the form of a given partition $\rho_0$. 
The model also has the
ability of up- or down-weighting different subsets in $\rho_0$, thus effectively
reflecting various degrees of certainty on these subsets.
Some theoretical properties of the informed partition approach are developed, mostly related to the expected Rand Index, to quantify prior influence on a subsequent partition. As a consequence, we are able to guide prior specification and prior elicitation of certain key hyperparameters that drive the expected behavior of the random partition model.

Through extensive simulation studies, the model was shown to perform well with respect to different specifications of the prior partition. In addition, we offer some insights into the behavior of our informed partition model and loss-based alternatives that incorporate such prior information. 
Application to a dataset on ${\rm PM}_{10}$ measurements
in Germany highlights the benefits of our proposed model, showing that including reasonable prior information improves model fit. 

For the purpose of illustrating the approach (through simulations and data analysis), we have adopted simple Gaussian likelihood models. Nevertheless, the methods can be easily extended to cover other specifications, such as including the case of non-Gaussian responses and/or serial correlation over time points. Similarly, the EPPF in \eqref{eq:rhoprior2} may be easily extended to include covariate dependence, e.g., as in a PPMx-like model.

A possible avenue of further research would be to introduce some structure in the prior distribution of $\bm{\gamma}$.  That is,  consider the binary $\gamma_{ti}$ parameters and postulate an autologistic model such as binary CAR or a similar model. Here, the correlation structure of the CAR-like model would be such that units $i$ and $j$ are neighbors if and only if they belong to the same cluster.
\fq Another potentially interesting option for future research is to explore a different construction of a joint model $p(\rho_1,\ldots,\rho_T)$ for a sequence of partitions under the dependence framework we have proposed. In principle, a graphical modeling approach to connect partitions $\rho_{t_1}$ and $\rho_{t_2}$ say, when $t_1$ and $t_2$ are nonconsecutive integers could be attempted. Motivation for this could arise in the context of multi-view data. Such setting could still be accommodated by our proposed framework, redefining the role of the binary reallocation indicators $\{\gamma_{t_2 i}:\, i=1,\ldots,m\}$ to, for example, employ $\rho_{t_1}$ as an initial partition for $\rho_{t_2}$. In this context, the parameters $\alpha_{t_2 i}$ could be adjusted to represent the strength of connections between these two partitions, as described in Section~\ref{sect:model}. \qf 

\section*{Acknowledgments}

\fq We thank the Editors and Reviewers of this work for their valuable comments that have helped us to substantially improve our work. This research was partially funded by grant ANID Fondecyt Regular 1220017.
\qf

	\singlespacing
	\bibliographystyle{dcu}
	\bibliography{refs}
	
	\setcounter{section}{0}
	\setcounter{figure}{0}
	\renewcommand{\thefigure}{S.\arabic{figure}}
	\renewcommand{\thesection}{S.\arabic{section}}
	\newpage

\section{Additional Simulation Study Results}
\begin{figure}[H] 
  \centering
  \includegraphics[width=0.9\textwidth,height=0.7\textheight]{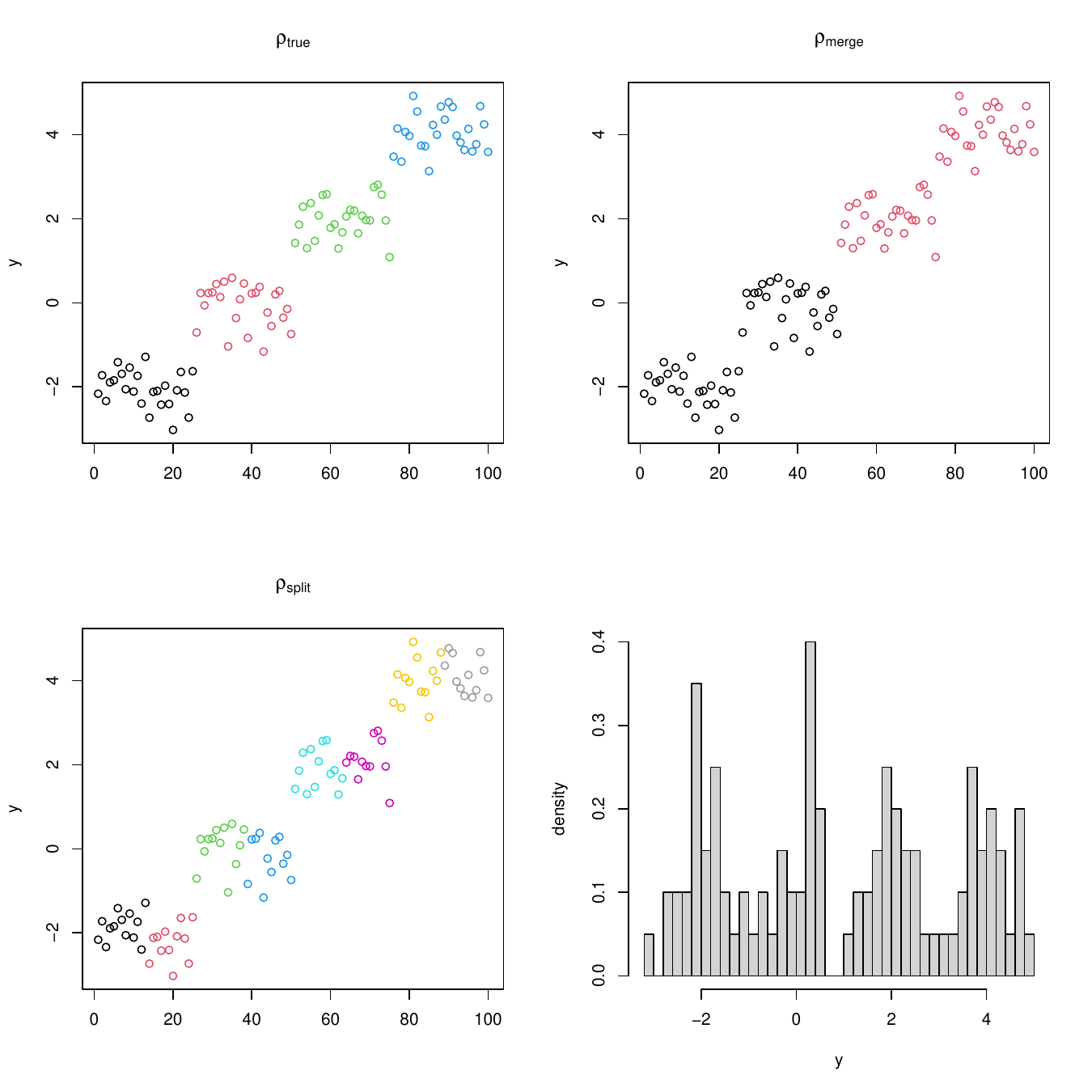}
    \caption{synthetic data (bottom-right panel) and the eight initial partitions employed.}
   \label{fig:init_part}
\end{figure}

\begin{figure}
  \centering
    \includegraphics[width=1\textwidth]{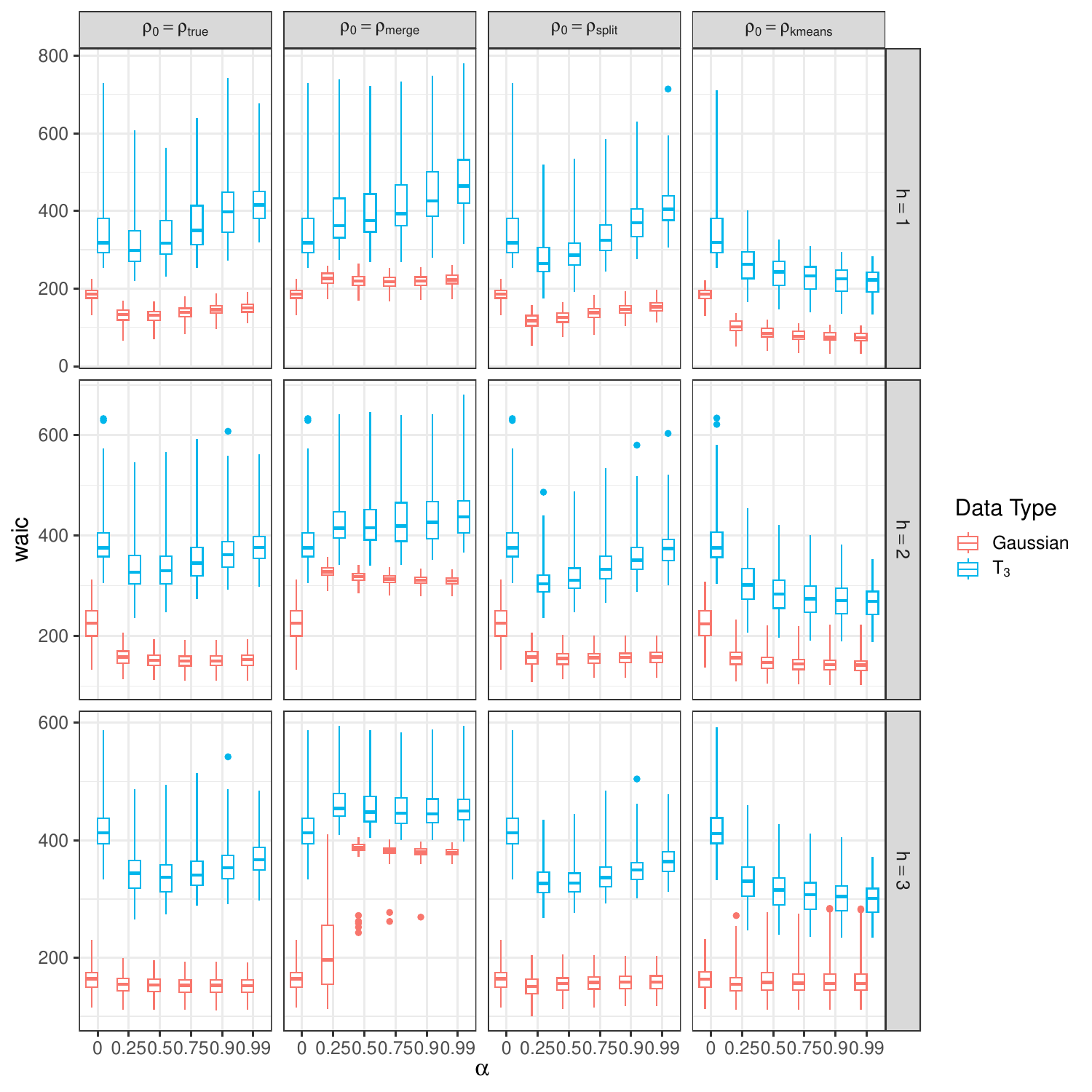}
    \caption{Results for fixed global $\alpha$. Distribution of WAIC across $100$ replicated data sets, for each value of $\alpha \in \{0, 0.25, 0.5, 0.75, 0.9, 0.99\}$ are displayed. Each panel shows results for different combinations of the cluster mean separation values used in the data-generating process and the type of initial partition $\rho_0$.}
   \label{suppmat_fig:waic_simstudy2}
\end{figure}

The top row of Figure \ref{fig:init_part} shows the initial partitions considered in the simulation study detailed in Section~\ref{sec:post.sim.onetime.point}. In addition to those, we also considered initial partitions displayed in the second and third rows of Figure \ref{fig:init_part}, the results of which are described shortly. First, however, we describe in a bit more detail WAIC results and then those associated with the unit local model for $\bm{\alpha}$.  Figure  \ref{suppmat_fig:waic_simstudy2} displays the WAIC values related to the different partitions as a function of $\rho_0$, $h$, and $\alpha$ (see 
Section~\ref{sec:post.sim.onetime.point} of the main document for more details).  Note that model fits are quite similar overall, with the initial partition $\rho_{\text{merge}}$ generally speaking resulting in the best fits, while initial partition $\rho_{\text{split}}$ performing similarity to not including an initial partition at all.   

\begin{figure}[H] 
  \centering
  \includegraphics[width=1\textwidth, page=2]{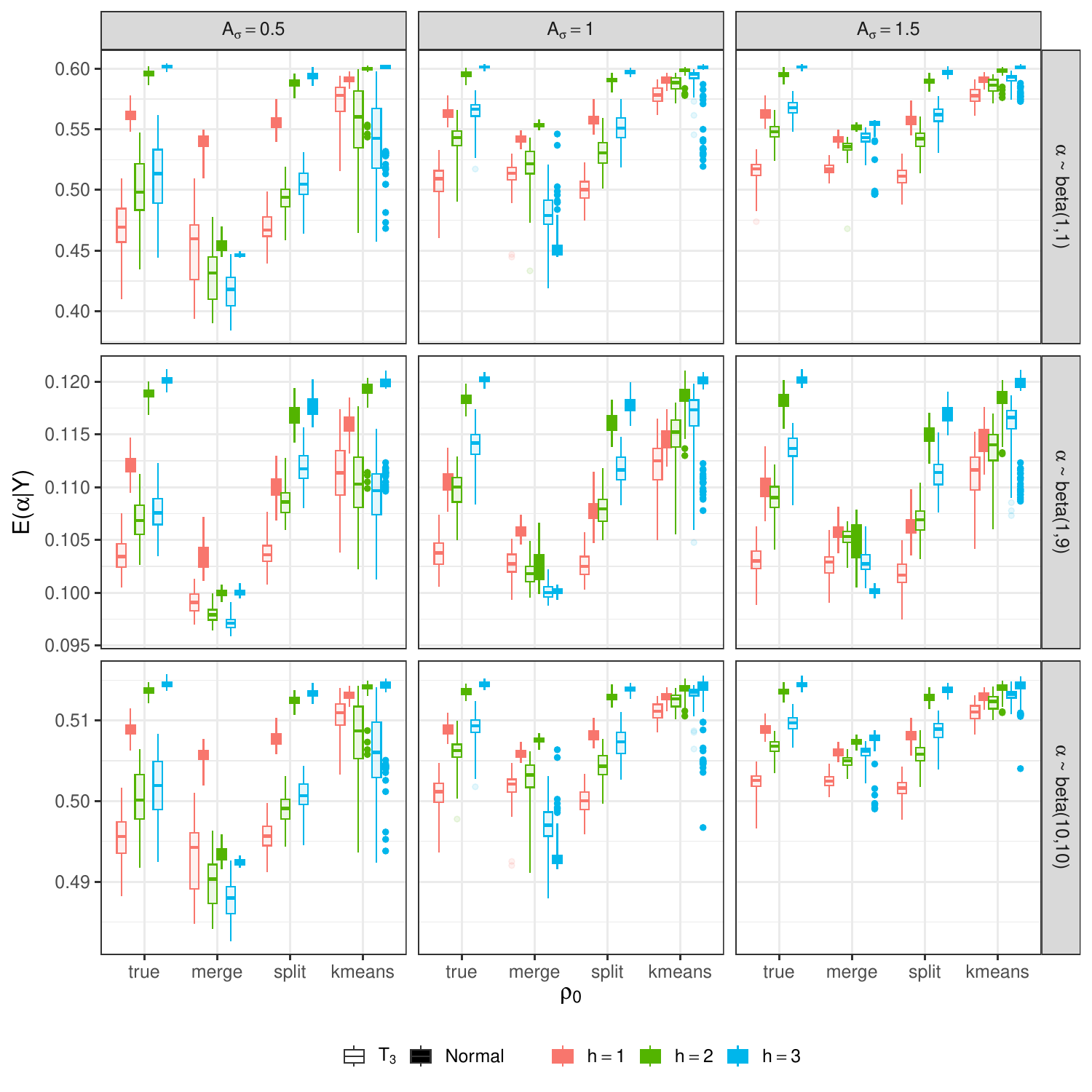}
    \caption{Results for unit local random $\alpha$. Distribution of the posterior mean of $\alpha$ for different choices of $\rho_0$ (x-axis) across $100$ replicated data sets using different values for the cluster means separation (boxplot colors). \fq Solid and transparent fill distinguish the data-generating scenario (Normal and $t_3$ distribution). \qf Each panel shows results for different combinations of prior choices for $\alpha$ and $A_{\sigma}$.}
   \label{fig:mean_alpha_local}
\end{figure}

Figure  \ref{fig:mean_alpha_local} displays the average of the posterior means for $\alpha_i, i = 1, \ldots, m$ under the unit local model.  As expected, there are differences between the two.  This is a consequence of the number of $\gamma_i$'s used in the unit local model. Notice that for the global $\alpha$ model, the posterior means of $\alpha$ are more concentrated towards 0 or 1 and are less influenced by the prior distributions. This is because the model borrows strength across all $m$ parameters $\gamma$s when estimating $\alpha$.
Regarding the partitions, it seems that when formulating an initial partition, one should err on the side of being parsimonious.  If the initial partition contains more clusters than the true partition, one must apply quite a bit of prior weight away from the initial partition, or it remains heavily weighted. This is less so for the local $\alpha$ model.

\begin{figure}[H] 
  \centering
  \includegraphics[width=1\textwidth, page=4]{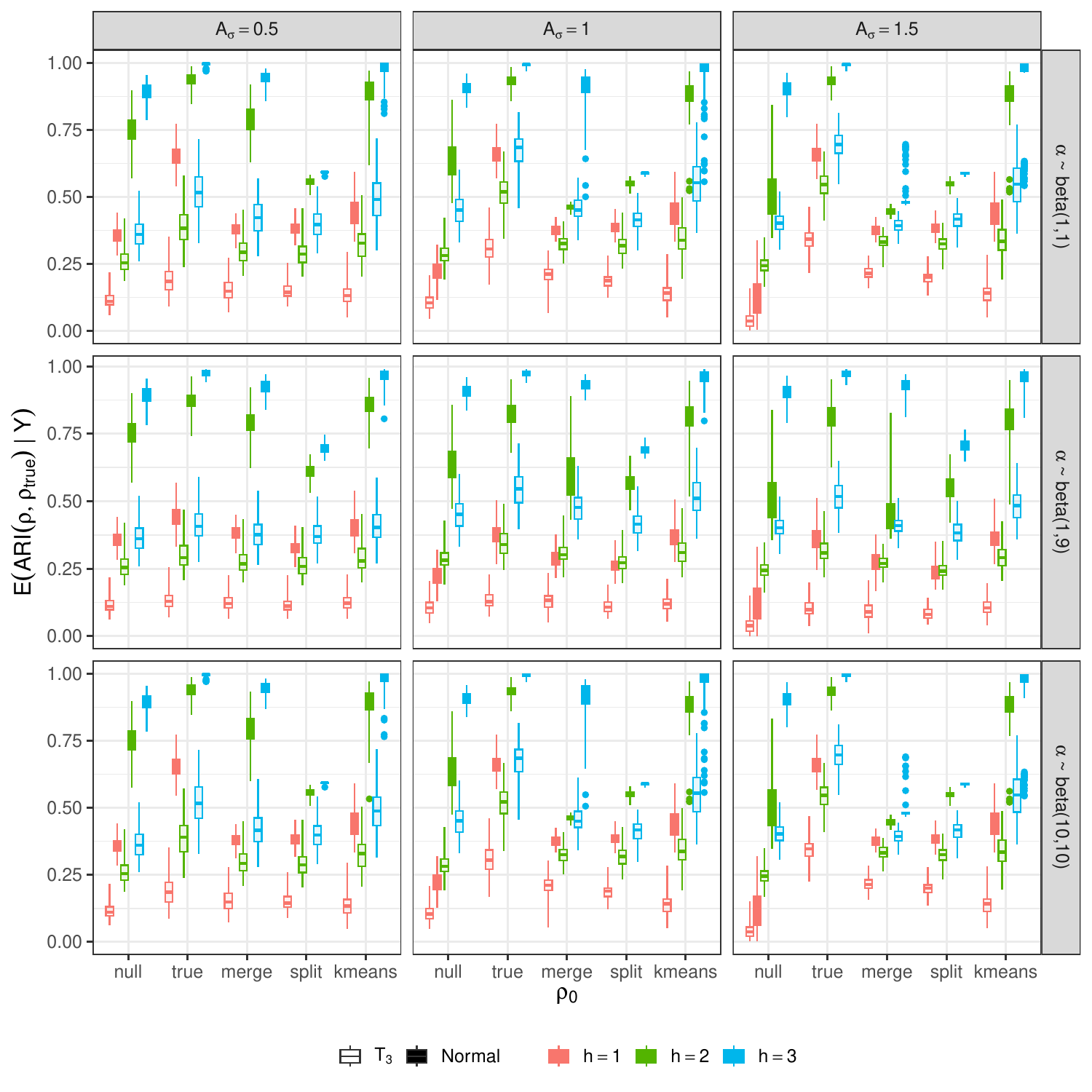}
    \caption{Results for unit local random $\alpha$. Distribution of $E(ARI(\rho, \rho_{true})\mid \bm{Y})$ for different choices of $\rho_0$ (x-axis) across $100$ replicated data sets using different values for the cluster means separation (boxplot colors). \fq Solid and transparent fill distinguish the data-generating scenario (Normal and $t_3$ distribution). \qf Each panel shows results for different combinations of prior choices for $\alpha$ and $A_{\sigma}$. Here $\rho_0 = \mbox{null}$ corresponds to a model that does not include an initial partition. Notice that when $\rho_0 = \mbox{null}$ results do not change for different values of $\alpha$, as that parameter is not included in the model.}
   \label{fig:ari_local}
\end{figure}

\begin{figure}[H] 
  \centering
  \includegraphics[width=1\textwidth, page=10]{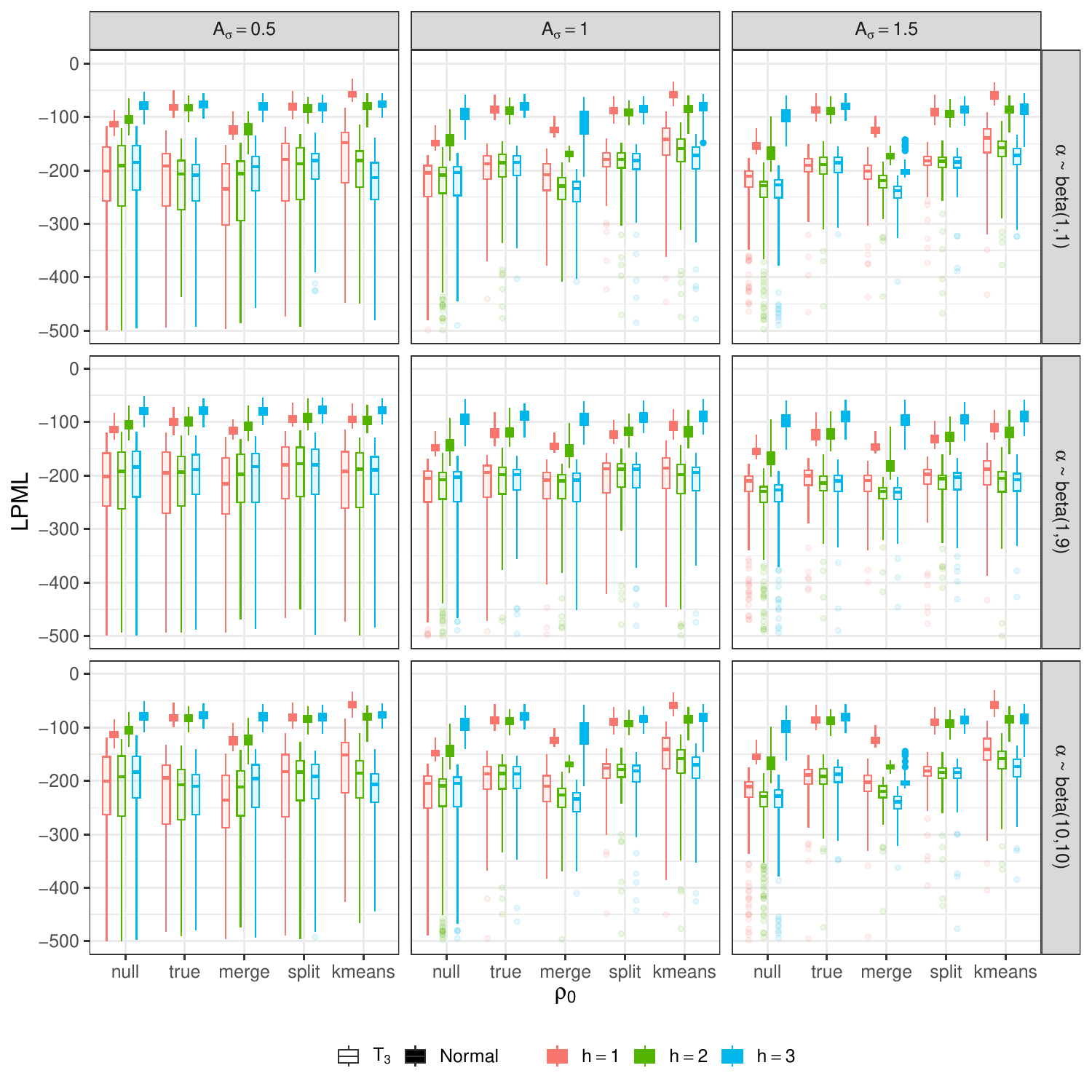}
    \caption{Results for unit local $\alpha$. Distribution of LPML for different choices of $\rho_0$ (x-axis) across $100$ replicated data sets using different values for the cluster means separation (boxplot colors). \fq Solid and transparent fill distinguish the data-generating scenario (Normal and $t_3$ distribution). \qf Larger values of LPML indicate a better fit.  Each panel shows results for different combinations of prior choices for $\alpha$ and $A_{\sigma}$. Here $\rho_0 = \mbox{null}$ corresponds to a model that does not include an initial partition. Notice that when $\rho_0 = \mbox{null}$, results do not change for different values of $\alpha$, as that parameter is not included in the model. }
   \label{fig:lpml_local}
\end{figure}

\section{Simulation with a Locally Weighted Prior Partition}

\subsection{Prior simulations}

\begin{figure}
    \centering
    \includegraphics[width = 0.32\textwidth]{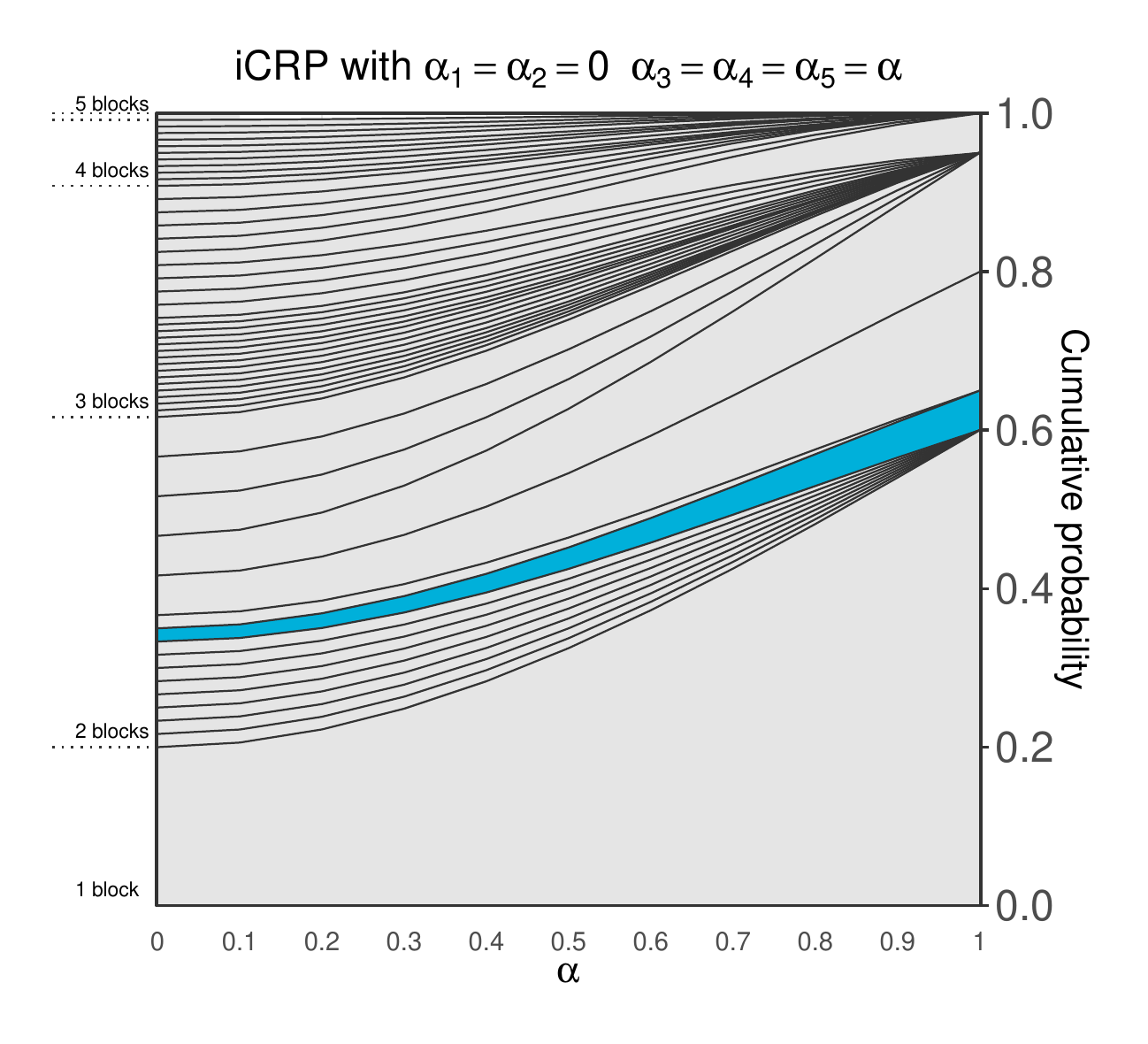}
    \includegraphics[width = 0.32\textwidth]{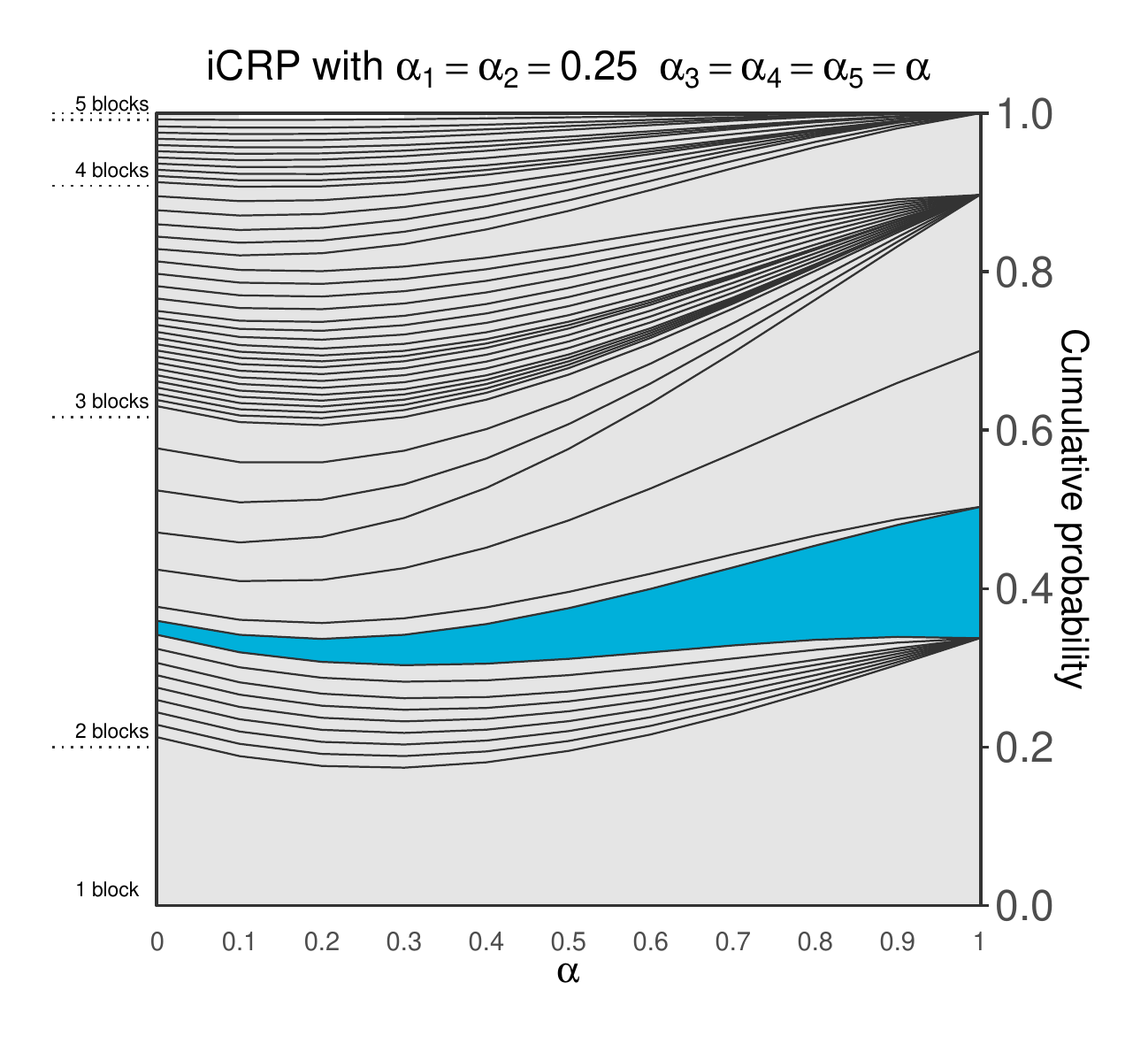}
    \includegraphics[width = 0.32\textwidth]{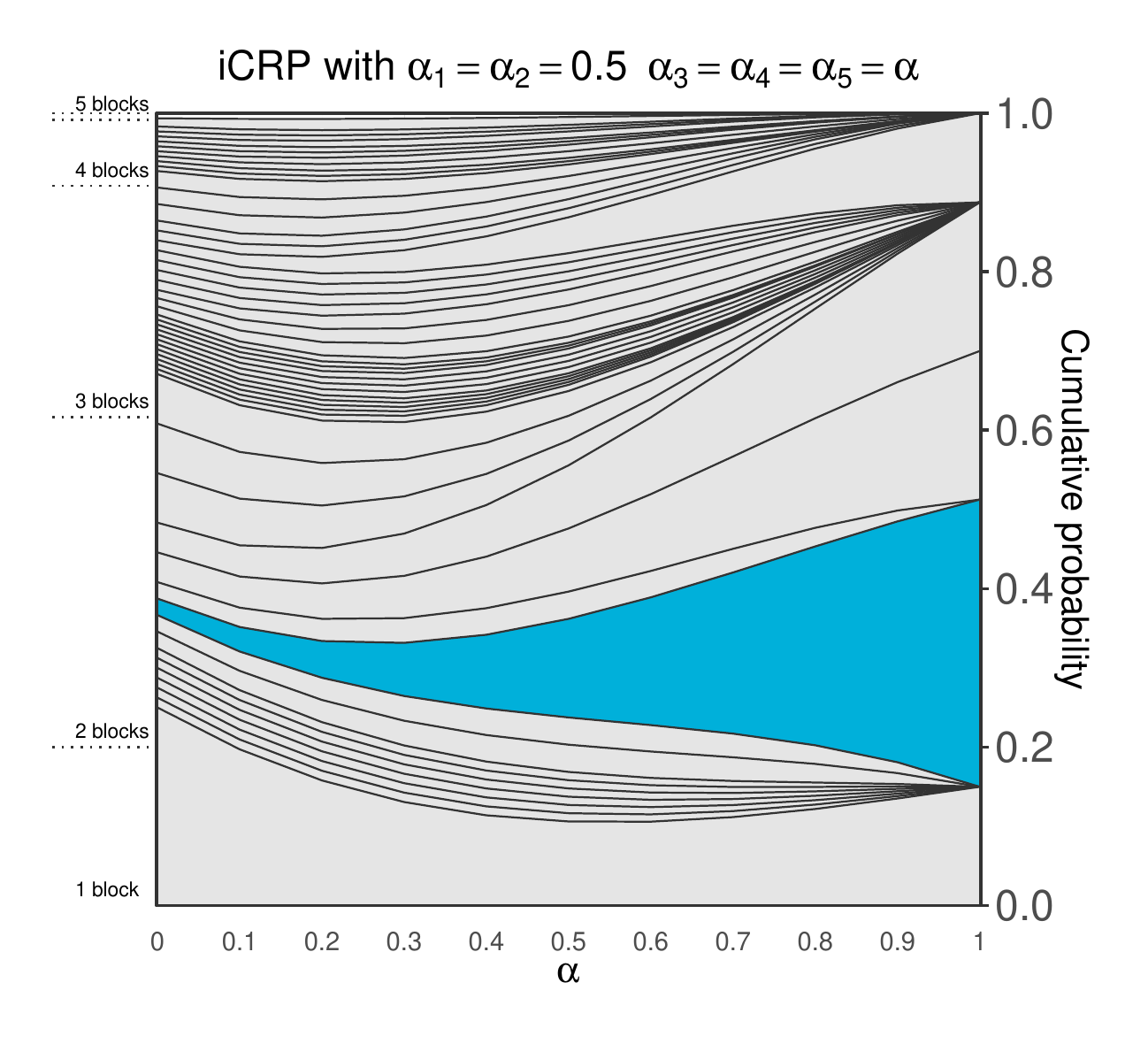}
    \includegraphics[width = 0.32\textwidth]{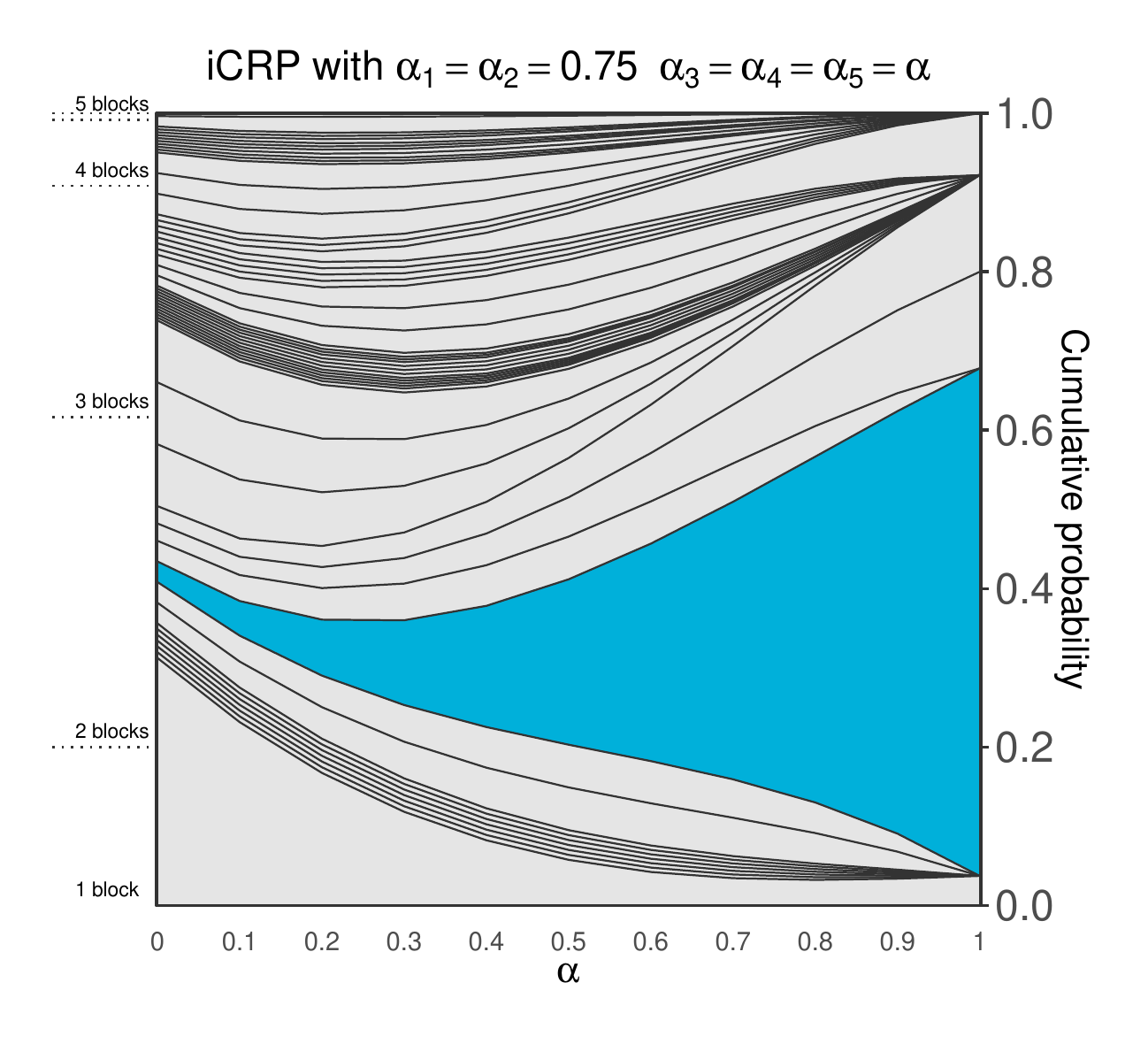}
    \includegraphics[width = 0.32\textwidth]{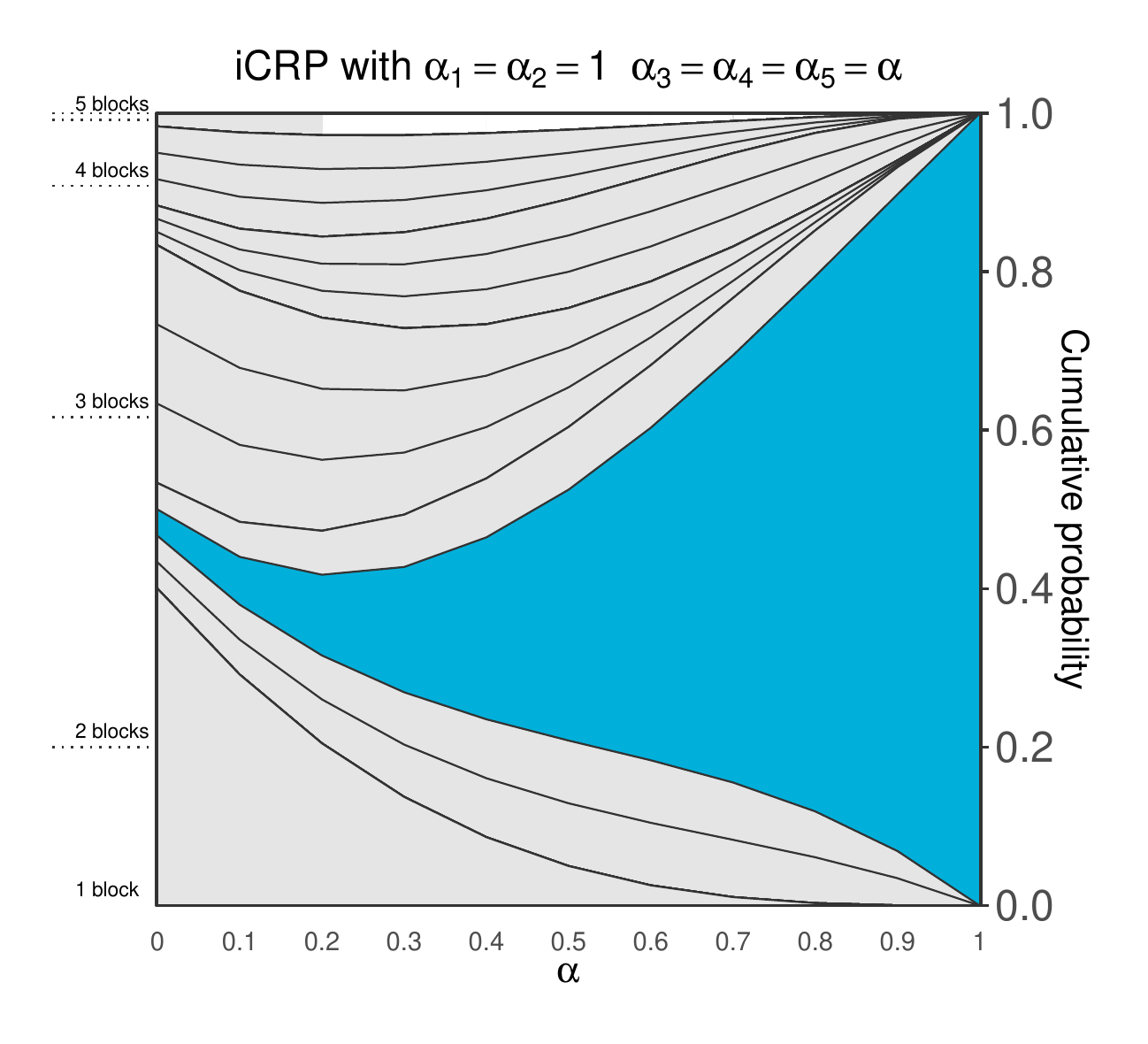}
    \caption{Prior probabilities for each of the $52$ partitions when the prior guess is ${\rho}_{0} = \{1,2\}\{3, 4, 5\}$ (highlighted in blue) for different specifications of the informed partition model using the CRP prior (iCRP). The cumulative probabilities across different values of the penalization parameters are joined to form the curves, while the probability of a given partition corresponds to the area between the curves. }
    \label{fig:enter-label}
\end{figure}

Figure \ref{fig:enter-label} displays the prior partition probabilities where $\rho_{0} = \{\{1,2\},\{3, 4, 5\}\}$ is the initial partition with cluster specific $\alpha$ values.  This provides the user with quite a bit of control over how to weigh the initial partition.

\subsection{Posterior simulations}

\begin{figure}
  \centering
  \includegraphics[page=1,width=0.95\textwidth]{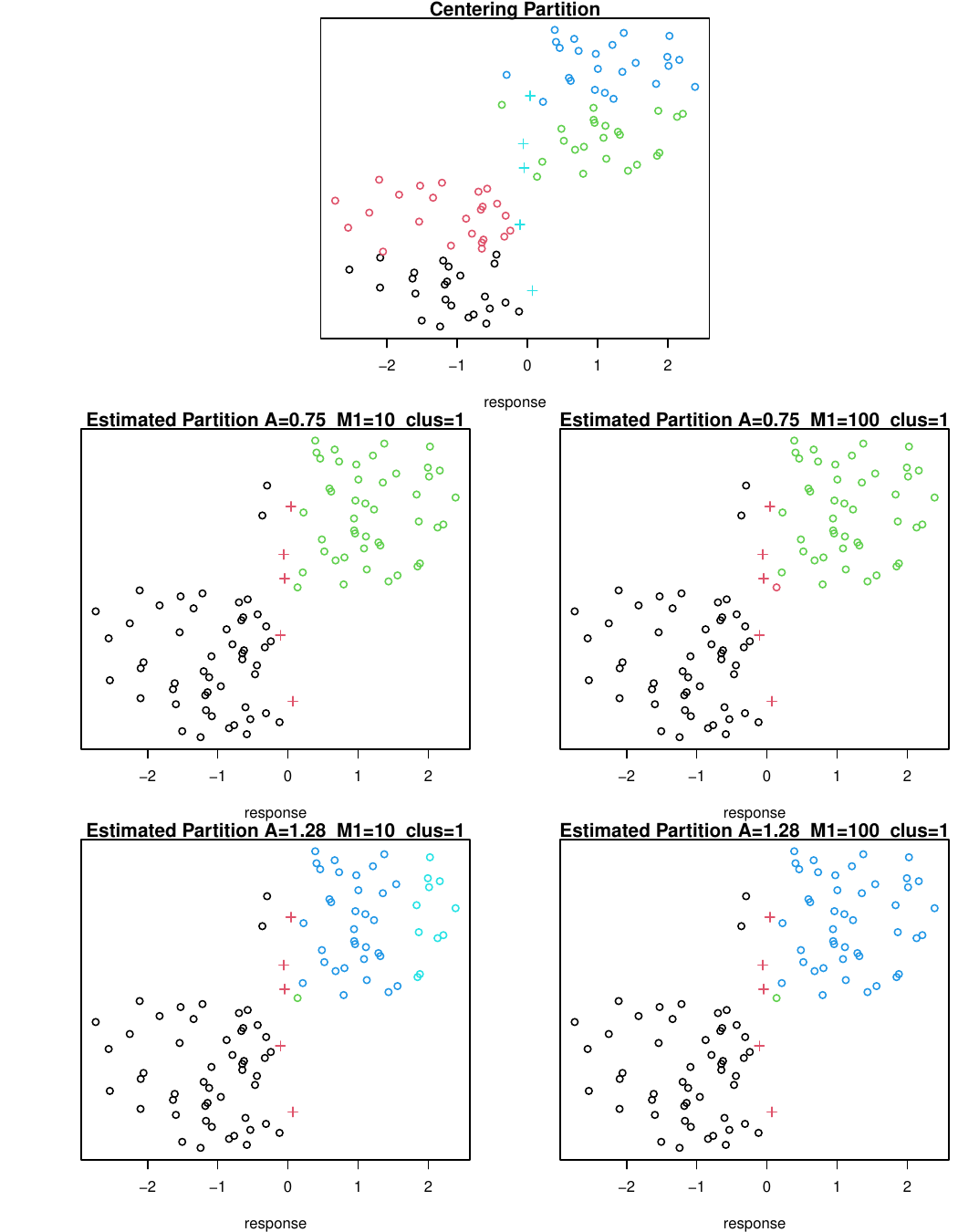}\\
    \caption{Results from a small simulation study that explores the impact that the prior on $\alpha_i$ has on posterior partition estimate.  The prior is specified such that the probability $\alpha_i$ of reallocating the ``+'' points is small. As a result, these points stay in their own cluster a posterior, even though the rest of the initial partition is overridden.}
   \label{post.sim2}
\end{figure}

\begin{figure}
  \centering
  \includegraphics[page=2,width=0.95\textwidth]{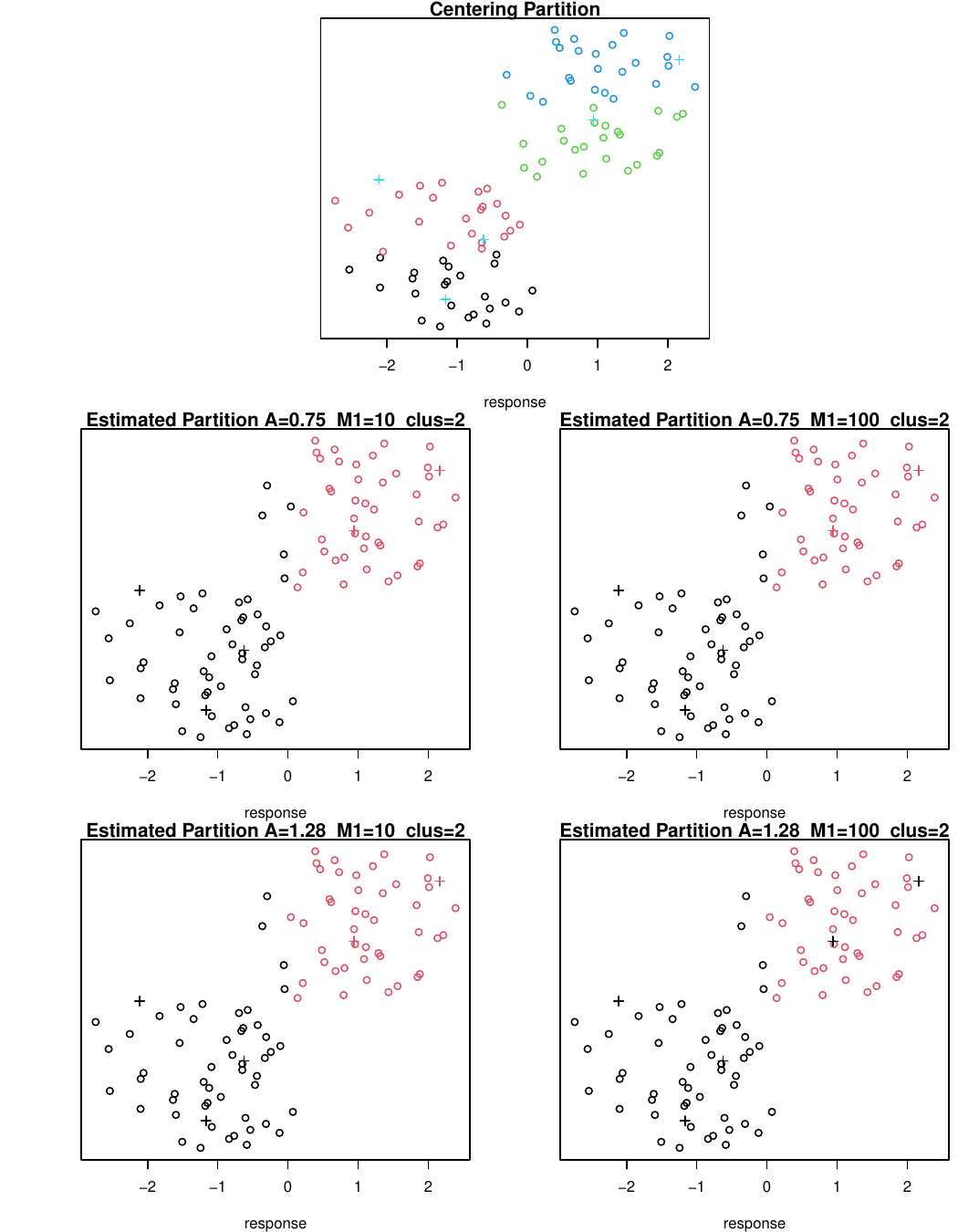}
    \caption{Results from a small simulation study that explores the impact that the prior on $\alpha_i$ has on posterior partition estimate.   The prior is specified such that the probability $\alpha_i$ of  ``+'' the probability $\alpha_i$ of reallocating the ``+'' points is small.}
   \label{post.sim3}
\end{figure}

In this very small simulation, we explore the performance of the method when some units {\it a priori} cluster with high probability and the case when some units {\it a priori} belong to different clusters with high probability. 
It may be the case that there exists more prior uncertainty associated with particular sections of the initial partition than with others.  This can easily be incorporated in the prior construction by way of the unit-specific beta prior distributions.  In particular, for one partition, we have $\alpha_{i} \sim \Be(a_i, b_i)$ so that specif subsets of the partition are maintained unless the data strongly contradict. 
To include this information is enough to set $a_i$ and $b_i$ so that the probability of reallocating is small (e.g., $a_i=1$, $b_i = 10$).  To illustrate this, consider Figures \ref{post.sim2} and \ref{post.sim3}.  

Figure \ref{post.sim2} displays a synthetic dataset of 100 observations with two clusters, one centered at $-1$ and the other at $1$.  The top plot in Figure \ref{post.sim2} displays the initial partition employed.  Notice that the initial partition is such that the two clusters are split, and there are five observations on the edge shared by both clusters (the ``+'' points) that are assigned in the initial partition to their own cluster.  The bottom four plots in Figure \ref{post.sim2} are different fits of the model with different combinations of $A$ (the upper bound on $\sigma_i$) and $M$ (the scale parameter of the CRP).  Notice that a section of the initial partition that splits the two main clusters is overridden by the data, but the prior assigned to the ``+'' points is such that it forces those points to remain in their own clusters.  

Next, consider Figure \ref{post.sim3}.  The data-generating procedure is similar to that of Figure \ref{post.sim2}.  Now, however, the initial partition contains a group of five units that clearly should not be in the same cluster based on their response value (the ``+'' points).  Notice here that for specific hyper-prior values, these points do not remain in the same cluster, but they are absorbed by one of the two bigger clusters, since the data strongly contradicts the prior. 

\subsection{Posterior simulations: comparison with other informative pri-
ors}

\begin{figure}[h]
    \centering
    \includegraphics[width = \textwidth]{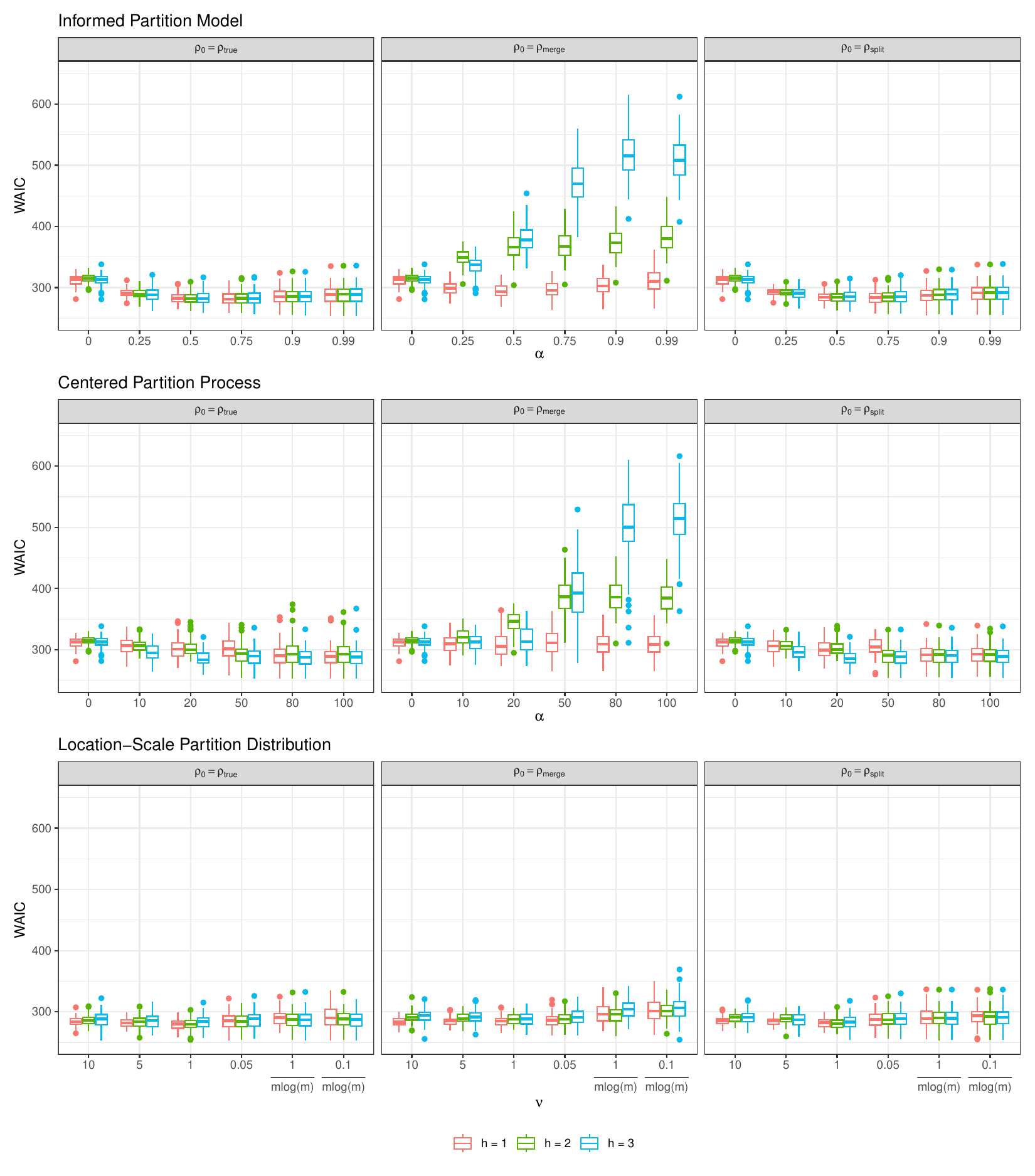}
    \caption{Comparison of posterior results using the iCRP, CPP, and LSP priors for different values of their tuning parameters. Each boxplot represents the distribution of WAIC across the $100$ generated data sets, with colors distinguishing between data-generating scenarios. Notice that the values of the tuning parameters are not directly comparable. }
    \label{fig:postComp_WAIC}
\end{figure}

\begin{figure}[h]
    \centering
   \includegraphics[width = \textwidth]{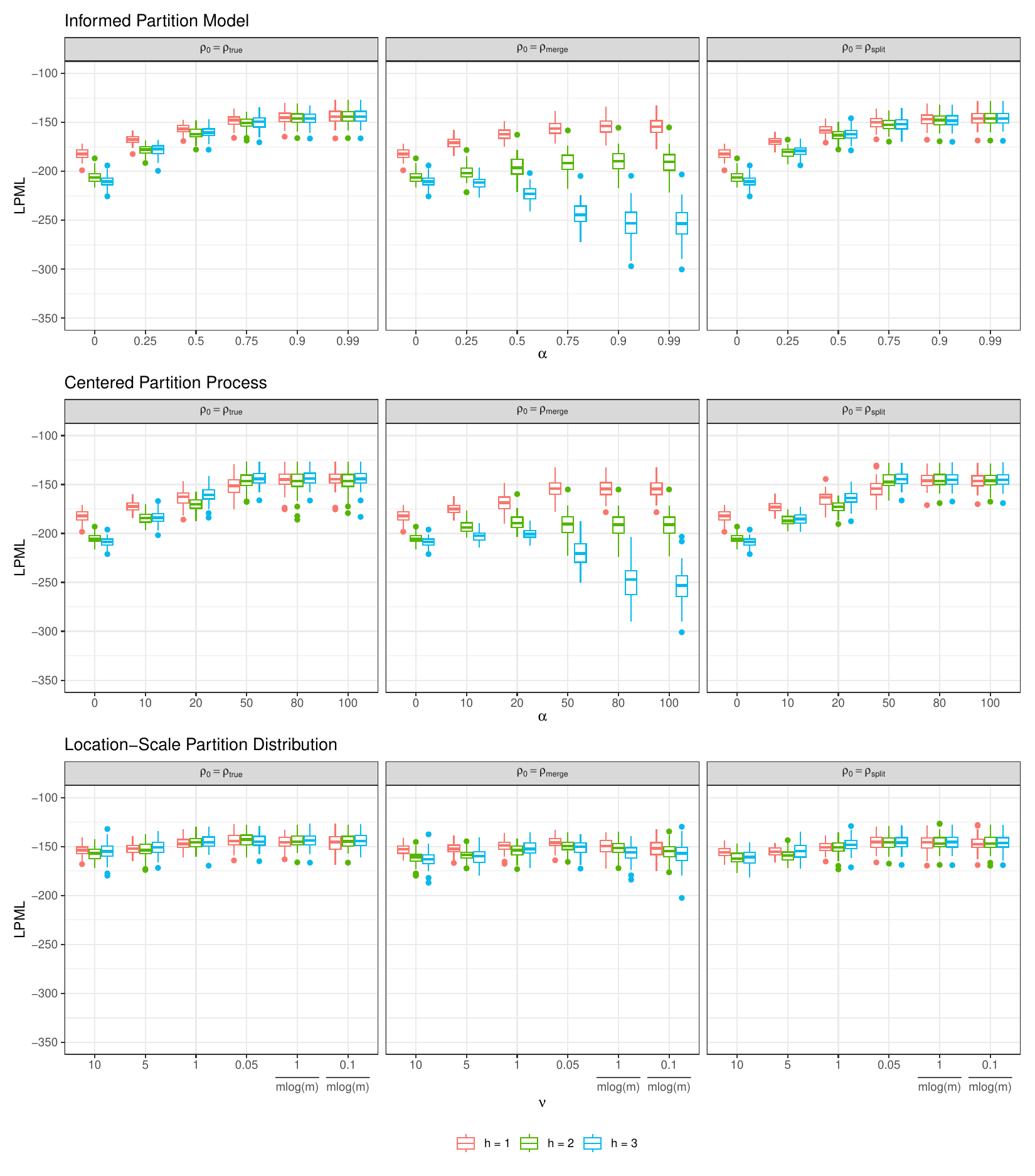}
    \caption{Comparison of posterior results using the iCRP, CPP, and LSP priors for different values of their tuning parameters. Each boxplot represents the distribution of LPML across the $100$ generated data sets, with colors distinguishing between data-generating scenarios. Notice that the values of the tuning parameters are not directly comparable.}
    \label{fig:postComp_LPML}
\end{figure}

Figures \ref{fig:postComp_WAIC} and \ref{fig:postComp_LPML} show the distribution of the WAIC and LPML across simulations for the different models. We can observe similar patterns for the Informed Partition Model and the Centered Partition Process. The best fit is obtained when the models are informed using $\rho_{\text{true}}$, while the worst fit is obtained when the initial partition is $\rho_{\text{merge}}$. Results under $\rho_{\text{split}}$ an initial partition are comparable $\rho_{\text{true}}$; even if the partition is not the correct one, the units are still informed towards coherent groups, as  $\rho_{\text{split}}$ divides each of the 4 clusters of the simulation in two. The LSP prior behaves differently as it seems to show a good fit overall. However, figures in Section~\ref{sec:prior.sim.mult.point} of the main manuscript show that this prior tends to induce a slight inflation in the prior probability for the initial partition unless the tuning parameter is really small. 

\newpage
\section{Further Exploration of the PM$_{10}$ Application}

\begin{figure} 
  \centering
   \includegraphics[width=1\textwidth]{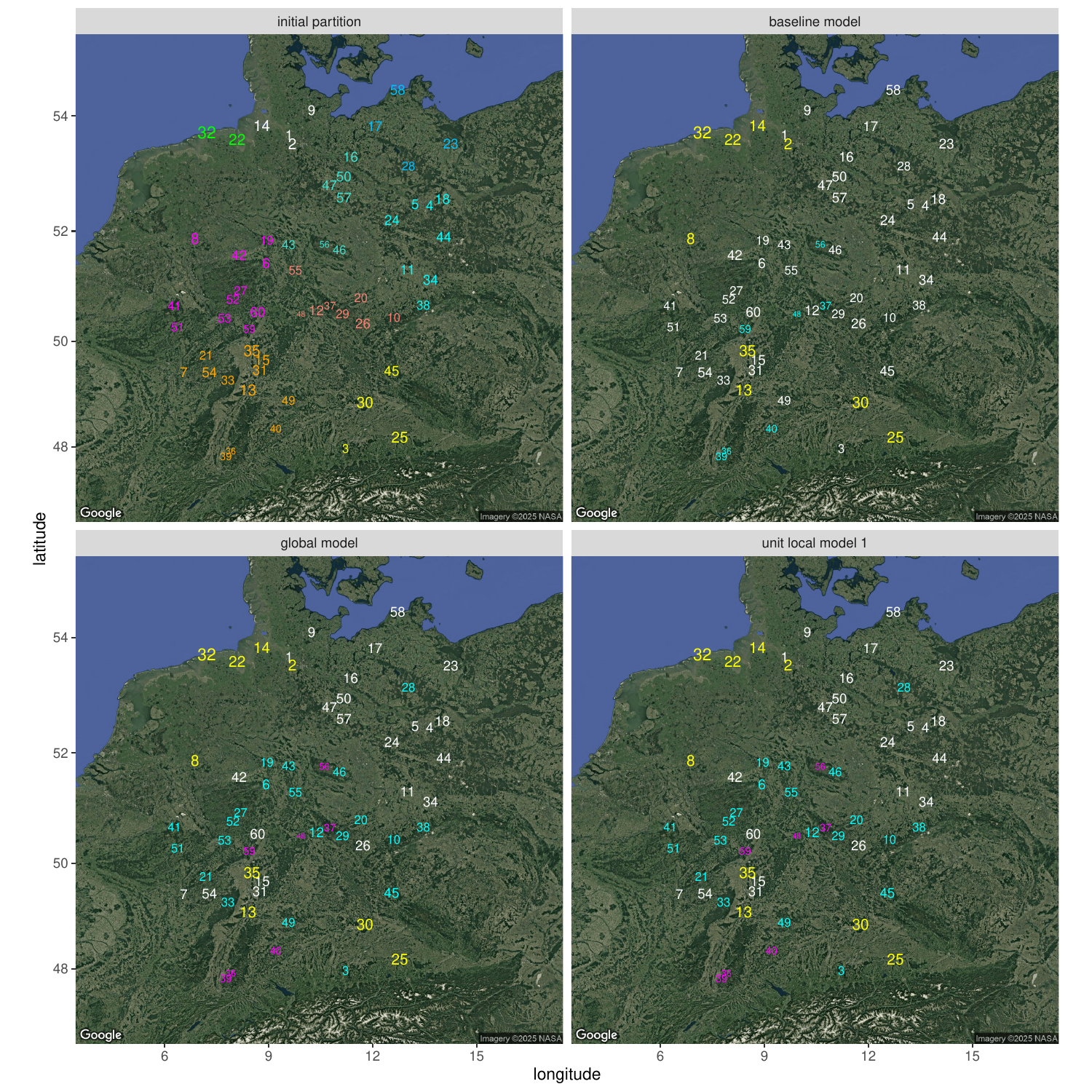}
    \caption{PM$_{10}$ data with number corresponding to station number.  The color in each figure corresponds to cluster membership.  The top left plot displays the initial partition (\fq $\rho_{\text{kmeans}}$\qf). The top right displays a partition estimate based on a model with no initial partition (baseline model). The bottom left displays a partition estimate based on a model using the initial partition and a global prior on $\mathbf {A}$ (global model).  The bottom right displays the partition estimate from a model that includes an initial partition and a local prior for $\mathbf {A}$ (unit local model 1). }
   \label{fig:pm10.ip_kmeans}
\end{figure}

\begin{figure}
  \centering
  \includegraphics[width=1\textwidth,height=0.9\textheight]{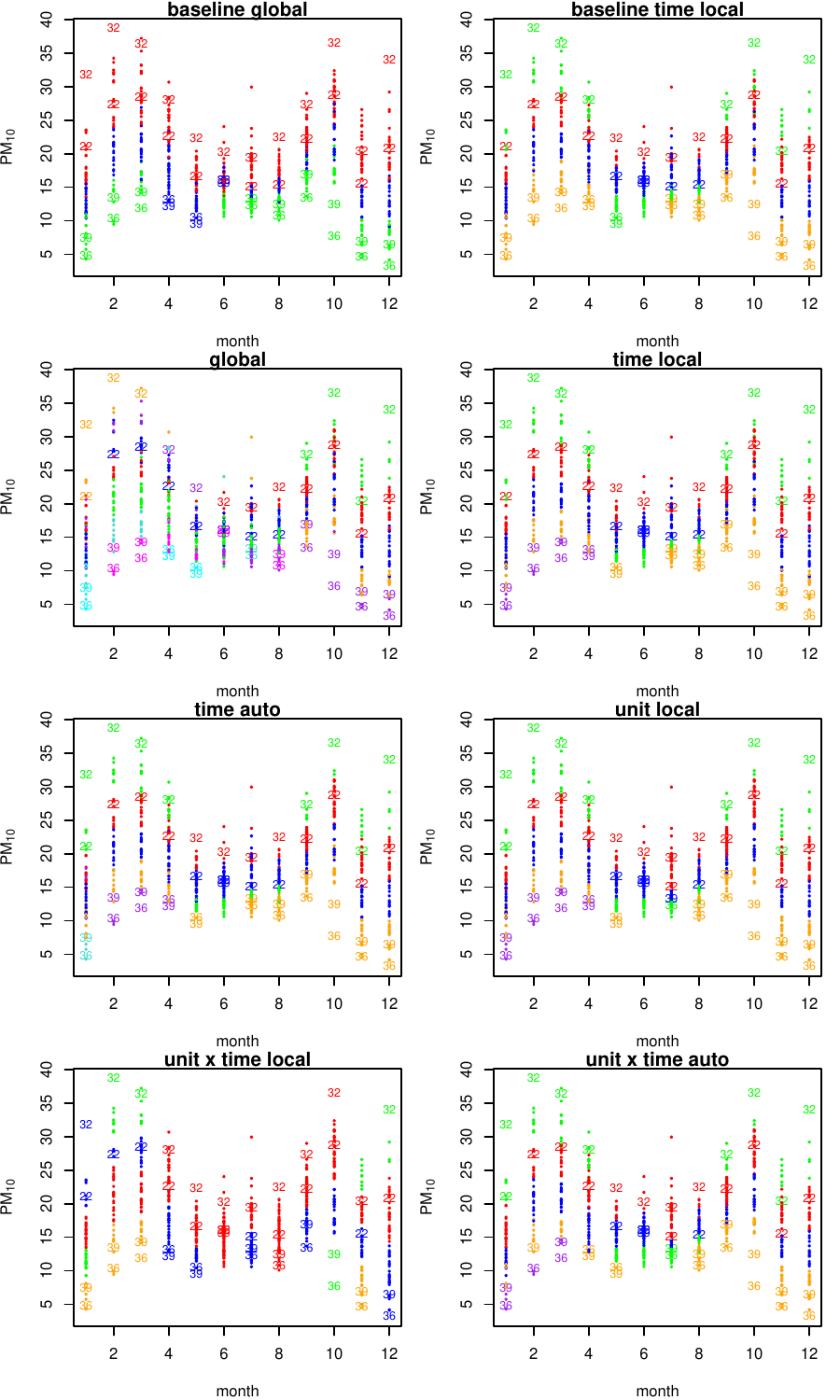}
    \caption{PM$_{10}$ measurements from each of the 60 stations across the 12 months.  Stations 22, 32, 39, and 36 are highlighted using their respective numbers.  The color indicates group membership \fq as estimated from each of the eight models fitted to the PM10 data. \qf}
   \label{fig:station_results}
\end{figure}

Figure \ref{fig:station_results} illustrates this along with partition estimates for each model at each time point (as before partition estimation is carried out using the {\tt salso} {\tt R}-package by \citealt{dahl_etal:2021}).  The first thing to notice from this figure is that the baseline model results in stations 22 and 32 being clustered together at each time point even though there are instances when their measurements are quite different.  This is due to a large estimated $\alpha$ value (95\% interval of (0.83,0.91)) which results in fairly static partitions over time that are based on a CRP type partition at time period 1.  Conversely, introducing $\rho_0$ in the global model results in stations 22 and 32 only being clustered at time periods 1 and 4.  The reason for this is that $\alpha$ is estimated to be smaller compared to the baseline model (95\% interval of (0.75,0.84)).  This is due to the fact that $\rho_0$ and the time 1's PM$_{10}$ measurements only marginally agree necessitating a larger percentage of units to be reallocated at time 1 which in turn requires a smaller value of $\alpha$.  Thus, it appears that when estimating partitions over time is desired, our method provides sufficient flexibility to produce partition estimates with more homogeneous clusters (compared to models that do not include $\rho_0$).

\end{document}